\documentclass[twocolumn]{aastex6}

\usepackage{graphicx}
\usepackage{natbib}
\usepackage{wrapfig}
\usepackage{amsmath}
\usepackage{appendix}
\usepackage{gensymb}
\usepackage{url}
\usepackage{hyperref}

\makeatletter
\newcommand*{\rom}[1]{\expandafter\@slowromancap\romannumeral #1@}
\makeatother

\newcommand{\Msun}{\rm M_{\odot}}
\newcommand{\Gizmo}{{\small\sc Gizmo}}
\newcommand{\Arepo}{{\small\sc Arepo}}
\newcommand{\Gadget}{{\small\sc Gadget}}

\newcommand{\Gasoline}{{\small\sc Gasoline}}

\newcommand{\SUBFIND}{{\small\sc Subfind}}
\newcommand{\FOF}{{\small\sc FOF}}

\newcommand{\RNum}[1]{\uppercase\expandafter{\romannumeral #1\relax}}

\begin{document}                          
\title{The Formation of a Milky Way - size Disk Galaxy --  \\
\RNum{1}. A Comparison of Numerical Methods}
                       
\author{
Qirong Zhu$^{1, 2}$ and Yuexing Li$^{1, 2}$ 
}

\affil{$^1$Department of Astronomy \& Astrophysics, The Pennsylvania State University, 
525 Davey Lab, University Park, PA 16802, USA}
\affil{$^2$Institute for Gravitation and the Cosmos, The Pennsylvania State University, University Park, PA 16802, USA}

\email{Email:qxz125@psu.edu}
\begin{abstract}

The long-standing challenge of creating a Milky Way-like disk galaxy from cosmological simulations has motivated significant developments in both numerical methods and physical models. We investigate these two fundamental aspects in a new comparison project using a set of cosmological hydrodynamic simulations of a Milky Way-size galaxy. In this study, we focus on the comparison of two particle-based hydrodynamics methods: an improved smoothed particle hydrodynamics (SPH) code {\Gadget}, and a Lagrangian Meshless Finite-Mass (MFM) code {\Gizmo}. All the simulations in this paper use the same initial conditions and physical models, which include star formation,  ``energy-driven'' outflows, metal-dependent cooling, stellar evolution and metal enrichment. We find that both numerical schemes produce a late-type galaxy with extended gaseous and stellar disks. However, notable differences are present in a wide range of galaxy properties and their evolution, including  star formation history, gas content, disk structure and kinematics. Compared to {\Gizmo}, {\Gadget} simulation produced a larger fraction of cold, dense gas at high redshift which fuels rapid star formation and results in a higher stellar mass by $20\%$ and a lower gas fraction by $10\%$ at $z = 0$, and the resulting gas disk is smoother and more coherent in rotation due to damping of turbulent motion by the numerical viscosity in SPH,  in contrast to the {\Gizmo} simulation which shows more prominent spiral structure. Given its better convergence properties and lower computational cost, we argue that MFM method is a promising alternative to SPH in cosmological hydrodynamic simulations.

\end{abstract}

\section{Introduction} 

Numerical simulations have been playing an indispensable role in theoretical studies of the formation and evolution of cosmic structures. The current state-of-the-art cosmological hydrodynamic simulations such as the Illustris Simulation \citep{Vogelsberger2014b} and the Eagle Simulation \citep{Schaye2015} have significantly advanced our understanding of the highly {nonlinear} processes of galaxy formation. However, while $N$-body simulations performed with different codes produced consistent properties of the dark matter halos (e.g., \citealt{Diemand2008, Springel2008, Stadel2009, Kim2014}) and large-scale structures (e.g., \citealt{Springel2012, Kuhlen2012}), hydrodynamic simulations using different numerical methods and physical models have not been able to achieve such a consensus. On the cosmic scale,  while the Illustris Simulation, which used the moving-mesh code {\Arepo}  \citep{Springel2010arepo}, and the Eagle Simulation, which used a variant of the smoothed particle hydrodynamics (SPH) code {\Gadget} \citep{Gadget1, Gadget2}, show overall agreement on the global star formation history \citep{Sparre2015, Schaye2015} and stellar mass functions \citep{Genel2014, Furlong2015}, significant discrepancies remain in a number of galaxy properties such as galaxy color bimodality \citep{Vogelsberger2014b, Trayford2015} and cluster gas fraction \cite{Genel2014, Schaye2015}. On the galactic scale, 13 hydrodynamic simulations using different numerical codes and physical processes but the same initial conditions in the Aquila Comparison Project \citep{Scannapieco2012} have been unsuccessful in producing a realistic disk galaxy such as the Milky Way (MW), and they showed large simulation-to-simulation variations in a wide range of galaxy properties including morphology, star formation, stellar mass, gas fraction and rotational curve. These studies highlight the challenges in galaxy simulations and uncertainties in hydrodynamic solvers and physical models.

In order to understand the effects of numerical methods and physical models on the resulting galaxy properties in cosmological simulations, it is critical to perform comprehensive and systematic comparisons. In a code comparison of ideal tests by \citep{Agertz2007}, it was reported that SPH failed both the ``blob'' and the Kelvin-Helmholtz (KH) instability tests.  For the ``blob" test, SPH codes are unable to disrupt the dense cloud exposed to a supersonic wind. For the KH test, SPH codes tend to suppress the growth of {nonlinear} structure on the contact surface of two fluids and prevent the fluids from mixing. The presence of a ``pressure blip'' in across contact discontinuities in SPH was shown to be responsible for the erroneous behavior in these two tests. These problems of SPH originate from errors in the discreteness and smoothing processes. The approximation to derive the SPH formulation of Euler equations exhibits zeroth order errors in estimates of density and momentum \citep{Zhu2015}. In the subsonic regime where characteristic gas velocity is less than the sound speed, it was reported by \cite{Bauer2012} that SPH dissipates the large-scale coherent motion into heat too quickly rather than {cascading} into smaller scale turbulences. Similar inertial ranges are not recovered in SPH simulations while the {\Arepo} produces results more consistent with analytical theory. Moreover, the errors from SPH gradient estimate are found to be many orders of magnitude higher than those from {\Arepo} simulation. The large errors in the pressure gradient estimate {are} responsible for the poor behavior of SPH in the subsonic regime.   

More recently, a number of studies have compared hydrodynamics in galaxy formation simulations with {\Gadget} and {\Arepo} codes (e.g., \citealt{Sijacki2012, Keres2012, Vogelsberger2012, Torrey2012, Nelson2013}). It was found that cosmological accretion shocks are less accurately resolved and there is substantial overheating in the post-shock region in SPH simulations \citep{Sijacki2012, Nelson2013}. The cooling rate of {the} hot gas is substantially underestimated in SPH simulations, which results in lower star formation rates in massive halos \citep{Nelson2013}. These differences are consistent with the behaviors of SPH identified by \cite{Bauer2012} in their subsonic turbulence tests. The gas disks in the massive halos are more clumpy with smaller scale {lengths} in {\Gadget} simulation \citep{Vogelsberger2012, Torrey2012}. Furthermore, a large population of  ``cold blobs'' fail to dissolve in hot virialized gas in {\Gadget} simulations due to the ``surface tension" in SPH. The ``cold blobs" problem is a serious shortcoming of SPH since it is not present in simulations with {\Arepo} or other static grid-based codes. SPH was also found  by \cite{Okamoto2003} to cause spurious angular momentum transfer between hot- and cold-phase gas. Special treatments have been proposed by  \cite{Okamoto2003, Marri2003} and \cite{Scannapieco2006} to circumvent the spurious angular momentum transportation by decoupling hot- and cold- phases and by \cite{Read2010, Saitoh2013} and \cite{Hopkins2013} to prevent this numerical artifact by using density independent formulation of SPH. 

In order to solve these problems of SPH, a number of fixes have been proposed in the literature. One common feature of the improved SPH codes is to use a smoother kernel function with a larger number of neighbors to reduce its low-order errors. More sophisticated switches for dissipation were proposed to improve its shock capturing ability \citep{Cullen2010, Read2010}. On the other hand, more radical changes  have been proposed by \cite{Gaburov2010} and \cite{Hopkins2015} to form a new class of Lagrangian meshless methods. The new scheme is designed to cure the error in the partition of volume by means of slope-limited reconstruction and Riemann solvers. The new meshless code {\Gizmo} developed by \cite{Hopkins2015} shares great similarities with the tradition finite-volume methods, it eliminates the need for traditional SPH and has been shown to give great performance over a large body of test problems.
  
The formation of a realistic MW-like disk galaxy provides a critical test of the numerical methods and physical processes in galaxy simulations. Recently, exciting progress has been made in the simulations of disk galaxies in the standard dark energy-cold dark matter ($\rm{\Lambda}$CDM) cosmology  \citep[e.g.][]{Guedes2011, Agertz2011, Scannapieco2012, Stinson2013a, Aumer2013, Marinacci2014}. In particular, the Aquila Comparison Project \citep{Scannapieco2012} demonstrated that the galaxy produced by  {\Gadget} and {\Arepo} codes differed in mass by a factor of two, among other different kinematic and structural properties.However, since the treatments of gas cooling, star formation and feedback were implemented differently in each code, these studies pointed to very different key ingredients for the creation of realistic disk galaxies. 

In a new comparison project, we aim to test the two fundamental aspects, numerical methods, and physical models, of galaxy simulations separately. Given the huge popularity of SPH codes in the astrophysics community (as reflected by 2000+ citations of {\Gadget} and 500+ citations of {\Gasoline}), it is extremely useful and urgent to test these developments on the formation of a realistic galaxy. In addition, certain errors originated from mesh noise have been reported recently in {\Arepo} simulations \citep{Pakmor2015, Mocz2015}, it is important to compare the performance of different numerical schemes on galaxy formation. To this end, we have significantly improved the {\Gadget} code with a number of implementations including time-dependent conduction, time-dependent artificial viscosity, higher order smoothing kernel, and adaptive time step limiter. On the other hand, we have also implemented a comprehensive list of physical processes in both {\Gadget} and {\Gizmo} codes, including ``energy-driven'' outflows, {metal-dependent} cooling, stellar evolution and metal enrichment from stellar evolution. 

We perform a series of cosmological hydrodynamic simulations of the same MW-size galaxy (halo C) as in the Aquila Comparison Project using the improved {\Gadget} and {\Gizmo} codes and various physical models, in order to (1) test the performance of the new numerical schemes; (2) identify important physical processes that determine the galaxy properties; and (3) cross check with other studies to understand limitations of both particle- and grid-based codes and minimize numerical artifacts in galaxy simulations.  

In this work (Paper I), we focus on the comparison of numerical methods using the improved {\Gadget} and {\Gizmo} codes, as well as those by \cite{Marinacci2014} using {\Arepo}  but with the same initial conditions and physical models. In a companion paper (Paper II, Zhu et al in prep), we will focus on the comparison of physical processes by systemically varying the model parameters {in order to identify the key} processes leading to the formation of a disk galaxy. 

The paper is organized as follows. In Section~\ref{sec:methods}, we describe the numerical methods including the {\Gadget} and {\Gizmo} codes and improvements made to {\Gadget}, physical models implemented in both codes, and the galaxy simulations. In Section~\ref{sec:results}, we compare galaxy properties from different simulations, including galaxy morphology, gas properties and evolution, star formation and assembly of stellar mass, and kinematics of the stellar disk. We discuss resolution studies and the future of SPH in Section~\ref{sec:discussion}, and summarize our findings in Section~\ref{sec:summary}.

\section{Methods}
\label{sec:methods}

In this study, we perform a set of cosmological hydrodynamic simulations of a MW-size galaxy using two different codes, the improved SPH code {\Gadget} over the version of \cite{Springel2005}, and the new meshless code {\Gizmo} developed by \cite{Hopkins2015}. We have implemented in both codes a list of physical processes following the galaxy formation model of \cite{Vogelsberger2013}.  In order to explore the effects and differences produced by different hydrodynamic solvers, both {simulations} are run with the same initial conditions and the same physical models. We also compare these simulations with the one obtained by  \cite{Marinacci2014} using the moving-mesh code {\Arepo}.  

\subsection{Hydrodynamic Codes}

\subsubsection{The Improved-{\Gadget} Code}

We have incorporated a number of important fixes of SPH into the original {\Gadget} code ({a} version of \cite{Springel2005}), and refer it as ``Improved-{\Gadget}". Here we list the major implementations and the purposes of these modifications. 

\paragraph{Smoothing Kernel Function}

The smoothing kernel $W(\mathbf{r}, h)$, which is critical to SPH technique, is not constrained to any specific curve but free to choose from a large set of curves. It is predominantly used in an isotropic form $W(|\mathbf{r}|, h)$ with exceptions such as \cite{Owen1998}. In general, one would choose a smooth bell-shaped curve for a good density estimation \citep{Price2012review}. \cite{Dehnen2012} further listed several important requirements for the smoothing functions such as the overall shape, the smoothness, and dynamical stability. Generally speaking, increasing the smoothness of the smoothing kernel can have {a} direct improvement on the accuracy of the results \citep{Rasio2000,Price2012review,Dehnen2012,Hu2014}. Both of the major two error terms, the smoothing error term $error_{\rm s}$ (from the smoothing of quantities in SPH) and the discretization error term $error_{\rm d}$ (from the discretization of the integral into finite sum), can be reduced with a smoother function. We use a bias-corrected Wendland $C^4$ function suggested by \cite{Dehnen2012}, as the smoothing kernel function. For the choice of {the} number of neighbors, we use $\rm{N_{nb}  = 200}$ in this work. We will return to a discussion of $\rm{N_{nb}}$ later.

\paragraph{Adaptive Time Steps}   

We use the time integration scheme in \cite{Durier2012} and find that it greatly improves the performance in test problems involving strong shocks.  This integration scheme limits the difference in time steps between neighboring particles within a factor of 4 and brings the inactive particles to active when their neighboring particles just have had a sudden change in their dynamical state. A similar approach was adopted in {\Arepo} code \citep{Springel2010arepo}. Without such modification, SPH cannot correctly handle the problem in the test involving strong blast waves \citep{Saitoh2009, Durier2012}. 

\paragraph{Improvement of SPH in Fluid Mixing}

A spurious pressure on the interface of multiphase gas in the KH test was found to be responsible for the suppression of the fluid mixing as it creates a repulsive force \citep{Agertz2007}. As demonstrated in \cite{Springel2010review}, this ``surface tension'' effect can have undesirable effects on the dynamics of the system, such as suppressing the fluid instabilities and placing the system into an artificial minimum energy state.  
 
To cure this artifact, \cite{Price2008} and \cite{Wadsley2008} proposed a diffusion term to actively smooth the internal energy on the boundary of two phases. We have implemented and tested the approach by \cite{Price2008} to use an artificial conduction (AC) term in the integration of {the} thermal energy of the SPH particles. However, we find that it introduces undesired dissipation in many test cases. Alternatively, \cite{Ritchie2001} suggested the use of a smoothed density variable weighted by the internal energy of SPH to deal with multi-phase flows. \cite{Saitoh2013}  and \cite{Hopkins2013} further developed this idea into an alternative formulation of the equation of motion based on gas pressure. Given the test problems presented in \cite{Saitoh2013}  and \cite{Hopkins2013}, this density independent form of Pressure-Entropy SPH (P-SPH) formulation shows {{much improved behaviors} in the multiphase flow tests. For the above reasons, we choose the Pressure-Entropy formulation in \cite{Hopkins2013} as the fix to fluid mixing in SPH. 

\paragraph{Time-dependent Artificial Viscosity}

SPH uses artificial viscosity to treat shocks by generating entropies in the shocks and to reduce the oscillations in the post-shock region. In most of the SPH galaxy formation simulations, a constant viscosity parameter is often used. Artificial viscosity has a large impact on the fluid dynamics in the subsonic regime \citep{Bauer2012} and on the angular momentum transfer in the shear flows as demonstrated by \cite{Okamoto2003} and \cite{Cullen2010}. \cite{Price2012turbulence} further argued that a time-dependent viscosity is critical to resolve subsonic turbulence with SPH, as also agreed by \cite{Bauer2012}. It is thus crucial to reduce artificial viscosity away from shocks in SPH while at the same time able to keep the SPH's ability to detect shocks. We therefore implement the time-dependent artificial viscosity scheme proposed by \cite{Cullen2010} for these purposes.

\subsubsection{Meshless  Code {\Gizmo}}

As pointed out by \cite{Zhu2015}, the most concerning aspect for SPH is its slow convergence rate. Other problems such as surface tension, artificial viscosity and fluids mixing can be improved with the {above mentioned} SPH fixes developed over the past few years. However, the noisy behavior of SPH has a deeper root in its formulation. Various test problems indicate the convergence rate of SPH is quite slow, indicating some Monte Carlo behavior. We showed in \cite{Zhu2015} that the cause of this slow convergence rate is the violation of partition of unity. The ``volume'' in SPH is not conserved so that a constant scalar field cannot be reproduced due to particle disorder. This introduces {zeroth-order error term}. As a result, SPH shows some degree of fluctuation, even with a regular particle distribution, around the true value. Of course, such fluctuation can be reduced with a larger number of neighbors. However, this indicates that in order to achieve true convergence in a consistent manner, one needs to increase the number of neighbors systematically as the resolution increases. We showed that such requirement is a strong function of the total SPH numbers. (This can be achieved with Wendland functions such that pairing instability will not appear.) Otherwise, a constant number of neighbors will inevitably give only statistical converging rates. 

Thus, a necessary step to improve the traditional SPH method is to include a more accurate volume estimate to ensure a partition of unity. It is also highly desirable for galaxy formation and evolution simulations to follow the mass in a Lagrangian manner as in SPH. This can be achieved with an unstructured moving Voronoi mesh as in {\Arepo} code \citep{Springel2010arepo}. Alternatively, one can also explicitly correct the zeroth order error in SPH with renormalization. Since the kernel function is symmetric, the error in the volume (density) estimate would now be reduced to second order. \cite{Lanson2008} presented a meshless finite volume scheme by noticing the similarities between {the} partition of unity method and SPH formulation. Now  $W(|\mathbf{r}|, h)$ essentially becomes a partition of unity function. Subsequently, \cite{Gaburov2010} implemented this method into an astrophysical MHD code demonstrating its ability solving complex astrophysical problems even in the presence of magnetic field. Dissipation can be treated with the same Riemann solvers in traditional grid codes without using any artificial viscosity or artificial conductivity terms. The control of dissipation is handled in reconstruction on the interface states and in Riemann solvers. Moreover, ``surface tension" in SPH is no longer present as the treatment of density and internal energy are consistent in the context of finite volume method with {a} slope-limited reconstruction of left/right states. 

\cite{Hopkins2015} implemented this finite volume method into {\Gizmo} code (Meshless Finite-Volume) and develops {a} variant of this method by enforcing no mass flux between two mass points (Meshless Finite-Mass). These two methods solve the Riemann problem in a local co-moving frame of particle pairs. As a result, it inherits all the major advantages of SPH method in mass, linear momentum, angular momentum and energy conservations. Advection error is greatly reduced and no longer a function of bulk velocity as in static grid codes. Galilean invariance is also guaranteed with such Lagrangian solvers. In many aspects, the new meshless code is very close to the moving mesh code {\Arepo} but with some minor differences in density and local gradient estimate. We refer the readers to the code papers for the underlying mathematics of these two methods and their outstanding performances in various test problems over the traditional methods including SPH and adaptive mesh refinement (AMR). 

\subsection{Physical Processes for Galaxy Formation}
\label{subsec: model}

\subsubsection{Gas Cooling and Star Formation Model}
The most important aspect of hydrodynamic simulations of is to follow the self-gravity, radiative cooling, star formation and feedback. We use the sub-grid multi-phase interstellar medium (ISM) model of \cite{Springel2003} for star formation and thermal feedback. For each {star-forming} gas particle, the pressure comes from {the} multi-phase model that the balance between the evaporation of cold gas and cooling of hot gas that is explicitly solved in a coupled equations. This approach naturally gives an additional pressure support against self-gravity. Moreover,  \cite{Springel2003} showed that this method has good numerical convergence properties. We adopt a Chabrier initial mass function (IMF) with the stellar mass ranging between $\rm{[0.1, 100]}  \rm{M_{\odot}}$. Stars more massive than 8.0$\rm{M_{\odot}}$ will explode as Type \rom{2} supernovae and return the mass and metals into the surrounding ISM. The number density threshold for star formation is 0.15 $\rm{cm^{-3}}$, which is commonly used in such cosmological simulations. These numbers are taken from the parameters listed in \cite{Vogelsberger2013} in order to have a direct comparison with the MW simulations by \cite{Marinacci2014}. 

We include {a} metal-dependent term in gas cooling calculations. \cite{Wiersma2009} showed that both metals and meta-galactic ionization background are critical to computing the cooling rate up to an order of magnitude. {A similar} approach was also adopted in recent hydrodynamic simulations by \cite{Shen2010}, \cite{Vogelsberger2013} and \cite{Schaye2015}. We adopt the metal cooling model of \cite{Wiersma2009} and use their tables to interpolate gas cooling rates based on a grid of specific internal energy, gas density, {redshift, and overall metallicity. Following \cite{Wiersma2009b}, we calculate the ``metal mass density" based on the metal mass $m_{Z}$ as 
\begin{equation}
\rho_{Z} = \sum_{N_{\rm nb}} m_Z W(|\mathbf{r}|, h)
\end{equation}
and use a smoothed metallicity $\rho_Z/\rho$ as the overall metallicity to compute gas cooling rate.}

\subsubsection{Stellar Evolution and Chemical Enrichment}

\begin{figure}
\begin{center}
\includegraphics[scale=0.4]{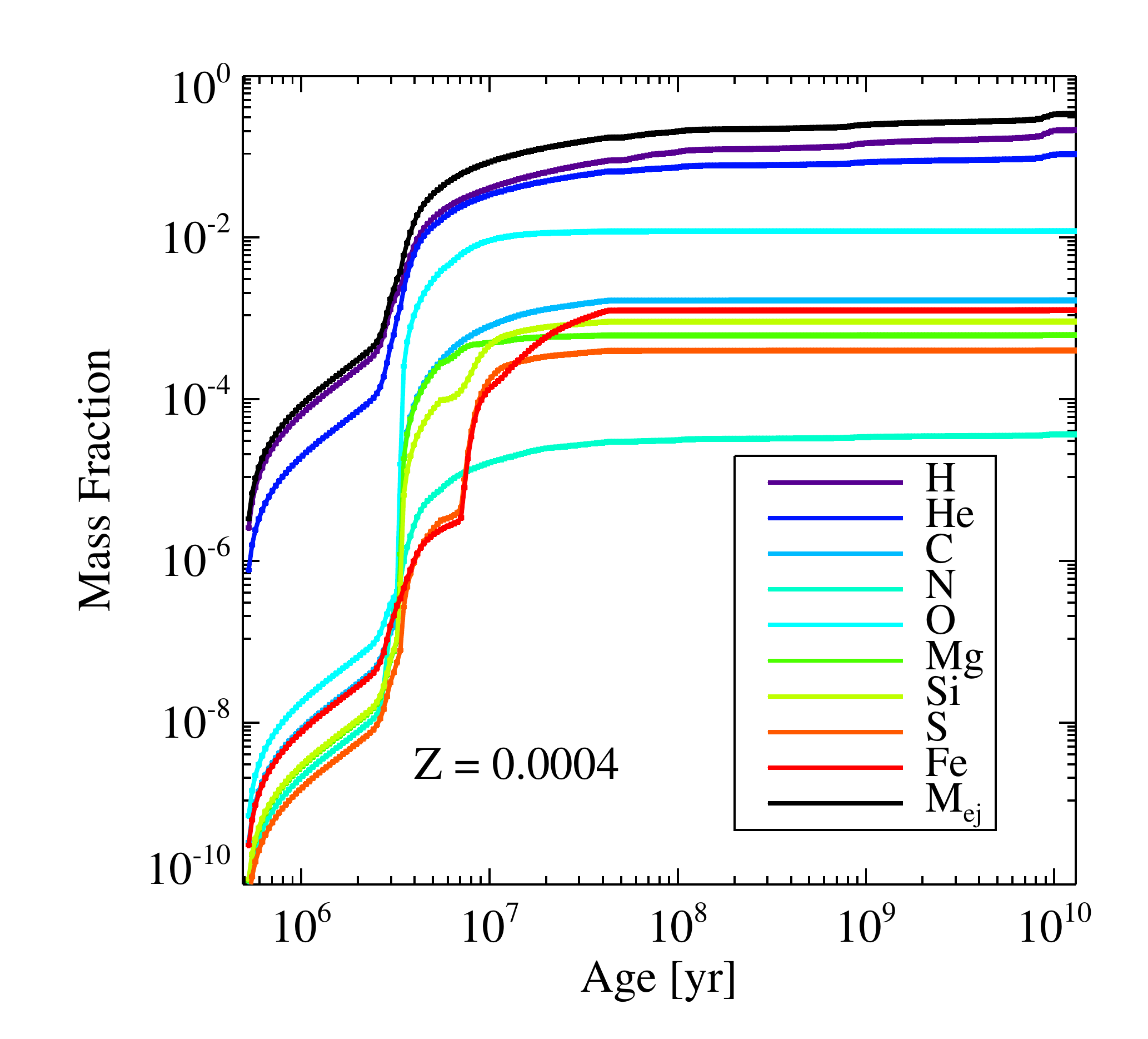}
\caption{\label{fig:enrichment} Cumulative mass  return fraction of individual element (colored curve) and the total ejecta (black curve) as a function of 
age for a stellar population with mass $\rm{M_{0} = 10^{6} M\odot}$ and metallicity $\rm{Z = 0.0004}$. 
Early stage of the mass return and chemical enrichment is from Type {\small\sc{II}} SNe and later stage is due to AGB stars. Note that Type Ia SNe is not included in this figure. For the chemical enrichment calculations in the galaxy simulations, a metallicity grid Z$=$0.0004, 0.004, 0.008, 0.02 to 0.05 is considered. } 
\end{center}
\end{figure}

We model the impact of stars from their mass return and chemical enrichment onto ISM. Chemical enrichment would also enhance the gas cooling rates providing {coolants} such as oxygen and iron. As massive stars quickly evolve into their late stage within a few million years, they return a significant fraction of mass and heavy metals into the surrounding ISM in addition to the energy released from supernovae (SNe) explosions. For intermediate- and low-mass stars, the asymptotic giant branch (AGB) is an important phase to lose their mass and metals to the ISM. Overall, stellar evolution and chemical enrichment in the simulation are coupled processes involving the age, masses and metallicities of the stars, and they have a major, continuous impact on the ISM and star formation on the timescale of the entire simulation. When a new star particle is formed in hydro simulation, it inherits the metallicities of the parent gas particle. 

To model stellar evolution and chemical enrichment, we treat each star particle as a simple stellar population with an initial total mass of $\rm{10^6 M\odot}$, and a metallicity ranging from Z$=$0.0004, 0.004, 0.008, 0.02 to 0.05. A Chabrier initial mass function is assumed. Then each stellar population is evolved with stellar population synthesis  (SPS) code Starburst 99 \citep{Leitherer1999} with Padova models (including the contribution of AGB stars) to produce tables of enrichment from both Type {\small\sc{II}} SNe and AGB stars. Figure~\ref{fig:enrichment} shows an example of the cumulative mass fraction for various element returned to ISM as a function of stellar age for Z$=0.0004$.  In this example, about 30$\%$ of the initial mass of the stellar components is returned to the ISM within {1 Gyr}. 

At each time step, we tabulate the returned  mass and chemical enrichment from active star particles by interpolating on the stellar age--metallicity grid. {The returned gas mass and metal of individual and total elements} from the evolved stars is then redistributed into the neighboring gas particles based on {an SPH weight $$w = \frac{W(|\mathbf{r}|, h)}{\sum_{N_{\rm nb}}{W(|\mathbf{r}|, h)}}, $$ where $h$ is the smoothing length of each star particle which we also include during the normal density calculation routine for gas component. When searching for neighbors of each active star particles, we do not distinguish between normal gas particles or wind particles such that mass return and metal enrichment are treated equally for normal gas and wind particles, which is simpler than the reduced metal loading model used in \cite{Vogelsberger2013}.} Type Ia SNe is included as an important channel of iron production. Similarly to \cite{Vogelsberger2013}, we assume a simple power-law (with a slope $s = -1.12$) delay time distribution for Type Ia mass return and chemical enrichment. Type Ia SNe start to contribute 40 {Myrs} after the onset of star formation. The total mass and individual element mass from Type Ia SNe follow the calculation of  ``W7'' model in \cite{Iwamoto1999}. Running Starburst99 code, we also obtained broadband photometries as a function of metallicity and stellar age, which allows appropriate color rendering when making the projected stellar images. 

\subsubsection{Feedback Processes}

The galactic outflow has a profound impact on the galaxy properties. The complex interplay between the outflows and the inflows crucially determines the gas supply to the galaxy and the amount of gas available for star formation. From abundance matching between $N$-body simulations and observed galaxy mass functions, \cite{Moster2013} found that star formation efficiency (SFE) peaks in halos with {a} mass at $\rm{10^{12}  M_{\odot}}$ and it drops at both higher- and lower-mass ends. The exact origin of such a SFE--mass relation remains poorly understood, but it is generally believed that it is due to supernovae feedback in the low-mass end and active galactic nuclei (AGN) feedback in the massive end. However, explaining of this efficiency dependence on the mass of dark matter halo from the ``first principles" with resolved physical processes are still computationally impracticable (but see recent efforts such as \citealt{Hopkins2013b}). 

For galaxies less massive $\rm{10^{12}  M_{\odot}}$, supernovae feedback is considered to play the most important role shaping galaxy stellar mass.  Often in cosmological hydro simulations, the mass resolution is not adequate to directly resolve the formation of giant molecular clouds and accurate the propagation of blast waves from SN explosions. Instead, parameterization of certain wind generating processes or other numerical techniques such as disabling gas cooling around SNe is widely adopted. In a subgrid multi-phase ISM model, an ``energy-driven" wind model proposed by \cite{Okamoto2010} and \cite{Puchwein2013} is  a phenomenological approach to address the observed low star formation efficiency in the low mass halos. We implement this model into our codes, assuming $70\%$ of the supernovae energy is available to launch the wind in two polar directions. \cite{Okamoto2010} and \cite{Puchwein2013} show that such model can well reproduce the mass function {at} the low mass end. Recent high resolution large scale simulation the Illustris project \citep{Vogelsberger2014b, Genel2014} also shows such model is able to reproduce observed galaxy populations from high redshift to the local universe. There is still some uncertainties on the low mass end \citep{Genel2014}. 

To compute the properties of the wind, \cite{Puchwein2013} assumes wind velocity has a direct dependence of the escape velocity of the dark matter halo.  Wind velocity $v_{w}$ is estimated from the local dark matter velocity dispersion $\sigma^{1d}_{\rm DM}$ computed a coefficient $v_{w} = 5  \sigma^{1d}_{\rm DM}$, which is slightly higher than $3.7$ used in \cite{Vogelsberger2013}. Once the desired wind velocity is known, we compute the mass loading parameter of the wind for each {star-forming} gas particle given the available driven energy from SNe and the probability of each particle turning into star or wind particle from the procedure outlined in \cite{Vogelsberger2013}. This modification helps to regulate the correct stellar mass and wind mass generation. 

Once {a} particle is flagged as {a wind particle}, we temporarily decouple it from hydrodynamics as in \cite{Springel2003}, \cite{Okamoto2013} and \cite{Vogelsberger2013} to allow the wind particles to travel up to a certain distance away from the {star-forming} region. Once the gas density around the wind particle is a factor of $0.1$ of the density threshold for star formation, we recouple the wind particles to hydrodynamics. The values of the available supernovae energy to power the outflow and the scaling between the wind velocity and the local dark matter velocity dispersion are slightly different from the ones in \citep{Vogelsberger2013}. Some quantitive difference between our simulation and \cite{Marinacci2014} is expected due to our choice of the outflow model parameters. Nevertheless, we keep the same parameters of the outflow for all of our simulations in this study. As a result, the impacts of resolution and the choices of hydro solver can be well studied.  

To address the interplay between numerical effects and physical processes, it is desirable to use a numerically {robust} model. One advantage of the above multi-phase ISM model treating star formation and its feedback, both in thermal and kinetic forms, is its very weak dependence on numerical resolutions. This subgrid model has been shown to have {a} good convergence of the cosmic star formation histories with varying resolutions \citep{Springel2003, Vogelsberger2013, Okamoto2013}. 

\subsection{Initial Conditions and Simulations}

We use the initial conditions of Aquila comparison project\footnote{\url{http://www.aip.de/People/cscannapieco/aquila/ICs/Gadget/}}. Besides the original Aquila comparison project, there are a number of other studies using the same initial condition of Aquila C-5 halo as in \cite{Scannapieco2012} including the investigation of the satellite galaxy properties \citep{Okamoto2010, Sawala2012, Wadepuhl2011} and the formation and evolution of the central galaxy \citep{Okamoto2013, Aumer2013, Marinacci2014}. We intend to compare our simulations with the simulations in Aquila project and by \cite{Marinacci2014}. As pointed by \cite{Aumer2013}, this halo itself is very sensitive to the numerical parameters they experimented. This is likely due to the fact that this halo is has a virial mass of $\sim 1.6\times10^{12} \Msun$, which is {at} the high mass end of recent hydrodynamic simulations of Milky Way-sized galaxies and has assembled most of its mass at very high redshift. 

The cosmological parameters used  in this study are the same as those in \cite{Scannapieco2012} and \cite{Marinacci2014}: $\Omega_M = 0.25,  \Omega_b=0.04,   \Omega_\Lambda=0.75,  \sigma_8=0.9,  n_s = 1$ and a Hubble constant  $H = 100 h = 73 \rm{km s^{-1} Mpc^{-1}}$. At level-5, the mass resolution is $0.4 \times10^6 \rm{M_{\odot}}$ for gas particles  and  $2.2 \times10^6 \rm{M_{\odot}}$ for dark matter particles in the high resolution zoom-in region. The mass resolution at level-6 is eight times coarser. For the gravitational softening length, we use $0.5 h^{-1} \rm{kpc}$ at level-5 and $1 h^{-1} \rm{kpc}$ at level-6 for dark matter particles, gas and star particles in the high resolution region in co-moving coordinates.  A lower limit at $0.20$ of the softening length is imposed on the gas particles when computing the smoothing length in both Improved-{\Gadget} and {\Gizmo} simulations. We evolve the simulations from the initial conditions from $z = 127$ to $z = 0$. For each snapshot, a friend-of-friend ({\FOF}) group finder is firstly used to identify the groups and then {\SUBFIND} is used to identify the gravitationally bound structures.

\section{Results}
\label{sec:results}

\subsection{The Assembly of A Disk Galaxy}
\label{subsec:morphology}

\begin{figure*}
\begin{center}
\begin{tabular}{cccc}
\resizebox{1.35in}{!}{\includegraphics{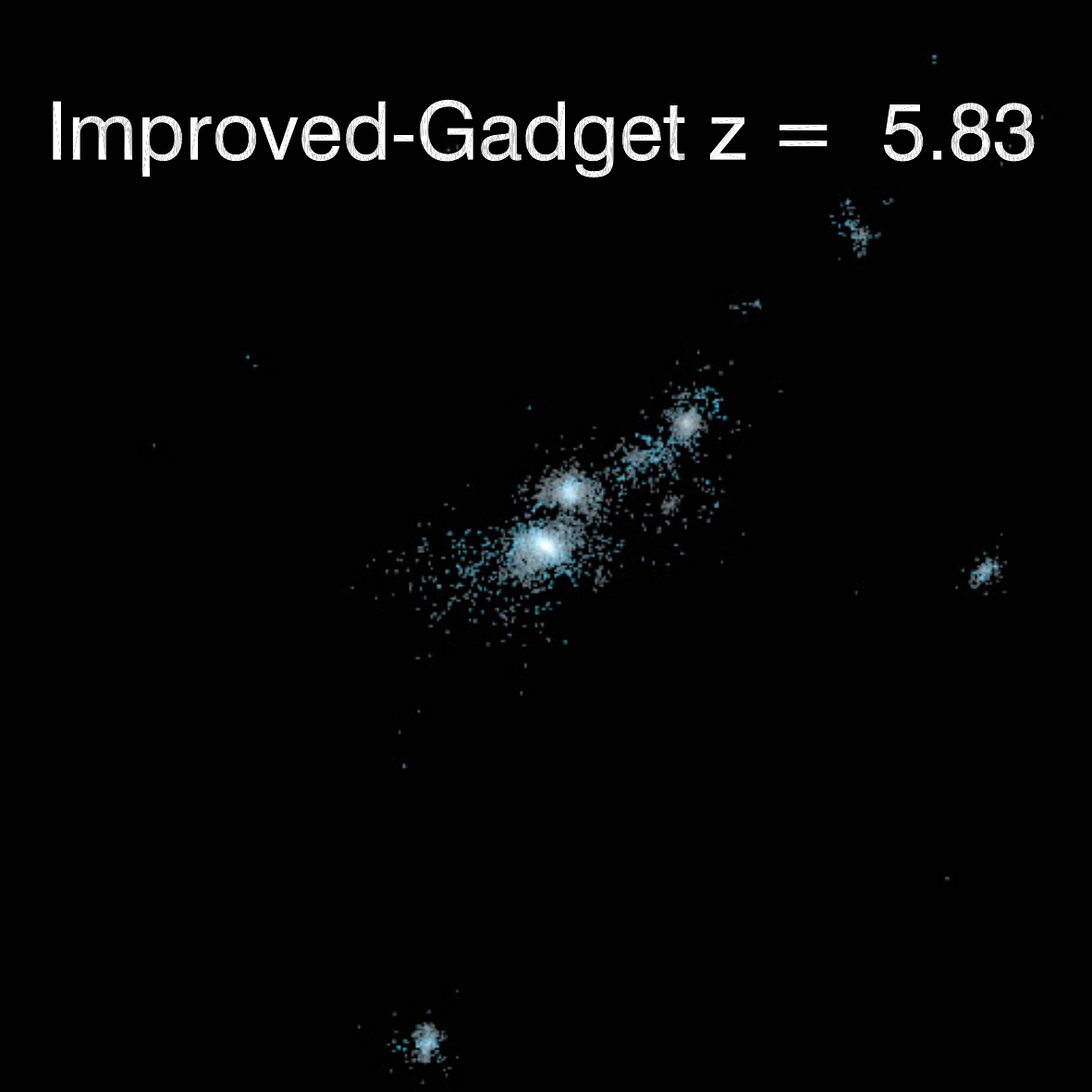}}
\resizebox{1.35in}{!}{\includegraphics{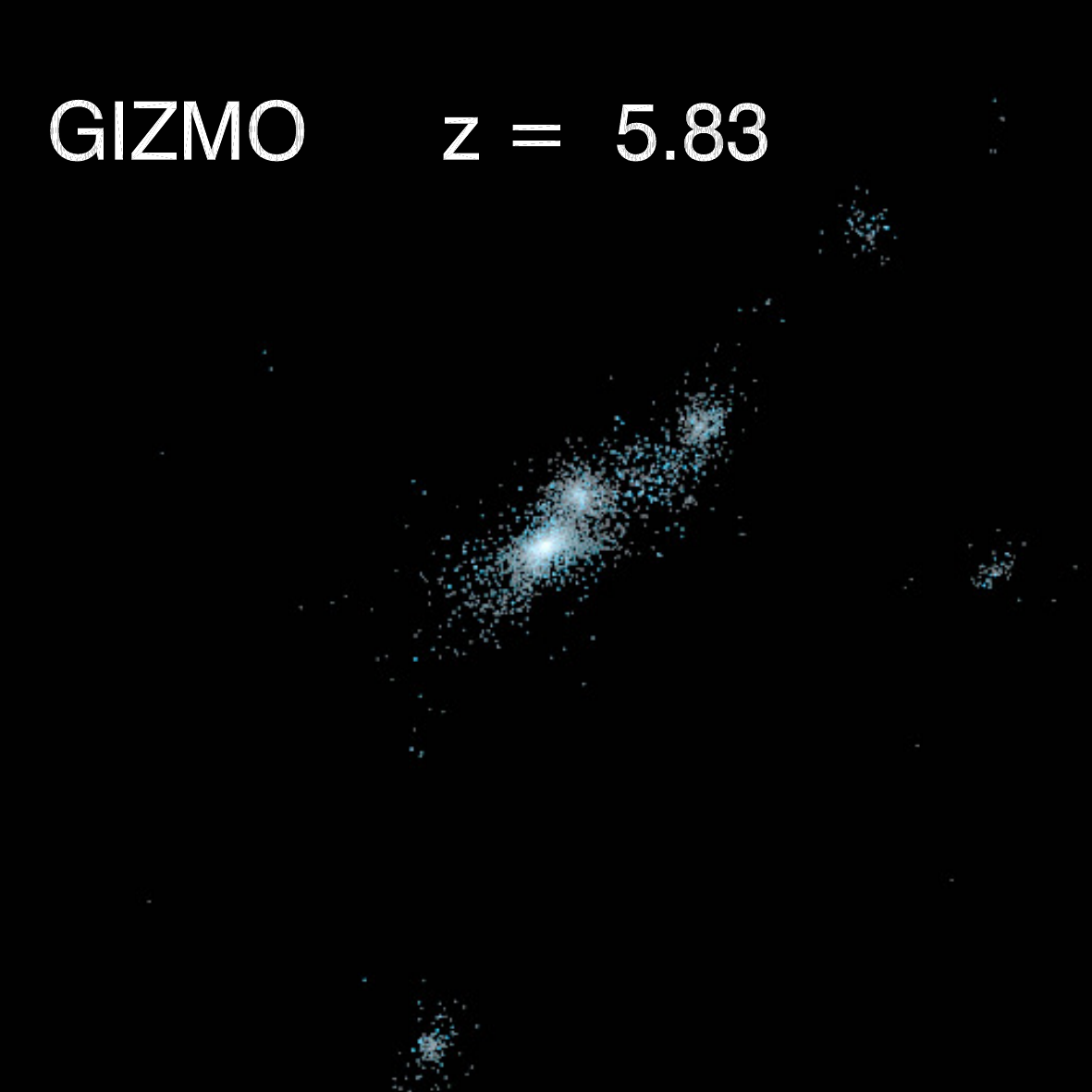}}
\resizebox{1.35in}{!}{\includegraphics{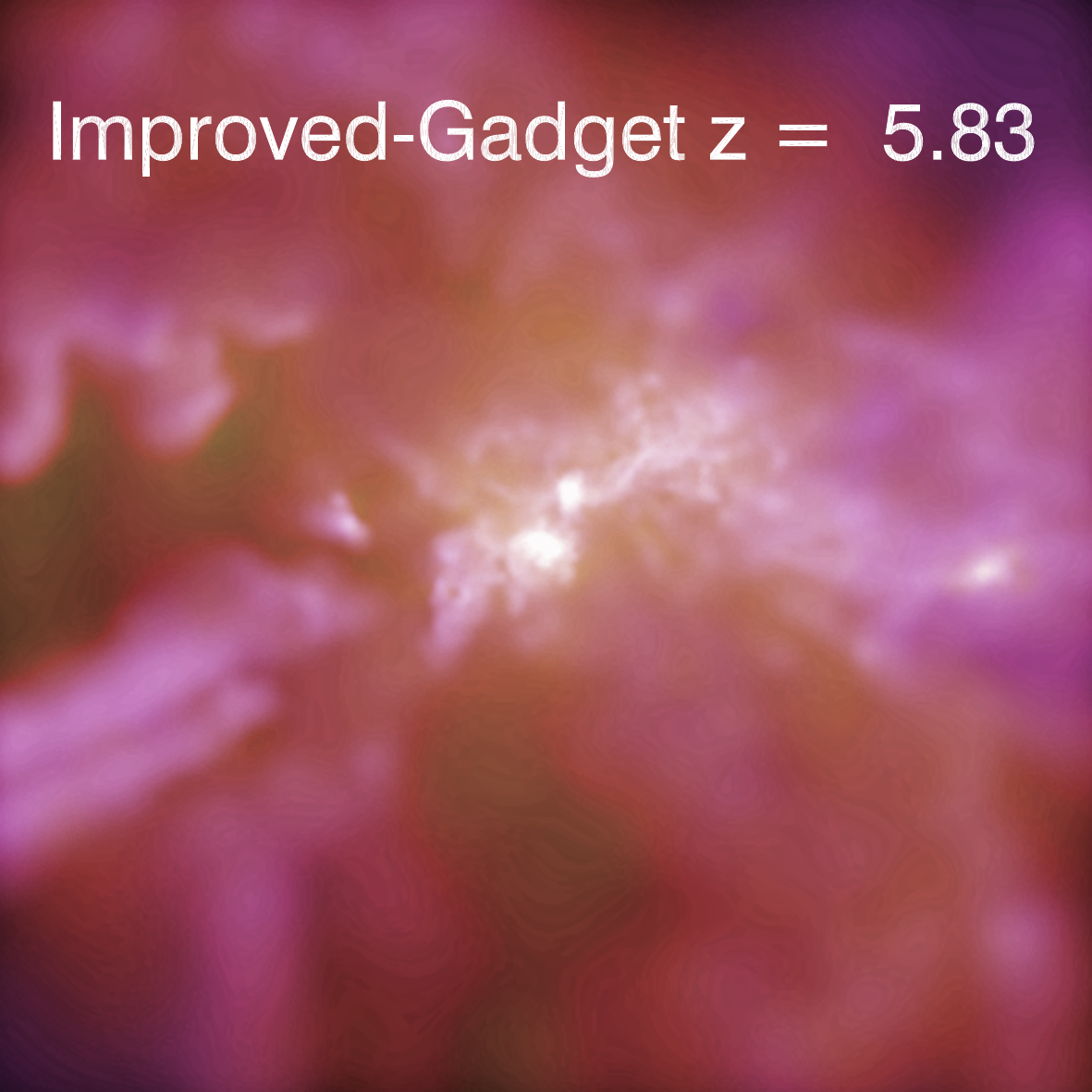}}
\resizebox{1.35in}{!}{\includegraphics{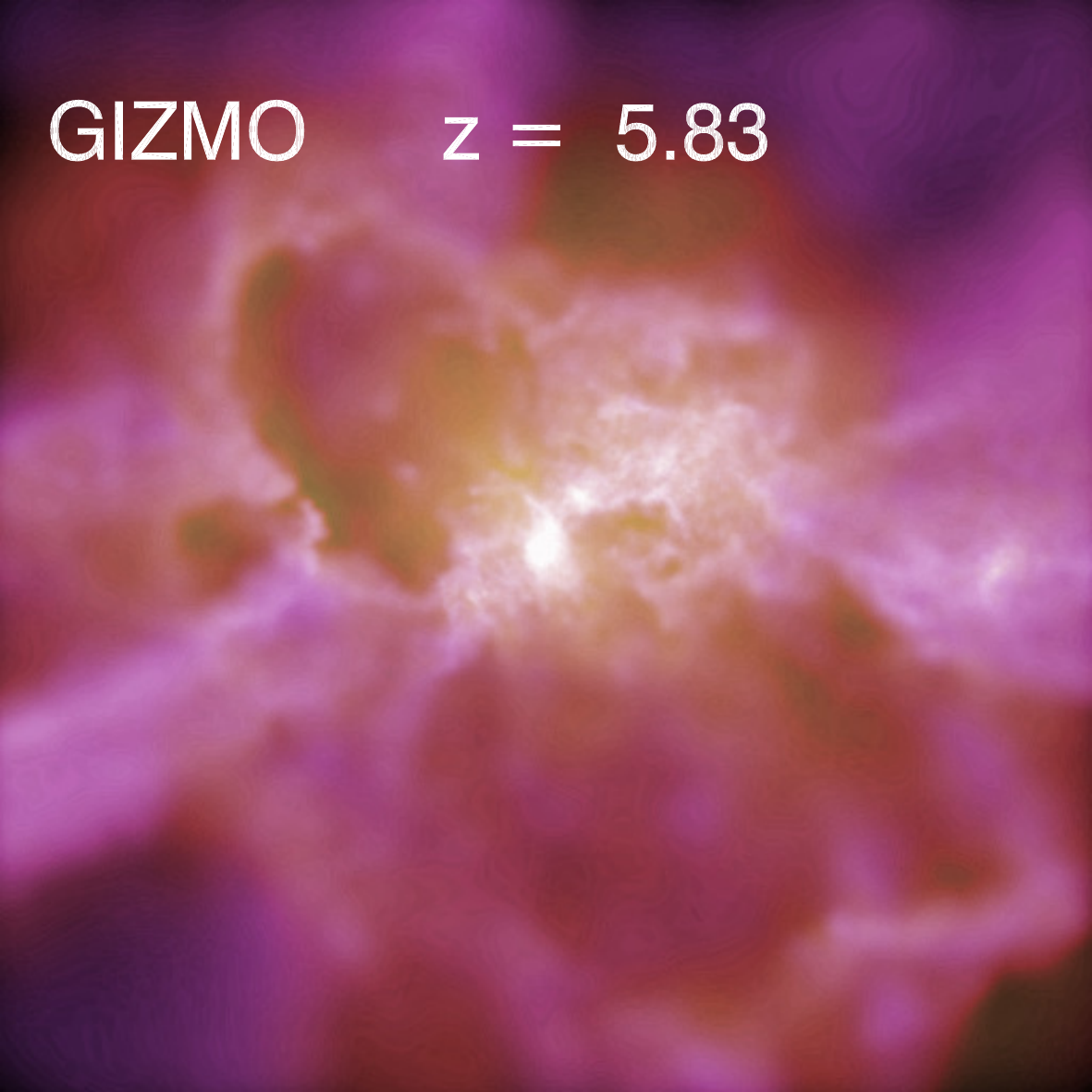}}\\
\resizebox{1.35in}{!}{\includegraphics{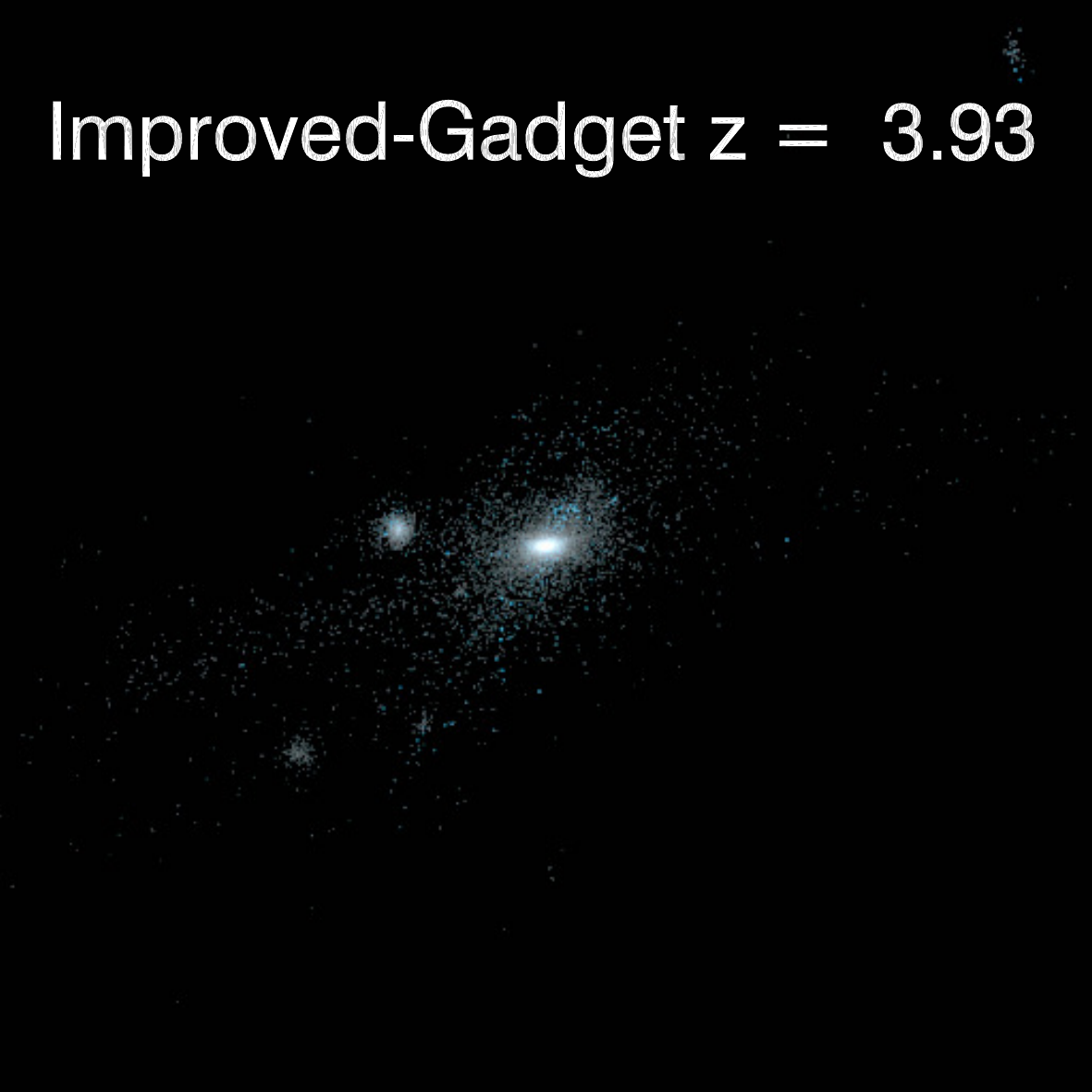}}
\resizebox{1.35in}{!}{\includegraphics{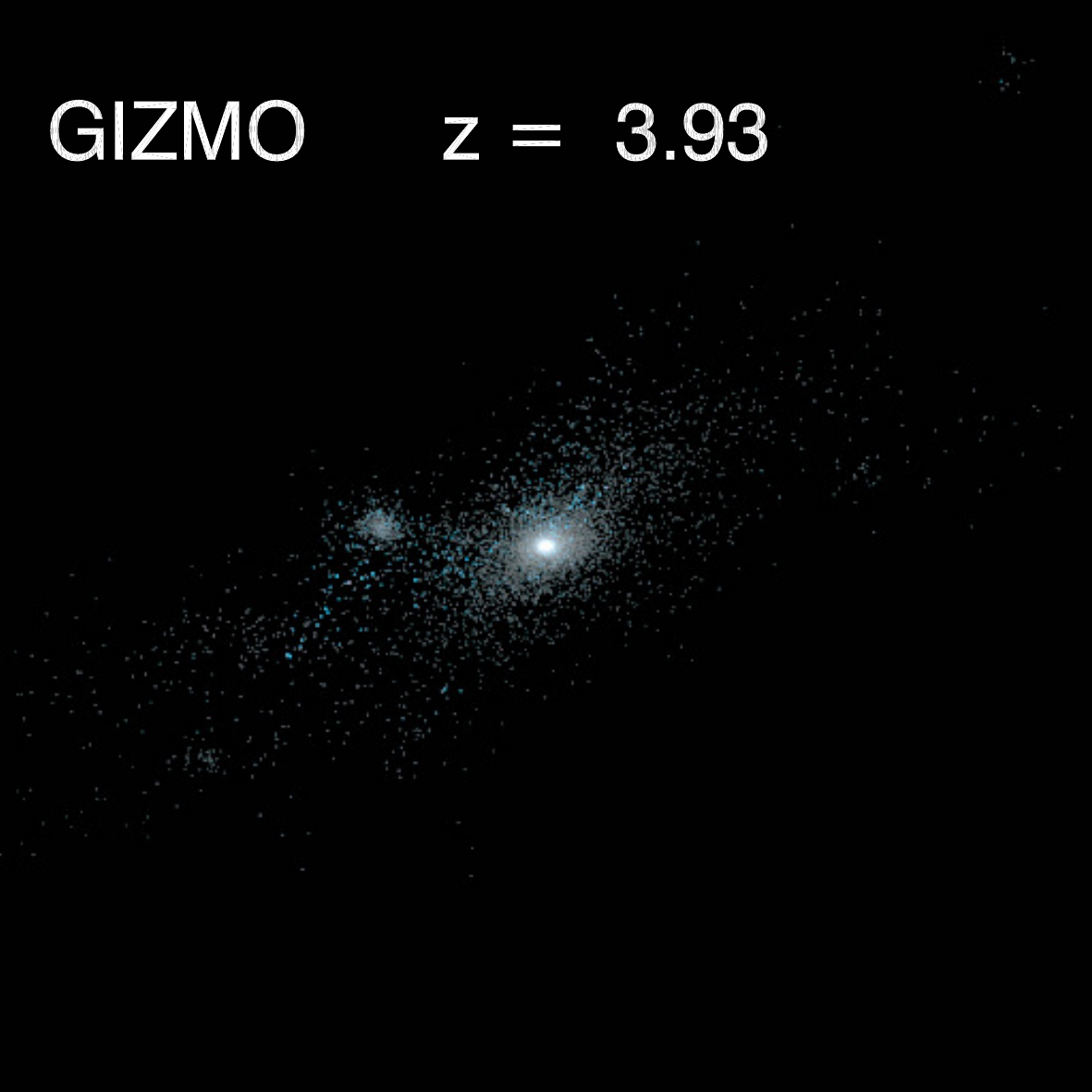}}
\resizebox{1.35in}{!}{\includegraphics{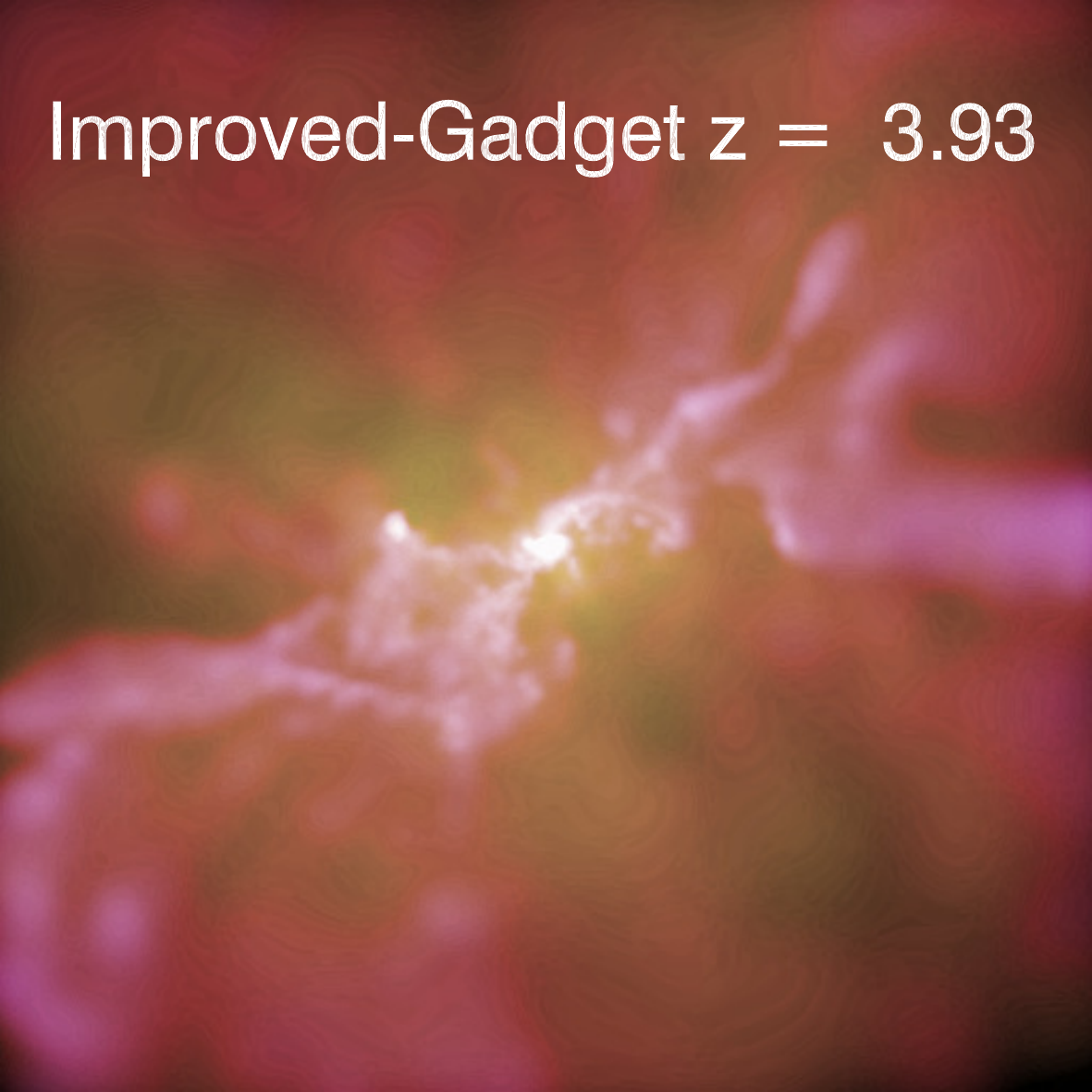}}
\resizebox{1.35in}{!}{\includegraphics{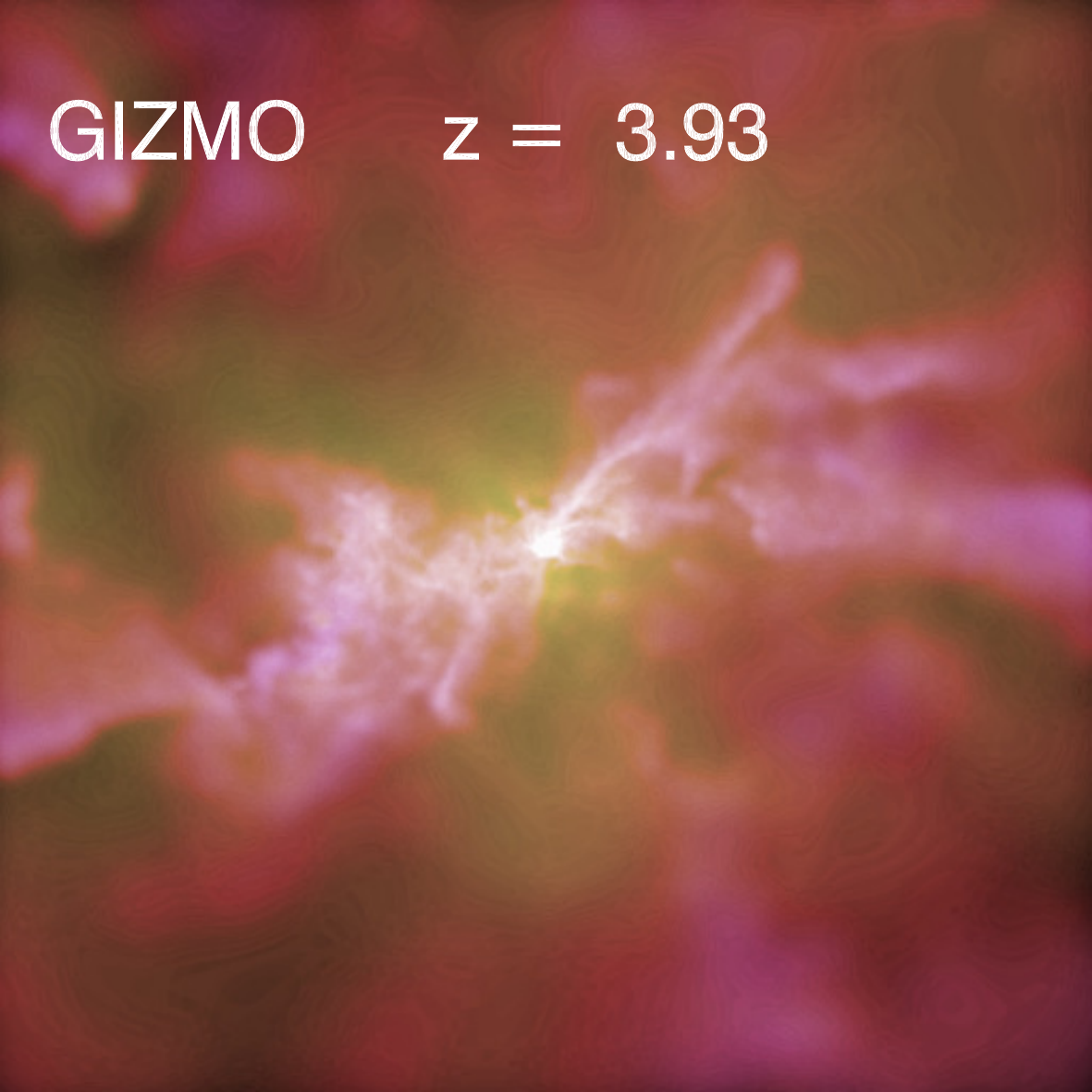}}\\
\resizebox{1.35in}{!}{\includegraphics{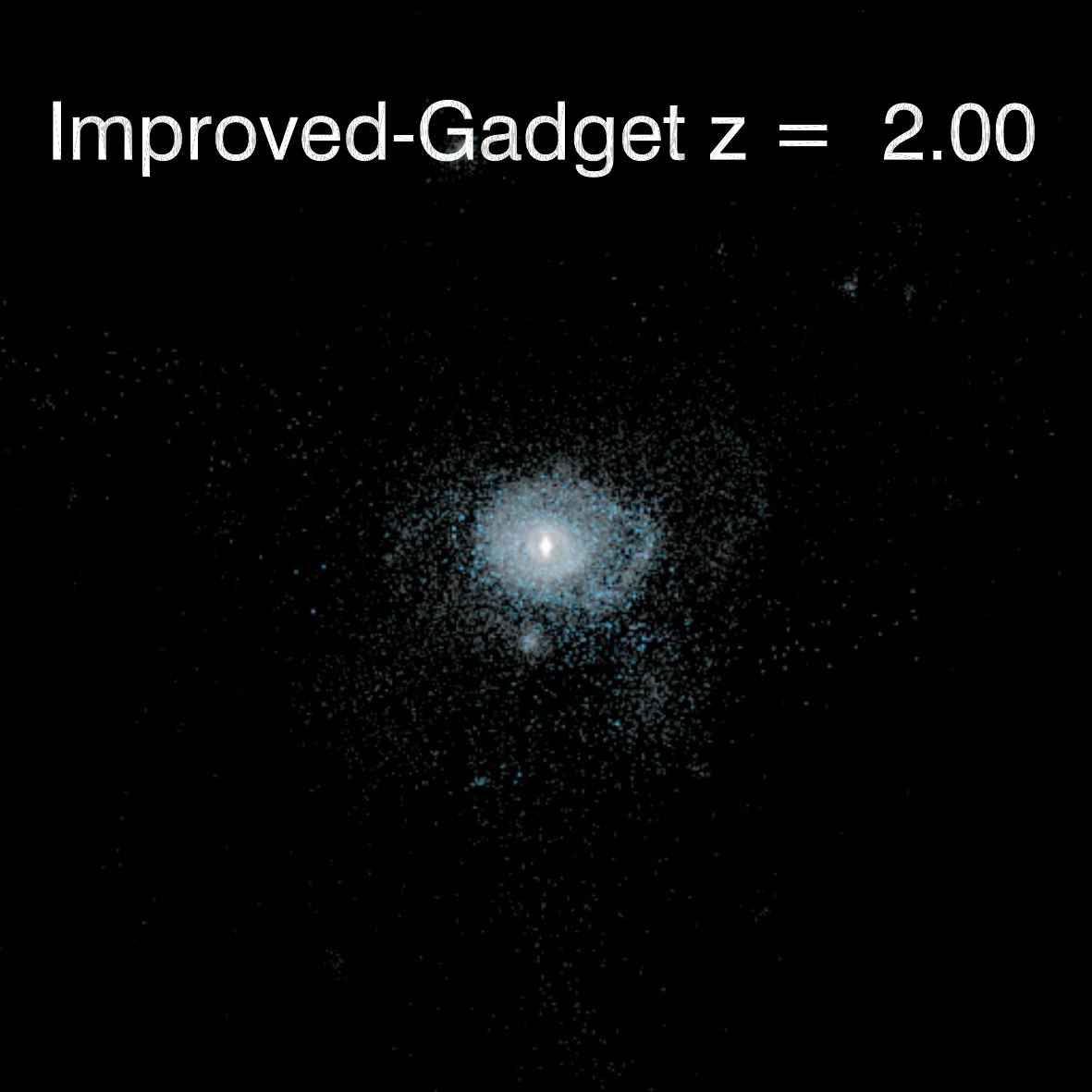}}
\resizebox{1.35in}{!}{\includegraphics{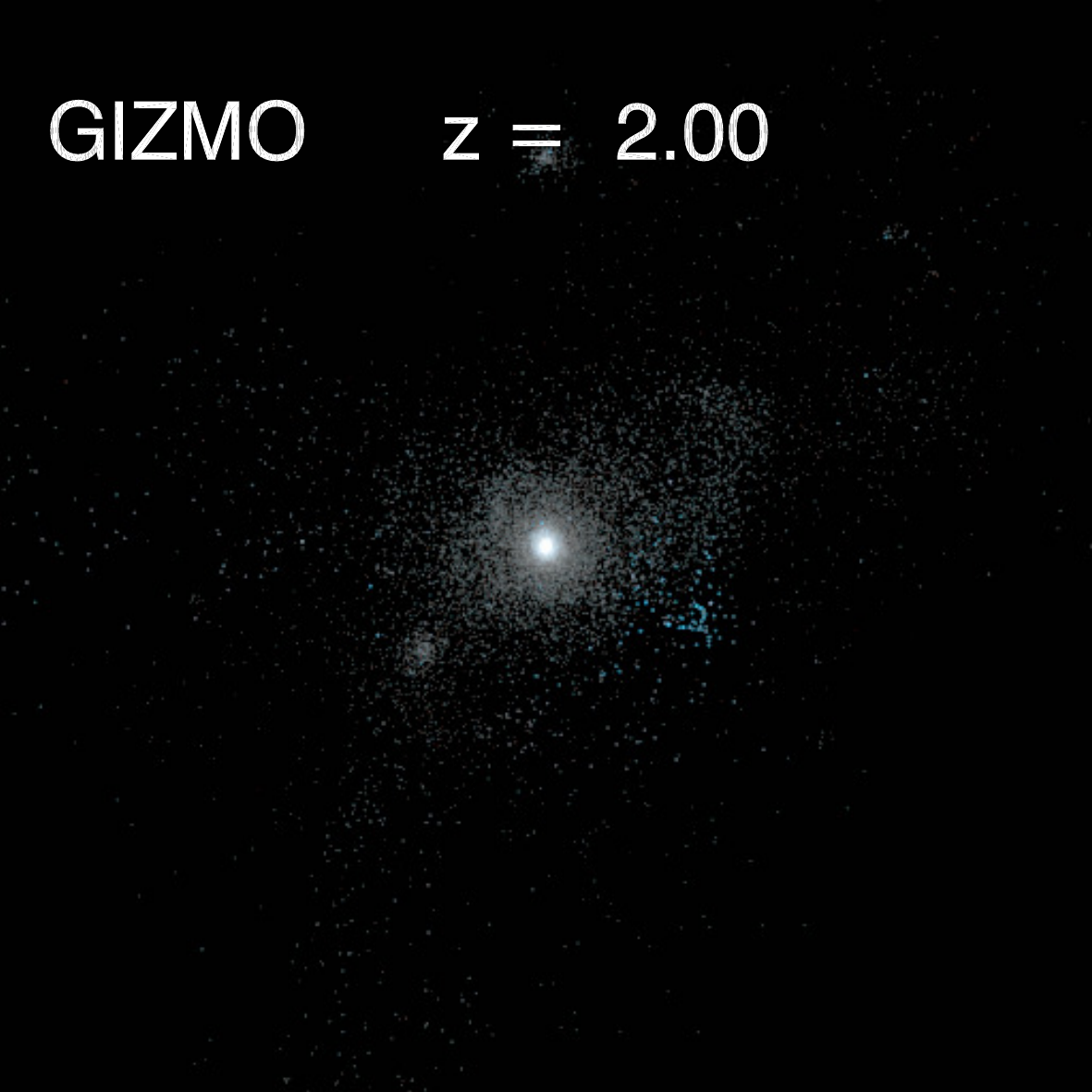}}
\resizebox{1.35in}{!}{\includegraphics{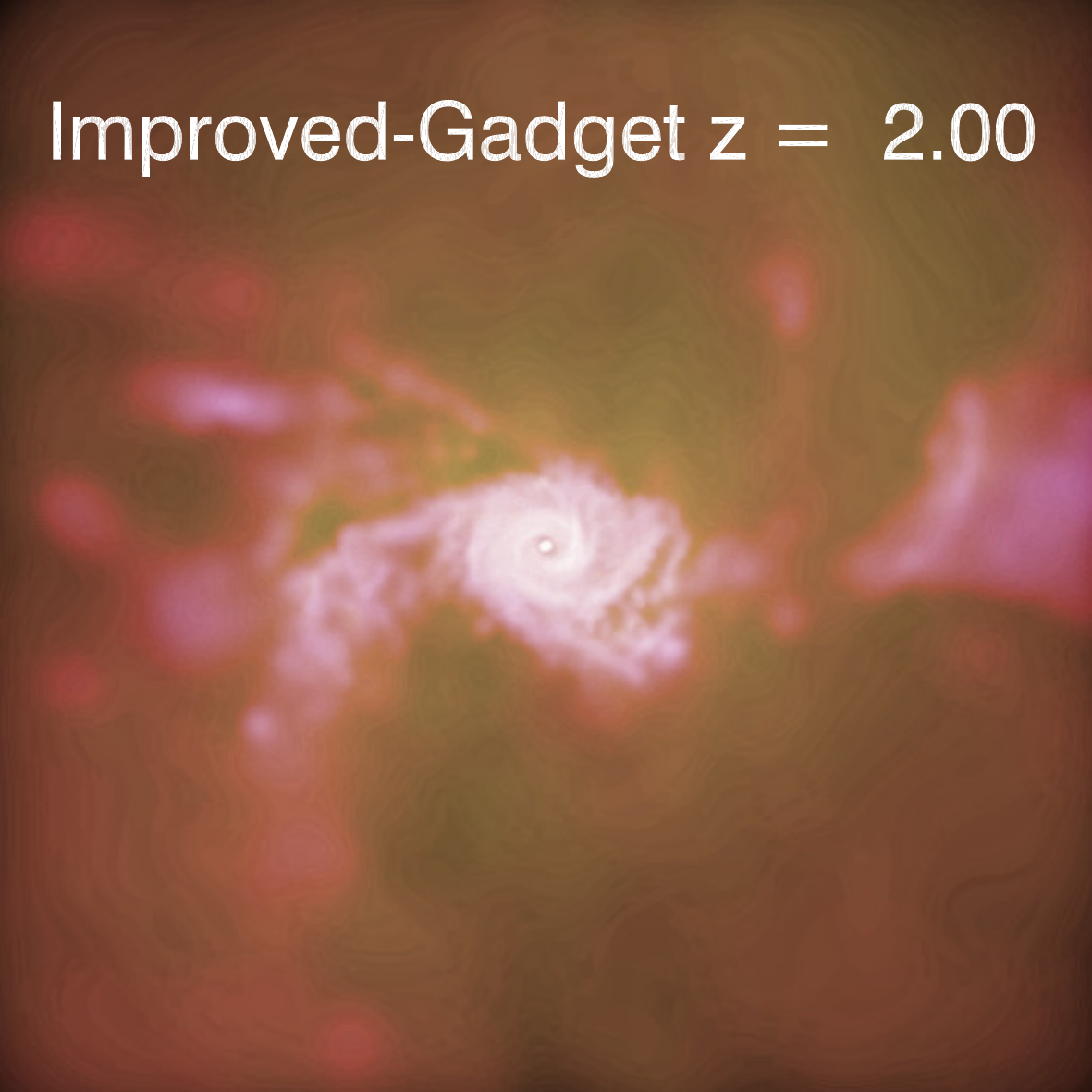}}
\resizebox{1.35in}{!}{\includegraphics{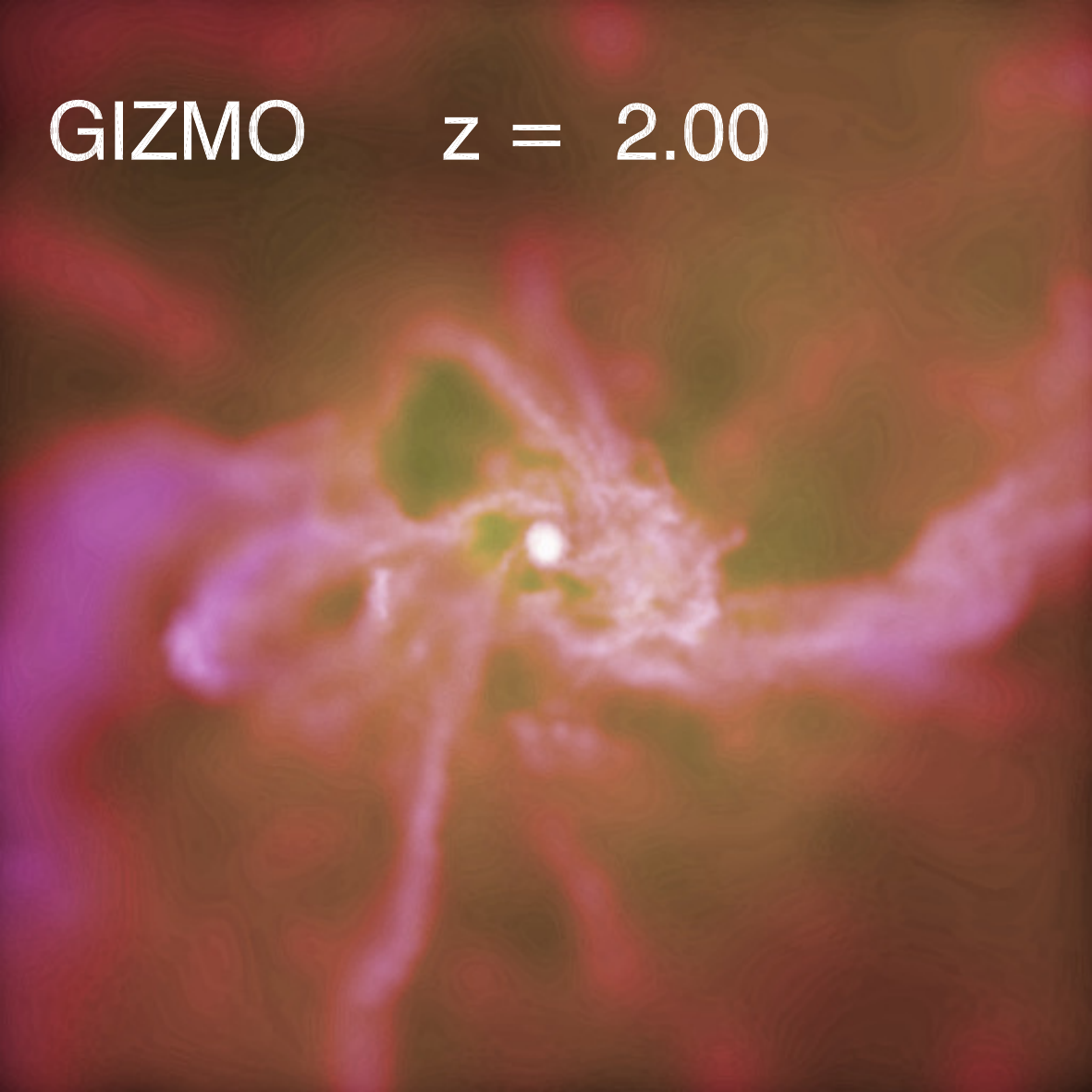}}\\
\resizebox{1.35in}{!}{\includegraphics{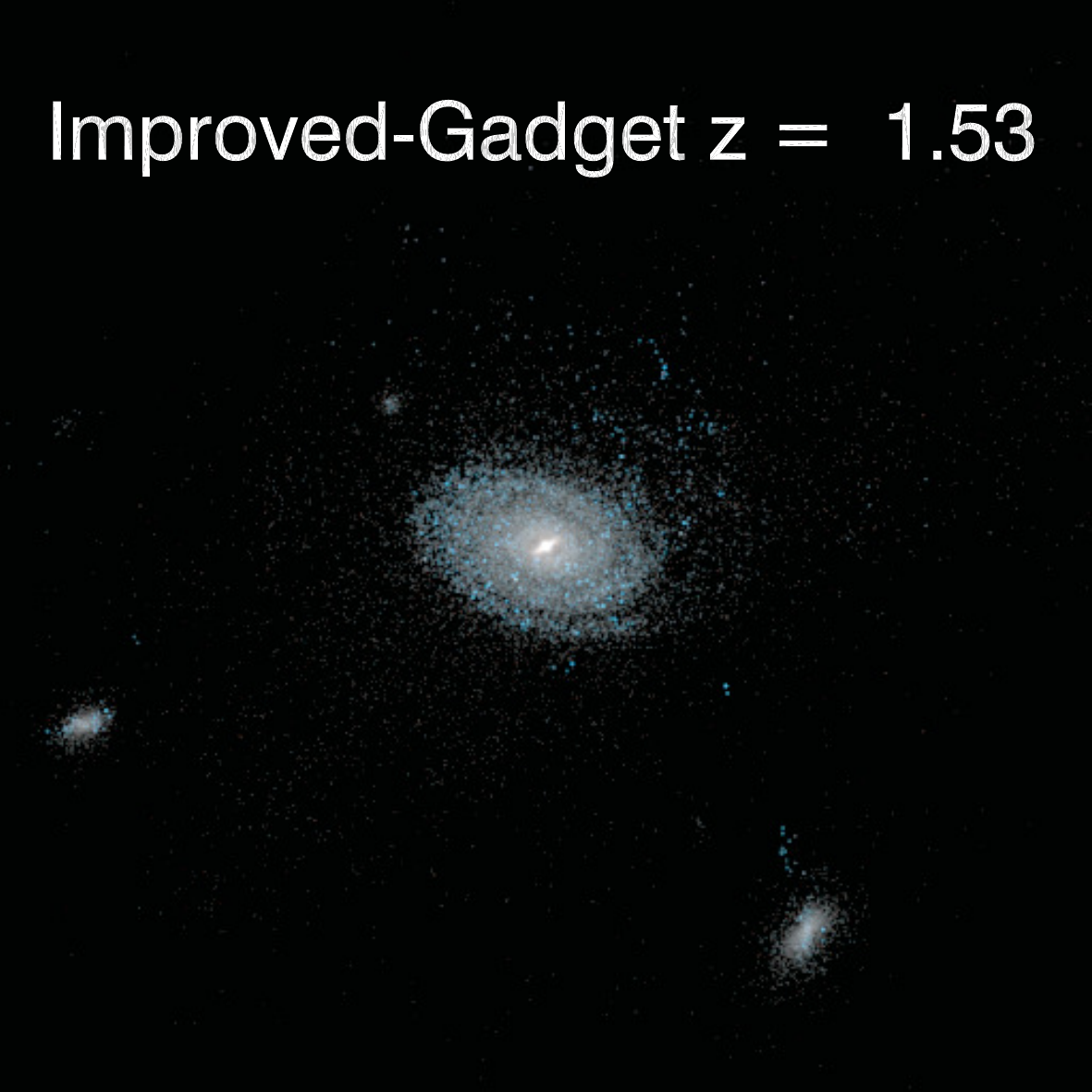}}
\resizebox{1.35in}{!}{\includegraphics{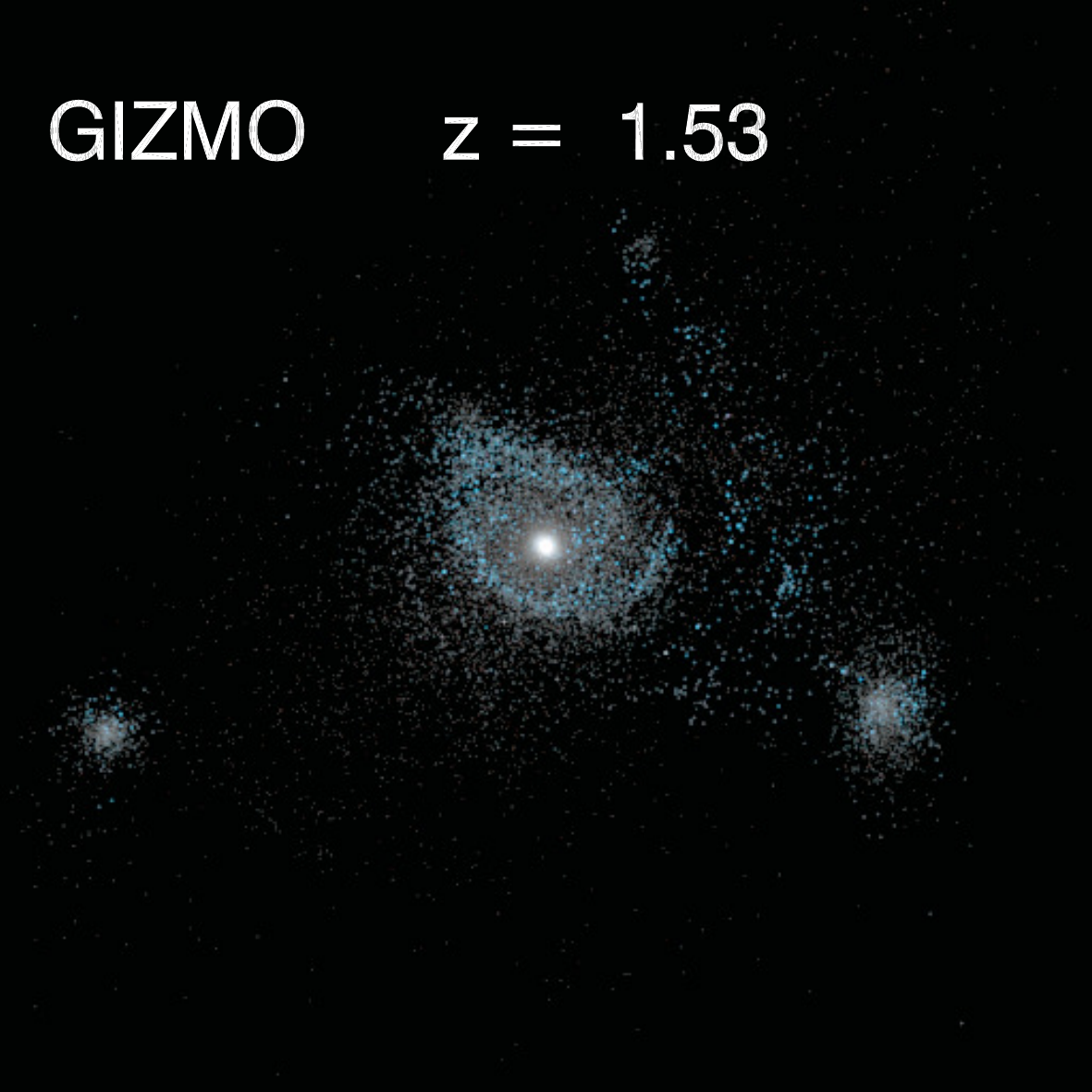}}
\resizebox{1.35in}{!}{\includegraphics{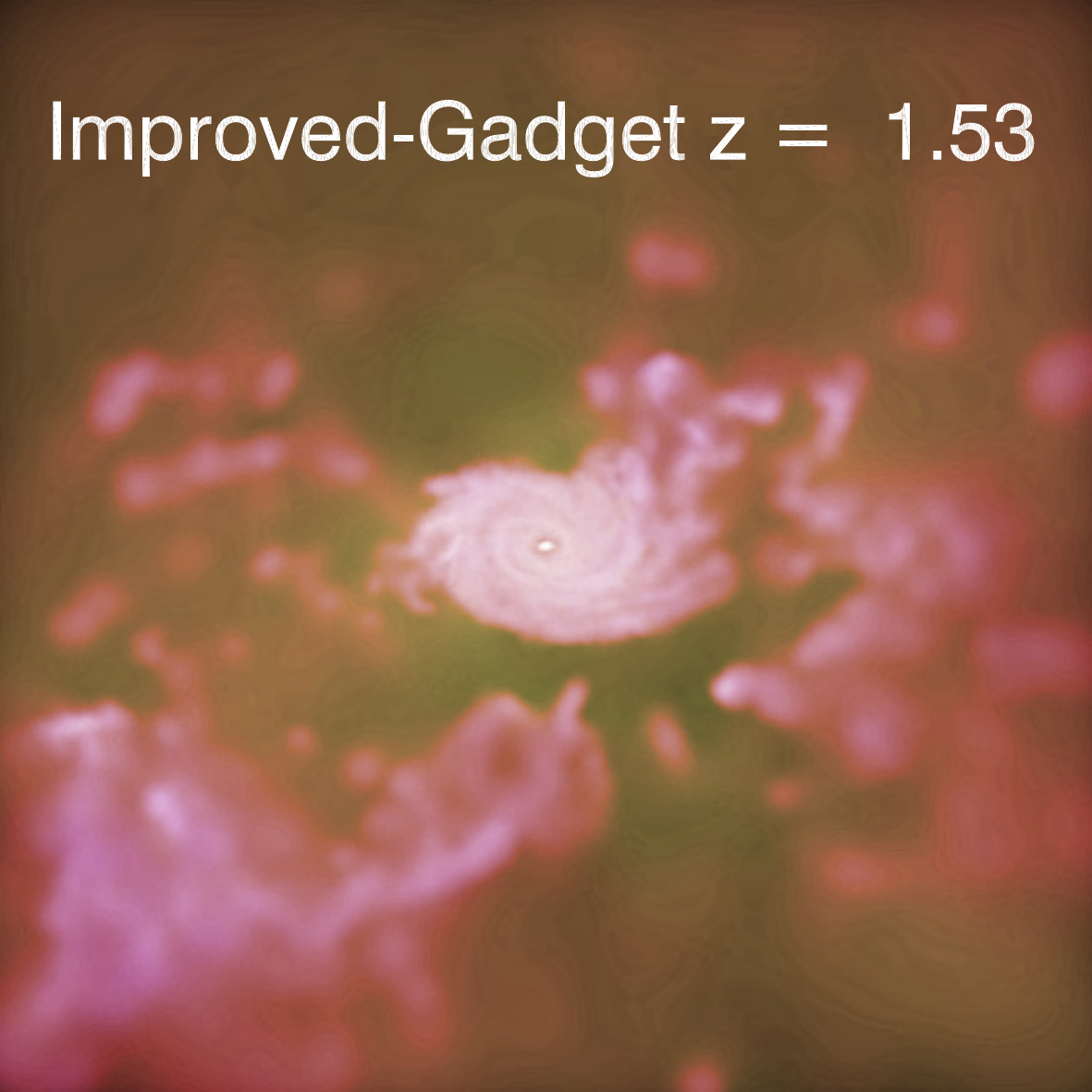}}
\resizebox{1.35in}{!}{\includegraphics{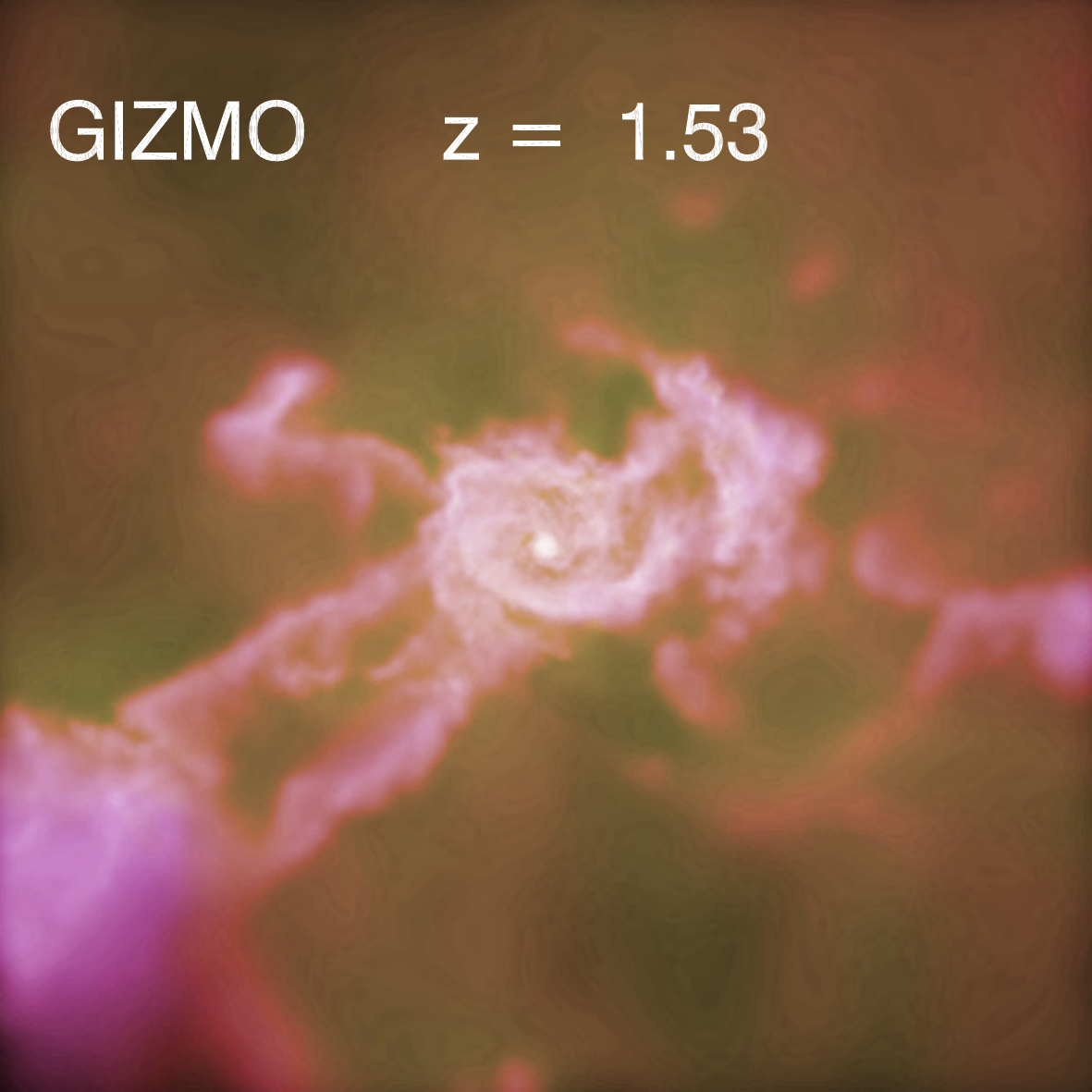}}\\
\resizebox{1.35in}{!}{\includegraphics{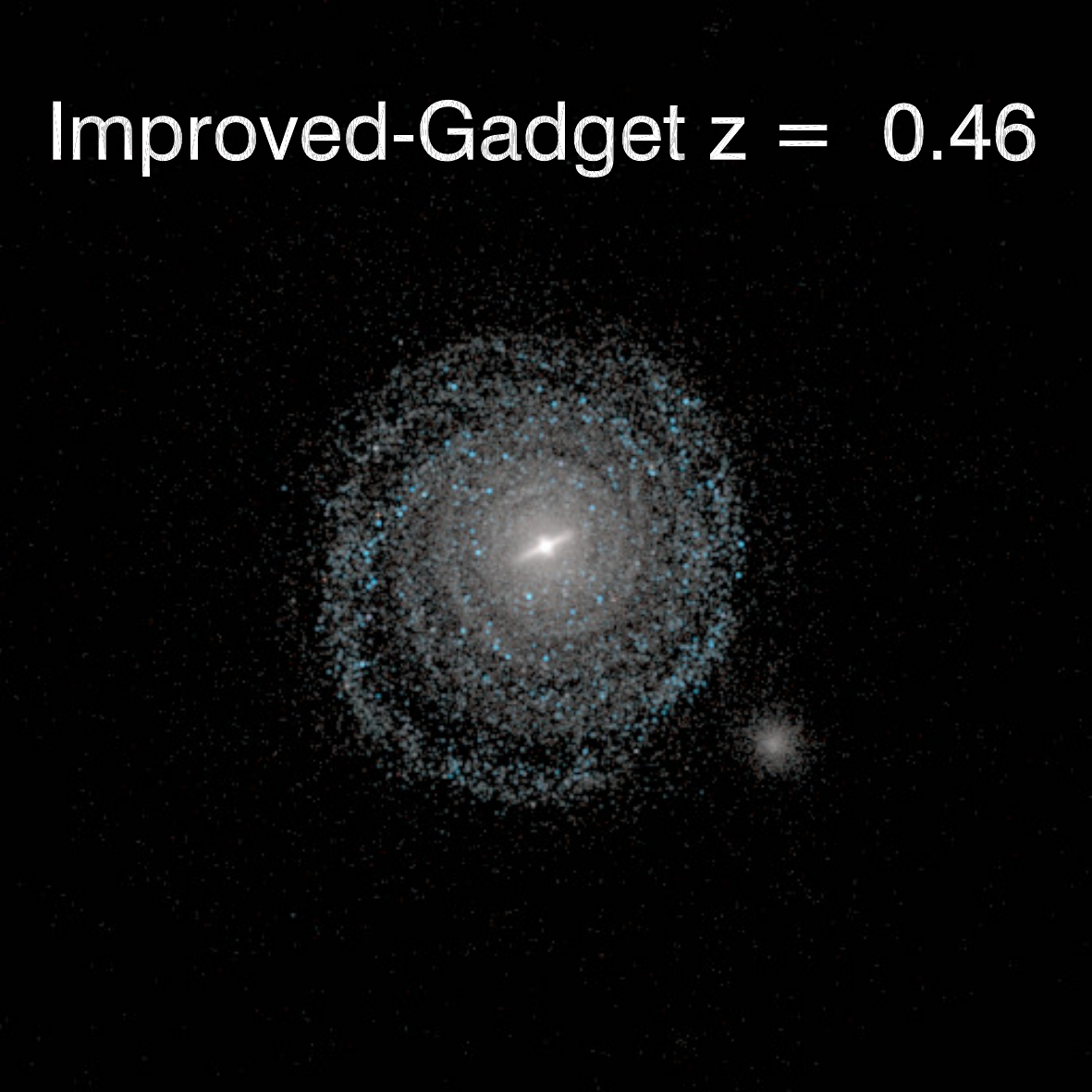}}
\resizebox{1.35in}{!}{\includegraphics{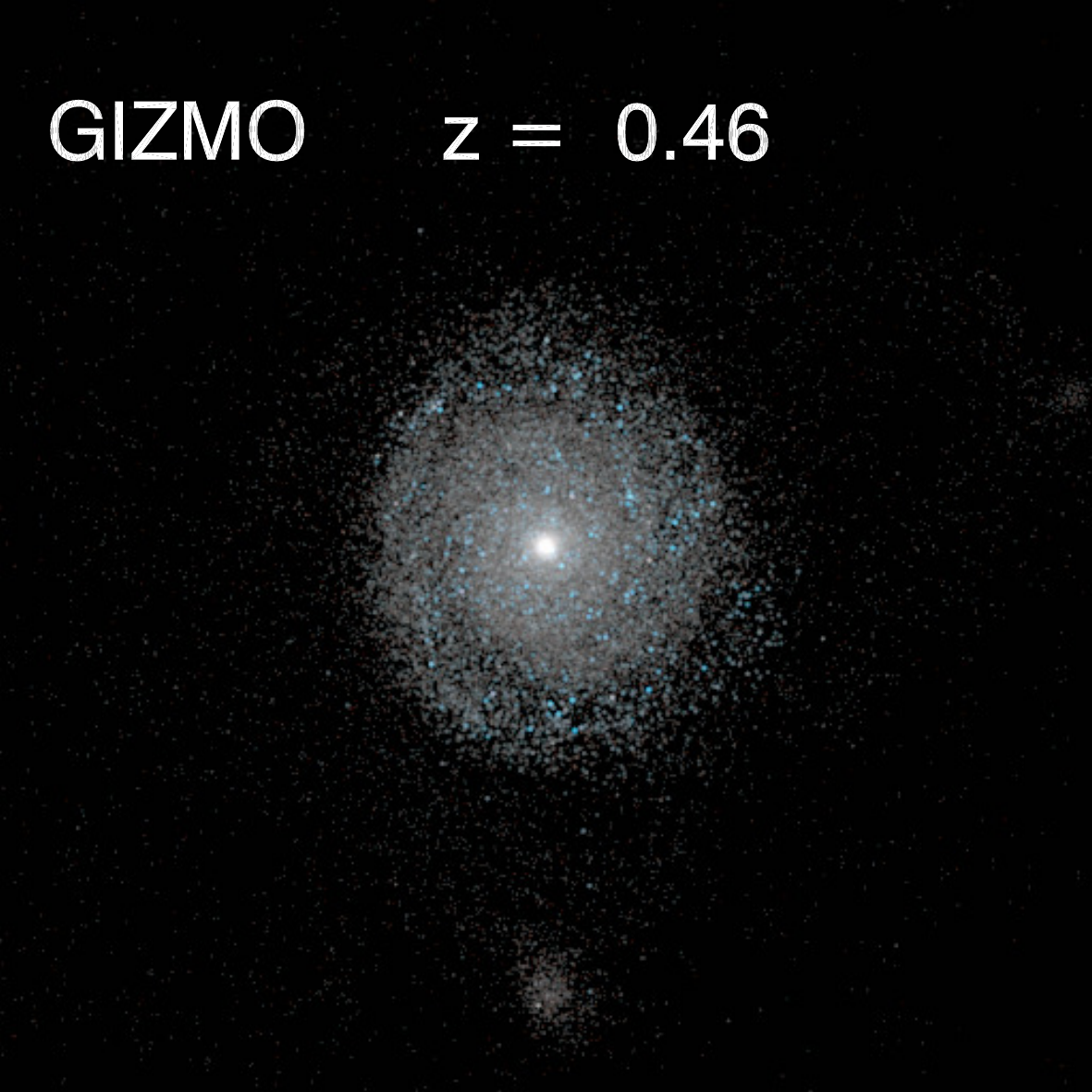}}
\resizebox{1.35in}{!}{\includegraphics{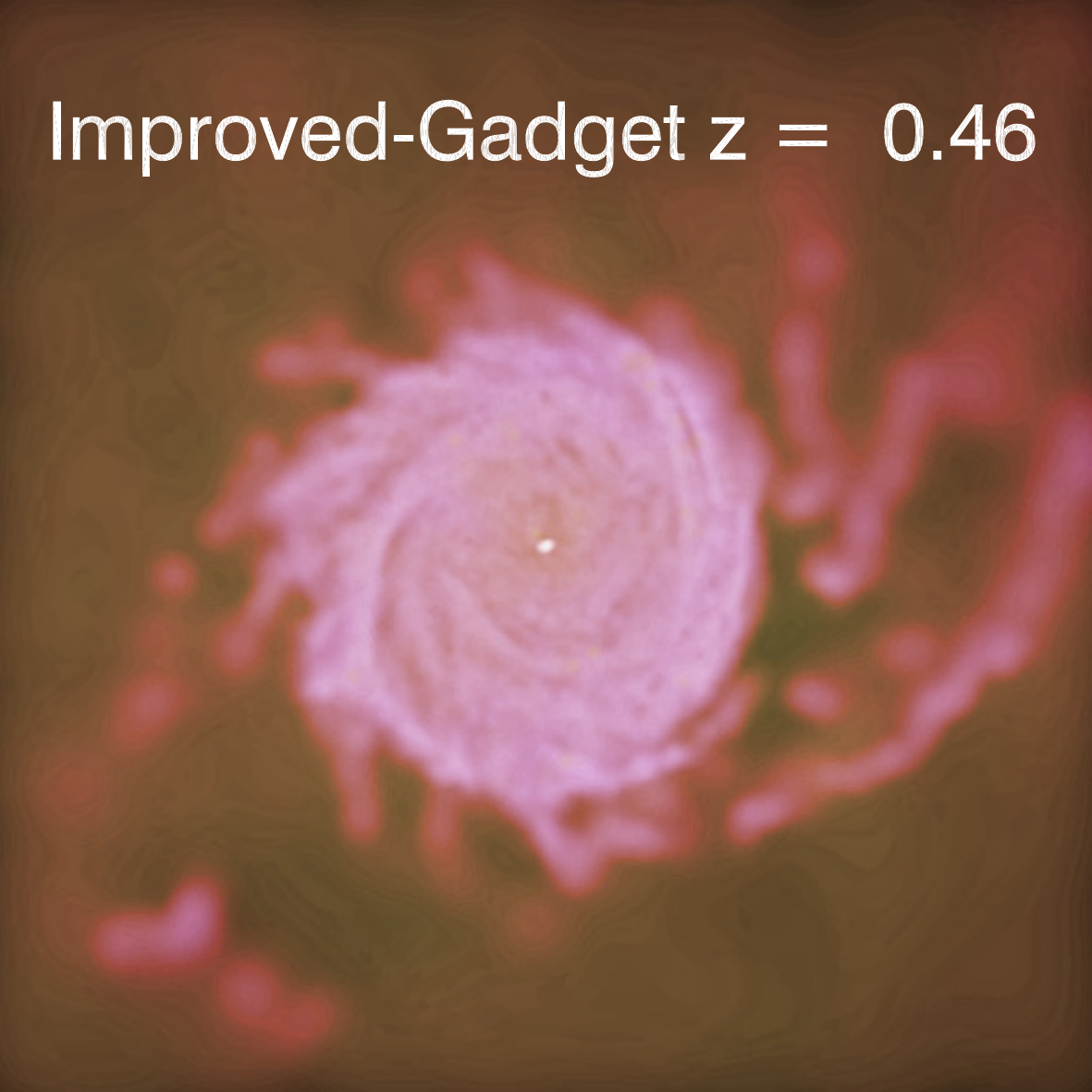}}
\resizebox{1.35in}{!}{\includegraphics{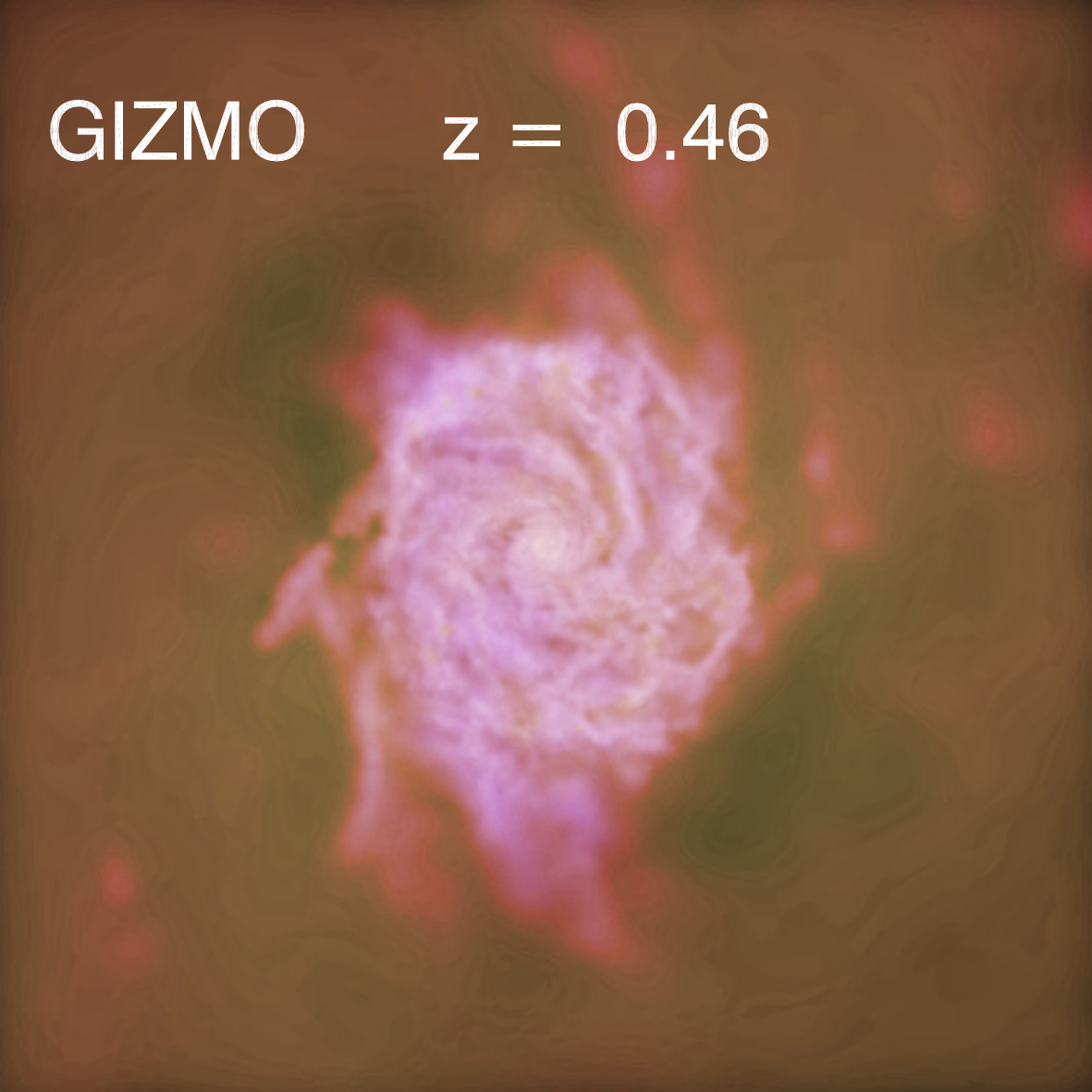}}\\
\resizebox{1.35in}{!}{\includegraphics{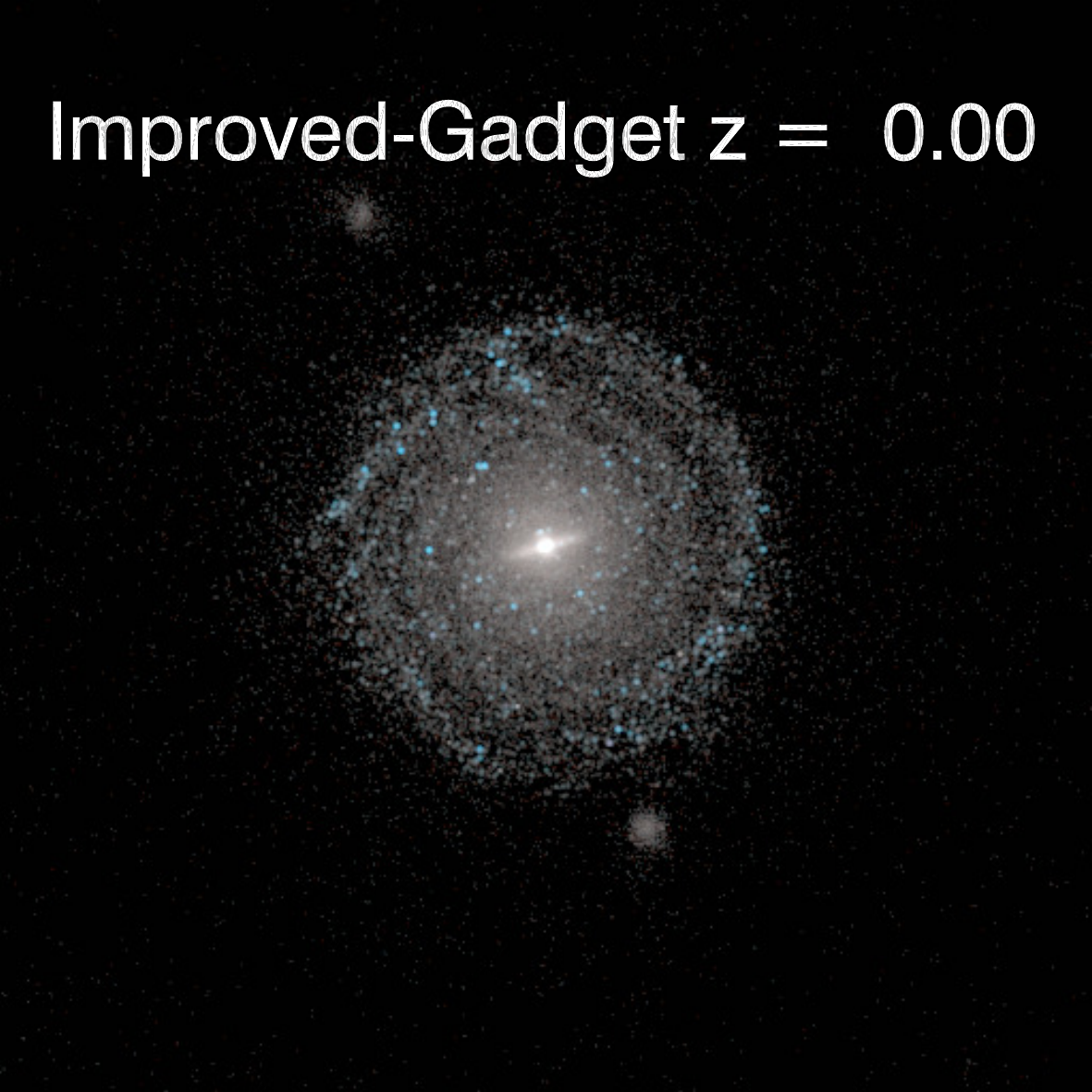}}
\resizebox{1.35in}{!}{\includegraphics{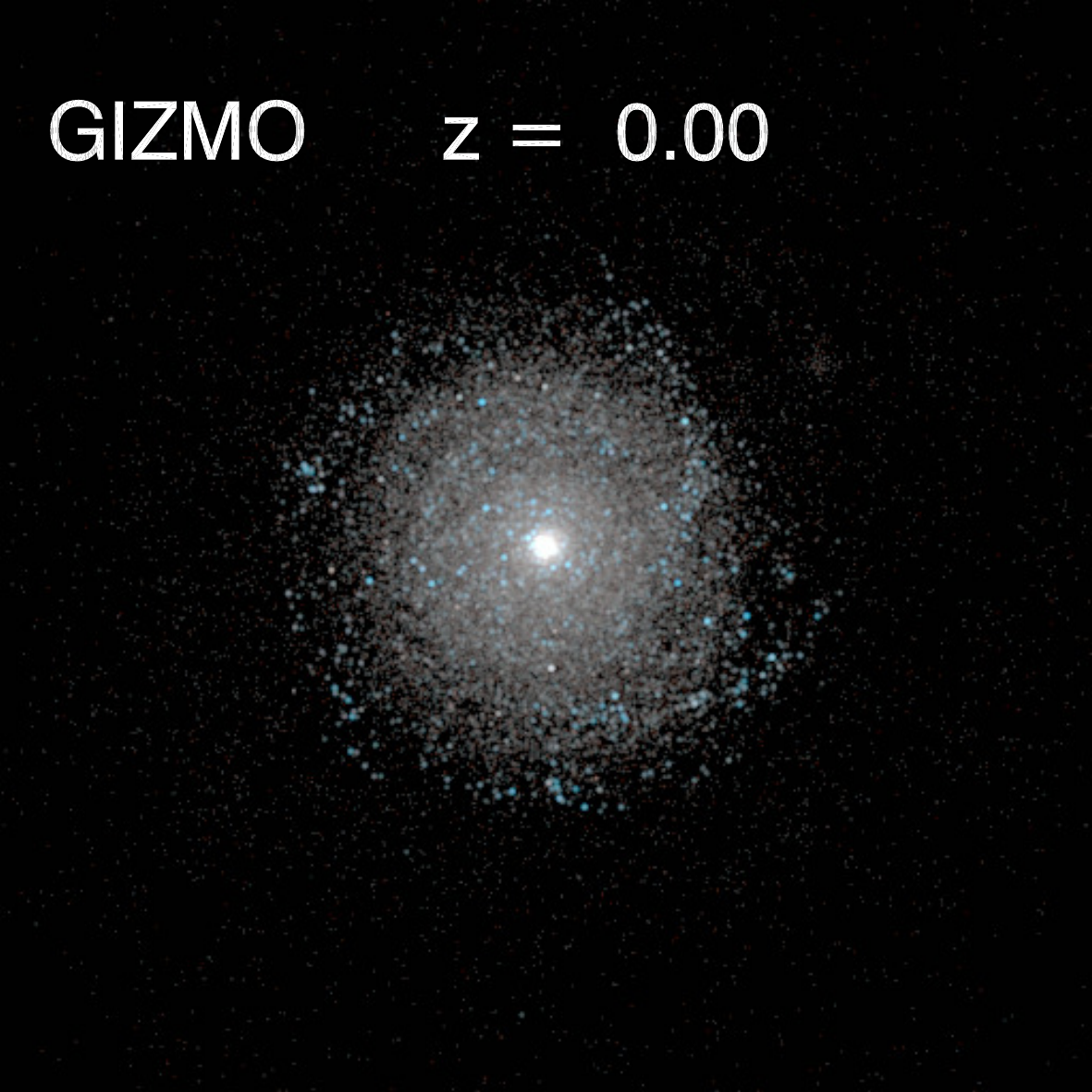}}
\resizebox{1.35in}{!}{\includegraphics{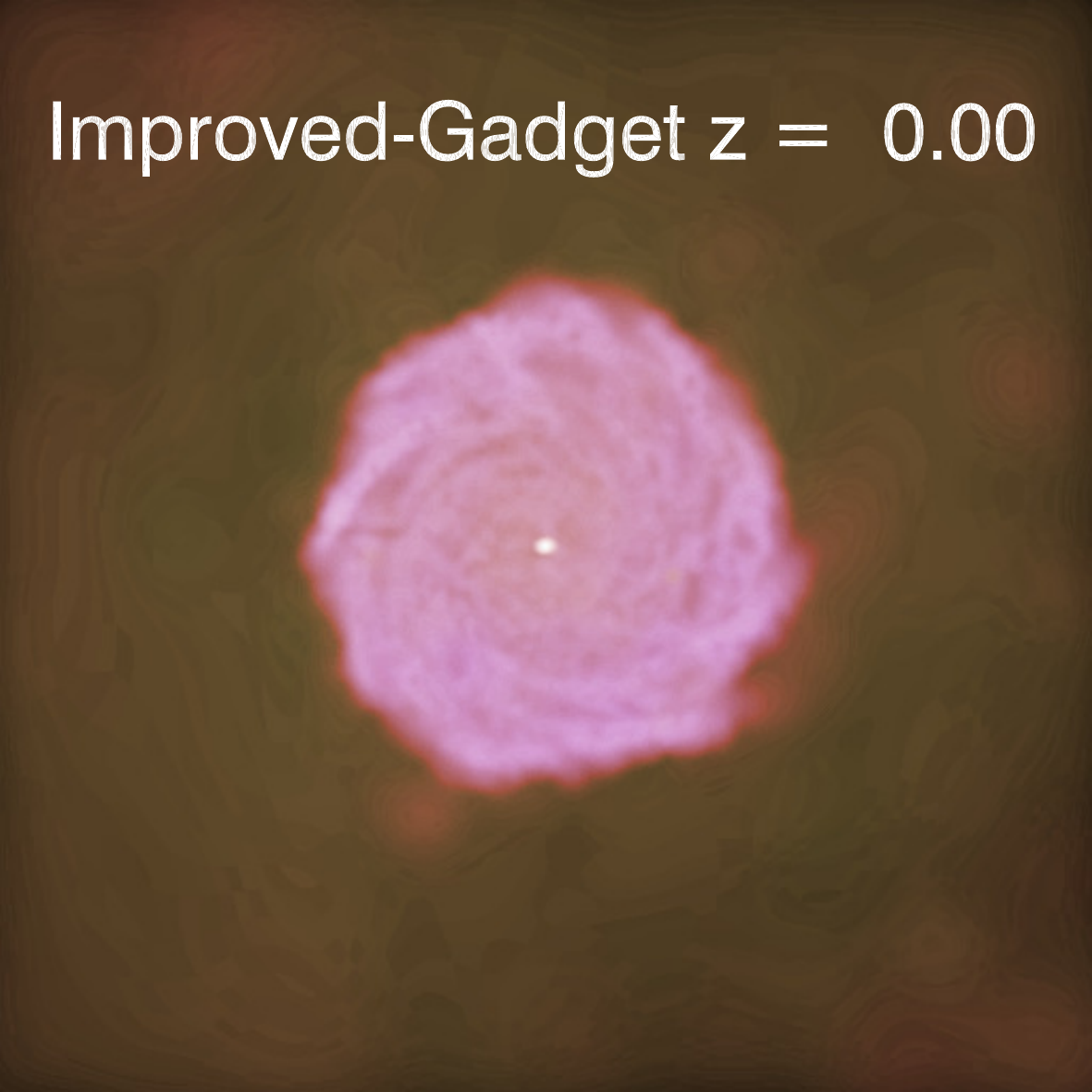}}
\resizebox{1.35in}{!}{\includegraphics{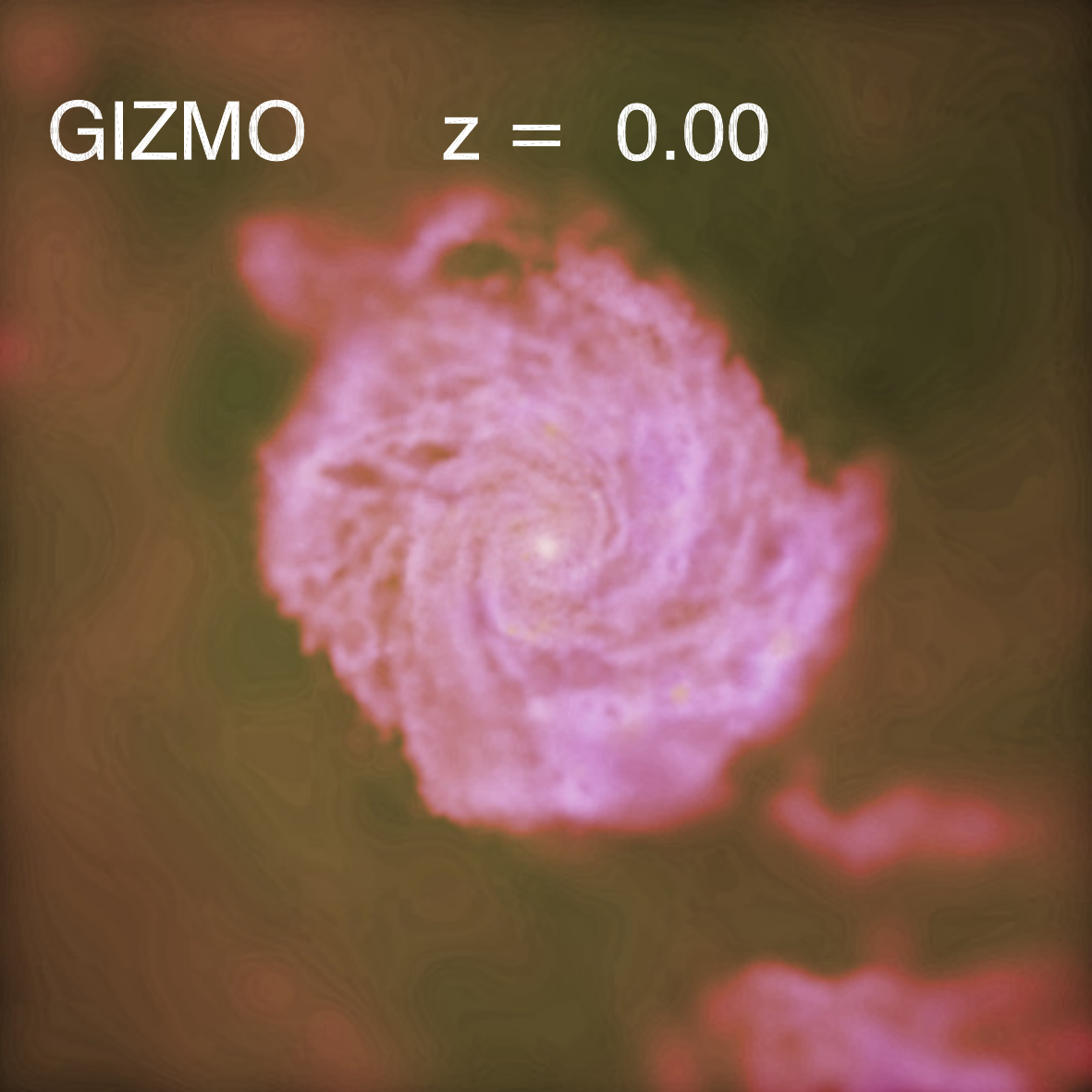}}\\
\end{tabular}
\caption{\label{fig:star_gas_images} 
A comparison of the formation and dynamical evolution of the modeled galaxy between the two simulations using Improved-{\Gadget} and {\Gizmo}, respectively. {\it Left  panels}: Projected stellar density at different redshifts. The luminosity of the stars is calculated from Starburst99 based on its mass, age and metallicity, while the colors are assigned based on rest-frame $K$, $V$ and $B$-band magnitudes in $rgb$ channels. Newly formed stars appear in blue color, while old ones appear in red. {\it Right  panels}: Projected gas density at corresponding times. The image is color coded with gas temperature from cold (in blue) to hot (in yellow). The size of each panel is 100 kpc in physical coordinates.}
\end{center}
\end{figure*}

The modeled galaxy was selected as a close match to the MW \citep{Springel2008, Scannapieco2012} in terms of mass and merging history. It builds the bulk of its mass in an early stage through multiple mergers by redshift $z \sim 1$, as illustrated in Figure~\ref{fig:star_gas_images}. Both simulations using the Improved-{\Gadget} and {\Gizmo} show similar assembly history of the disk galaxy, in which an extended disk was formed  at $z \sim 2$ and it continued to grow into an extended disk of stars and gas until the present day.  

The growth of the stellar disk appears to follow an ``inside-out" fashion. The strong galaxy interactions at early times produce strong shocks and gravitational torques which draw the cold gas to the deep potential well at the galactic center and trigger starbursts. As the gas spirals in, it rotates around the galactic center and forms a disk. The newly formed  stars inherit the angular momentum from the gas and they too rotate in the disk. The initially small disk grows rapidly outwards as more gas is being accreted and more stars are formed.   

In the Aquila comparison project \citep{Scannapieco2012}, misaligned gas disks were present in some models. In our simulation, however, the rotational directions of both gaseous and  stellar disks agree with each other very well since $z\sim2$. As a result, the new stars are mostly formed in a kinetically-cold, rotating structure and in a continuous fashion. The disk properties will be studied in detail in \S~\ref{subsec:disk}.

Our simulations produce similar formation history and dynamical evolution of a disk galaxy as those by \cite{Marinacci2014} (their Figure 12) using the same initial conditions and physical models. This is encouraging as it indicates that the role of numerical artifact is subdominant. However, remarkable differences are present between the Improved-{\Gadget} and {\Gizmo} simulations:  the stellar and gaseous disks started from at $z = 2$ in the Improved-{\Gadget} but at a later time at $z \sim 1.5$ in the {\Gizmo} simulation; there is a {bar-like} structure in the central region in the Improved-{\Gadget} simulation, which is absent in the {\Gizmo} one; at $z = 0$, the gas disk in the {\Gizmo} simulation is much more extended in size and it contains prominent spiral structures, in contrast to a more compact and  smoother disk in  the Improved-{\Gadget} simulation.  

In Figure~\ref{fig:star_gas_images}, strong interactions between the outflows and intergalactic medium (IGM) is visible at high $z$, where the hot gas from galactic outflow confronts  the cold gas from the filamentary structures, and the shock fronts are also visible around the hot bubbles. Although the overall pictures of the two simulations are similar, some  fine  structures differ significantly from each other. For example at $z = 5.8$, the strong outflow in the Improved-{\Gadget} travels beyond the 100 kpc box shown, while that in the {\Gizmo} simulation is confined to a smaller volume.  This can be understood as the shocking capturing is more accurate in {\Gizmo} thanks to its finite volume nature, while SPH, on the other hand, still relies on artificial viscosity which leads to {a} broadening of shocks over several smoothing lengths . 

Previous SPH simulations often produce large number of  ``cold blobs'' surrounding the central galaxies. From Figure~\ref{fig:star_gas_images}, it is evident that the distribution of gas is smooth thanks to its pressure-entropy formulation, and the ``cold blobs" are greatly reduced in the Improved-{\Gadget} simulation, although at some certain snapshots there are still some gas blobs present in the Improved-Gadget simulation, e.g. at $z = 1.53$ and $0.46$. For comparison, {\Gizmo} simulation shows more filamentary distribution connecting the cold gas in the disk to the filaments from the IGM. In addition, cold gas blobs are not found in {\Gizmo} simulation, demonstrating that ``cold blobs'' in the previous simulations are intrinsic to SPH formulations, and that the fluid mixing in {\Gizmo} greatly improves the treatment of multi-phase ISM. 

\subsection{Star Formation History}

Star formation histories from both Improved-{\Gadget} and {\Gizmo} simulations are shown in Figure~\ref{fig:sfr}. Both simulations show strong star formation peaks around $z\sim 4$, corresponding to a period of rapid mass assembly through galaxy mergers and accretions. {A} second peak in star formation at $z\sim 2$ marks a minor merger event during this time. These star formation histories are consistent with those from previous simulations of the same object \citep{Scannapieco2012, Okamoto2013, Marinacci2014}.

\begin{figure}
\begin{center}
\includegraphics[scale=0.4]{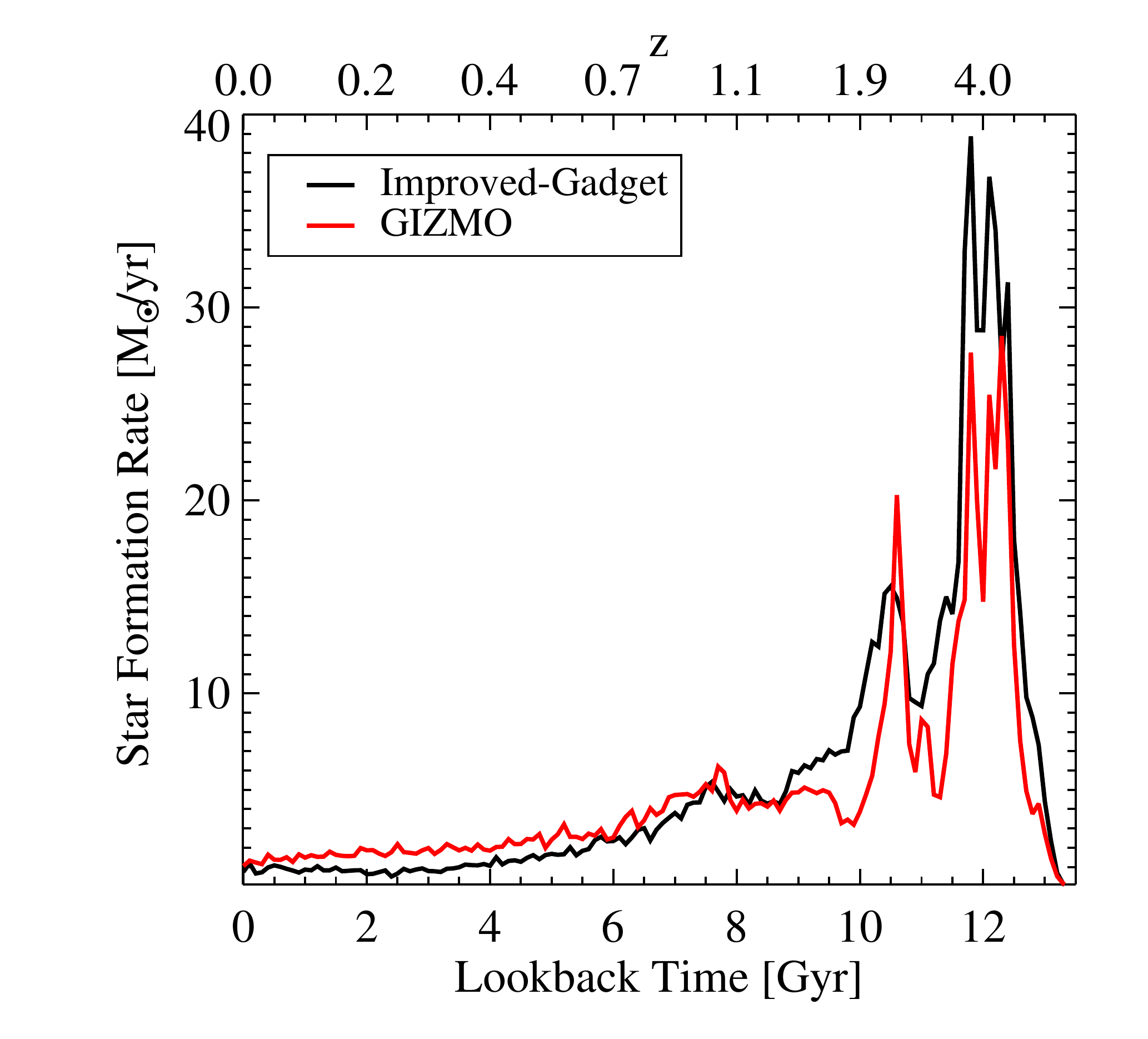}
\caption{\label{fig:sfr}  Star formation history as a function of cosmic time from both Improved-{\Gadget} (black line) and {\Gizmo}  (red line) simulations, respectively. The bottom x-axis shows the lookback time, while the top x-axis shows the redshift. }
\end{center}
\end{figure}

\begin{figure}
\begin{center}
\includegraphics[scale=0.4]{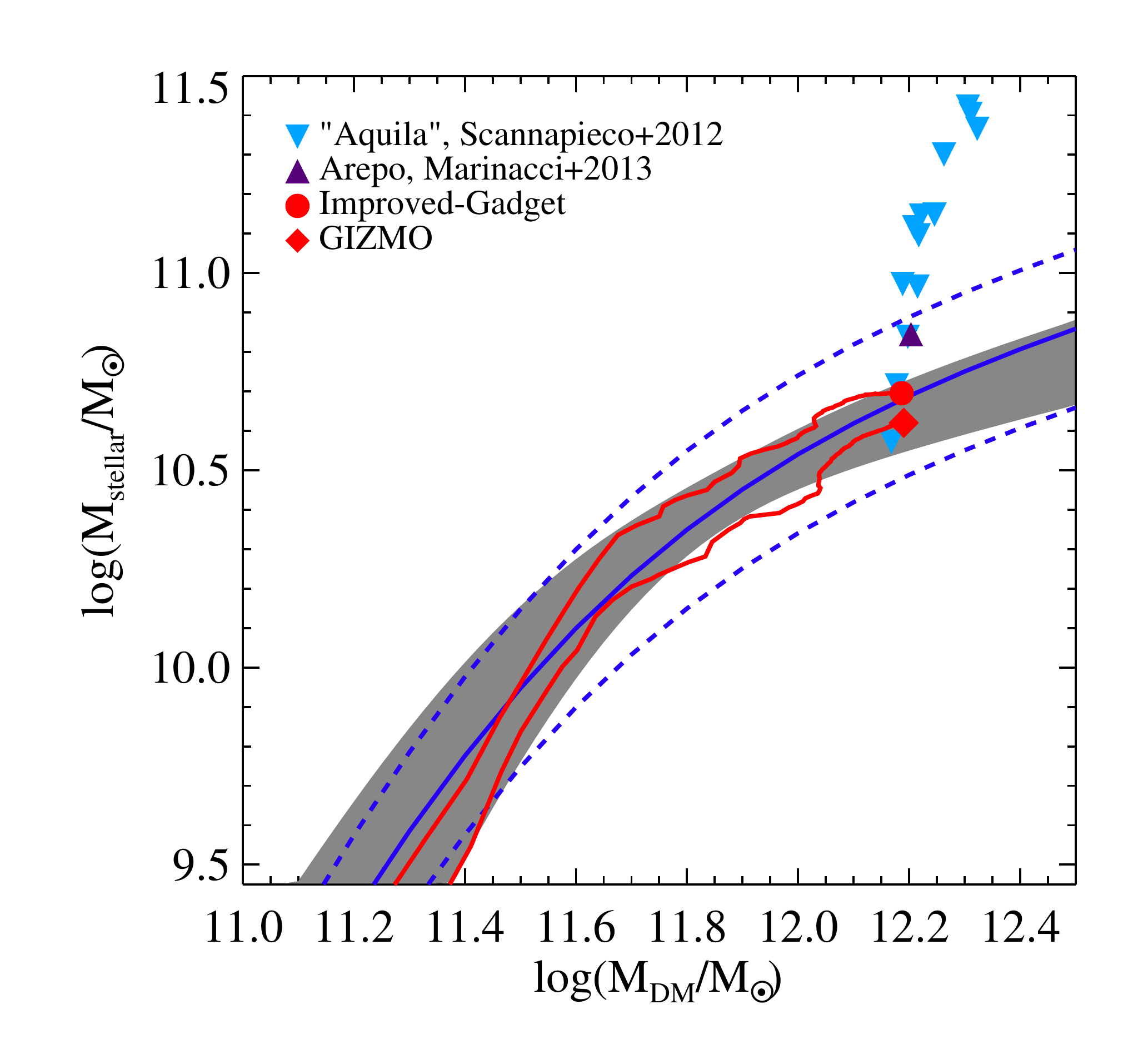}
\caption{\label{fig:halo_star} Stellar -- dark matter mass relations from both Improved-{\Gadget} and {\Gizmo} simulations, respectively, in comparison with previous simulation results from the Aquila Comparison Project \citep{Scannapieco2012}  (blue downward triangles) and \cite{Marinacci2014} (green upward triangles). The semi-analytic result from \cite{Moster2013} is indicated by the gray region, while that of \cite{Guo2010} is shown in solid line with a 0.2 dex deviation in dashed lines. The two red lines show the evolutionary tracks of the modeled galaxies from both Improved-{\Gadget} and {\Gizmo} simulations. } 
\end{center}
\end{figure}

However, there is {a} significant difference between the two simulations. Compared to {\sc Gizmo},  star formation rate produced by the Improved-{\Gadget} simulation is substantially higher by nearly a factor of $2$ from high redshift till $z \sim 1$, while it is $\sim50\%$ lower at  $z < 1$. As we will discuss in \S~\ref{subsec:gas}, the difference in star formation  is due to the different gas properties produced from the two simulations. The Improved-{\Gadget} produced substantially more cold dense gas with density $\rm{n_{H} > 10 cm^{-3}}$ than {\sc Gizmo} at high redshifts which fueled stronger starburst activity, while at lower redshift, it produced hotter gas which reduced star formation. 

{A} similar discrepancy in star formation histories was also reported by \cite{Vogelsberger2013} between {\Gadget} and {\Arepo} simulations. Such a difference can be understood from  the ``blob" test in \cite{Agertz2007} as a result of inefficient mixing between the cold and hot phases. As pointed out by \cite{Hopkins2015}, P-SPH simulations typically produce more cold gas than {\Gizmo} unless extra dissipation is applied \citep{Hu2014}. At $z < 1$, the gas properties is affected by the feedback  processes and the choice of cooling function. The cooling rate from hot halo gas in the Improved-{\Gadget}  simulation is underestimated due to the excess dissipation \citep{Bauer2012, Nelson2013}, which resulted in a lower star formation rate.  

The assembly of stellar mass from the simulations is shown in Figure~\ref{fig:halo_star}. As a result of {the} stronger star formation, the Improved-{\Gadget} simulation produced a higher stellar mass by $20\%$ at $z = 0$ compared to {\sc Gizmo}. The final stellar masses of our simulated galaxies, however, are within 1$\sigma$ deviation from {constraints} derived from a multi-epoch abundance matching technique by \cite{Moster2013}. Clearly, our results show better agreement  with abundance matching than the {\Arepo} simulation by \cite{Marinacci2014} and those from the Aquila Comparison Project \cite{Scannapieco2012}. 

Historically,  star formation efficiency in hydrodynamic simulations is commonly over-estimated and notoriously difficult to control \citep[e.g.][]{Agertz2011, Scannapieco2012}. Maximum star formation efficiency occurs around halo mass of $\sim10^{12} M_{\odot}$ for halos at $10^{12}\Msun$. This mass is the turning point in semi-analytic relations. Star formation efficiency quickly drops toward both higher and lower mass halos. The efficiency of star formation at $10^{12}\Msun$ is around 25\%. As shown by \cite{Moster2013}, previous hydrodynamic simulations produce a higher star formation efficiency. For example, disk galaxies with good morphologies in \cite{Agertz2011} show an efficiency of  $\sim 80\%$ to turn gas into stars. 

The discrepancy between hydrodynamic simulations and semi-analytic relations is more severe at high redshifts when most of the excess of stellar mass in previous hydrodynamic simulation is ashydrodynamicsembled \citep[][]{Scannapieco2012, Moster2013}. In Figure~\ref{fig:halo_star}, we also show the evolutionary tracks of two simulated galaxies in red solid lines. Overall, both of the tracks are within the $z = 0$ constrains imposed by both abundance matching relations. The total mass of the two simulations tracks very well with each other due to the same gravity solver. However, differences in the stellar mass assembly due to the choice of hydro solver is present. The Improved-Gadget simulation assembles its stellar mass faster than {\Gizmo} at early stage. At later time, {\Gizmo} run has faster growth rate in stellar mass while Improved-Gadget seems still suffering from excess heating from numerical artifacts mentioned above. It is interesting that the difference between their stellar mass is reduced by $z = 0$ as the stellar mass in {\Gizmo} catches up with the Improved-Gadget run. 

Compared to other SPH simulations in the Aquila Comparison Project \citep{Scannapieco2012}, we are able to reduce star formation in the early stage with the decoupled ``energy-driven" wind  model \citep{Puchwein2013, Vogelsberger2013} by imposing stronger mass loading in the wind. This outflow model is instrumental to reproduce the expected mass growth from  semi-analytical calculations. We note that AGN feedback from central supermassive black holes is not included in our simulations. It is proposed that AGN feedback might be responsible for the reduction of star formation efficiency in halos more massive than the one we considered in this study  \citep{Silk1998, DiMatteo2005, Schawinski2007}. 

\subsection{The Gas Properties}
\label{subsec:gas}

Gas plays a critical role in the formation and evolution of galaxies. In this section, we investigate important gas properties such as the density-temperature relation and the distribution of cold and hot gas, in order to understand the different star formation history and galaxy morphology between the Improved-Gadget and {\Gizmo} simulations discussed in the previous sections.

\subsubsection{The Multi-Phase ISM}

\begin{figure}
\begin{center}
\includegraphics[scale=0.4]{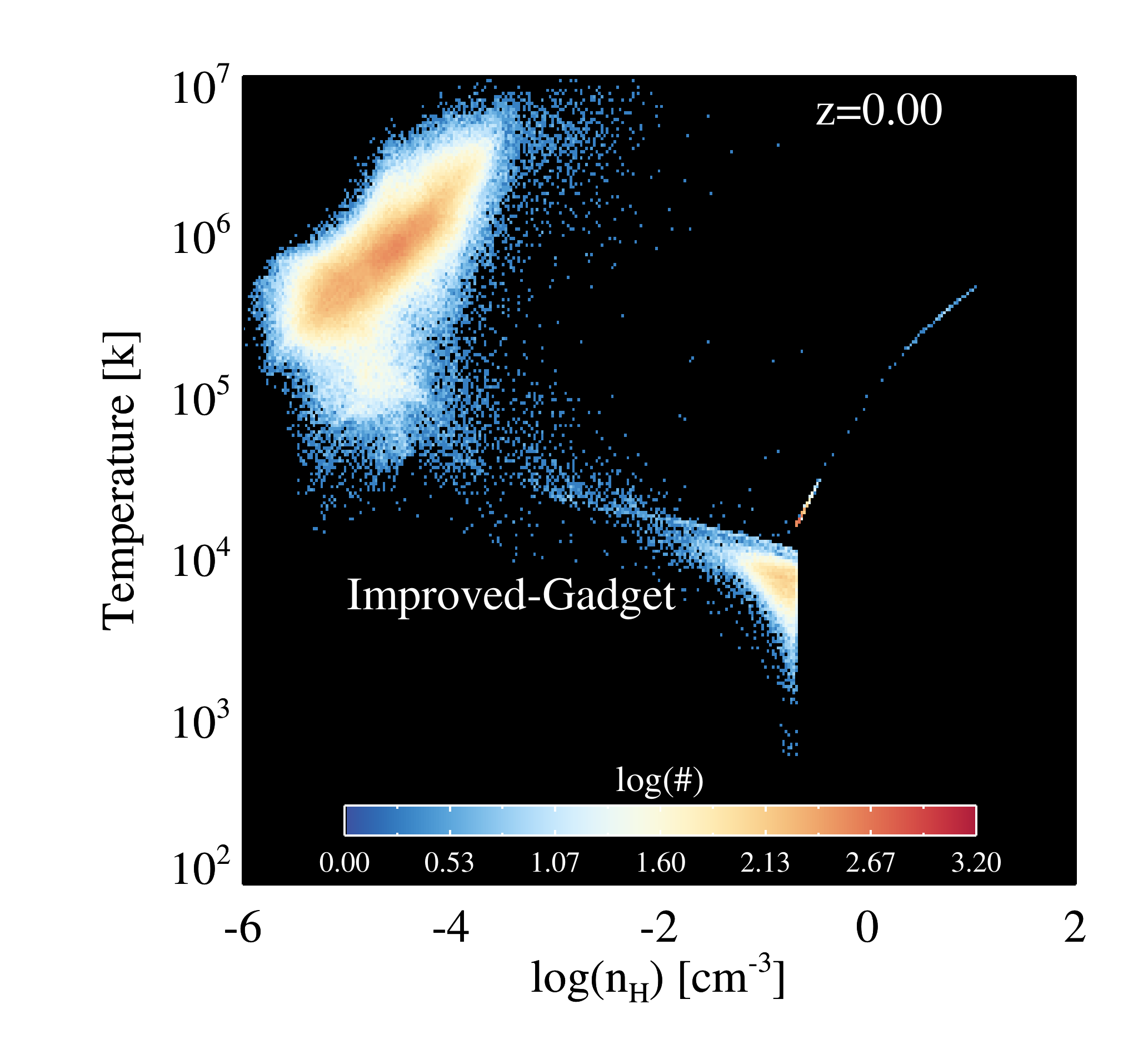}
\includegraphics[scale=0.4]{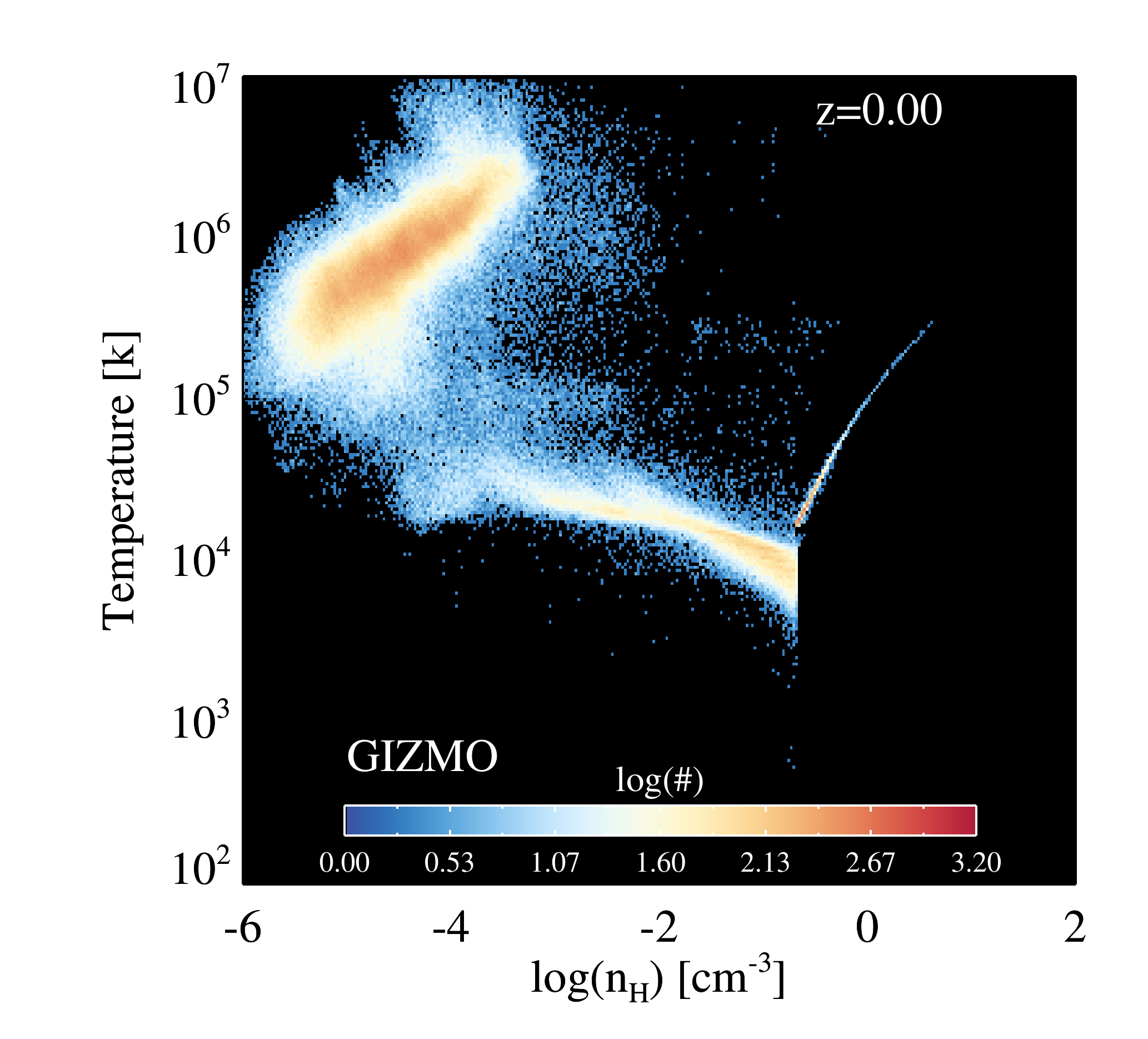}
\end{center}
\caption{\label{fig:gas_phase} Gas density-temperature phase diagram at $z = 0$ from both Improved-{\Gadget} and {\Gizmo} simulations, respectively. Both simulations use the same models for star formation and gas cooling and heating. Note that the gas with density above star formation threshold $\rm{n_{H} = 0.15\, cm^{-3}}$ is placed in an effective equation of state developed by \cite{Springel2003}, which appears to be a ``tail" in the phase diagram.}
\end{figure}

In our simulations, a sub-grid multi-phase ISM recipe developed by \cite{Springel2003} and {metal-dependent cooling functions} by  \cite{Wiersma2009} are used to model the gas, and star formation would occur when the gas reaches the threshold  density $\rm{n_{H} = 0.15\, cm^{-3}}$. In this model,  the ``cold'', ``warm'' and ``hot'' phase is defined by gas temperature only, and it refers to temperature $T < 2\times10^5$K, $\rm{2\times10^5\, K  < T < 10^6\, K}$, and $T > 10^6$K, respectively, while ``star-forming gas'' refers {to} gas with density above star formation threshold $\rm{n_{H} > 0.15\, cm^{-3}}$. 

Figure~\ref{fig:gas_phase} shows the resulting density-temperature phase diagram of gas at $z = 0$ from both Improved-{\Gadget} and {\Gizmo} simulations. While both simulations show {a} similar trend in the density-temperature relation, the Improved-{\Gadget} produced more hot, diffuse gas ($\rm{T > 10^6\, K}$ and $\rm{n_H < 10^{-3}\, cm^{-3}}$) and extremely dense gas clumps ($\rm{n_H > 10\, cm^{-3}}$), but showed a ``gap" in the density-temperature range of $\rm{10^{-4}\, cm^{-3} < n_H < 10^{-3}\, cm^{-3}}$ and $\rm{10^4\, K  < T < 10^6\, K}$, while the {\Gizmo} simulation produced more gas in that ``gap'' and no dense clumps with $\rm{n_H > 10\, cm^{-3}}$.

The existence of large amount of cold, dense gas in the Improved-Gadget simulation is mainly due to the inefficient mixing between the cold and hot phases in SPH. as demonstrated by the ``blob test" \citep{Agertz2007, Hopkins2015}. The different distribution of gas in the ``gap'' region is mainly due to {the} differences in radiative cooling {of the} hot halo gas. The excessive dissipation from artificial viscosity results in inefficient cooling from hot halo gas in the Improved-Gadget simulation. Similar result was also reported by \cite{Vogelsberger2012}, who found that {\Arepo} produced more pronounced distribution of gas around $\rm{T\sim10^{4-5}\, K}$ than Improved-{\Gadget}. {\Gizmo} shows {a} significant advantage over SPH on modeling multi-phase ISM due to its ability {to resolve} the Kelvin-Helmholtz instability and getting rid of the artificial viscosity. 

\begin{figure}
\begin{center}
\begin{tabular}{cc}
\resizebox{1.65in}{!}{\includegraphics{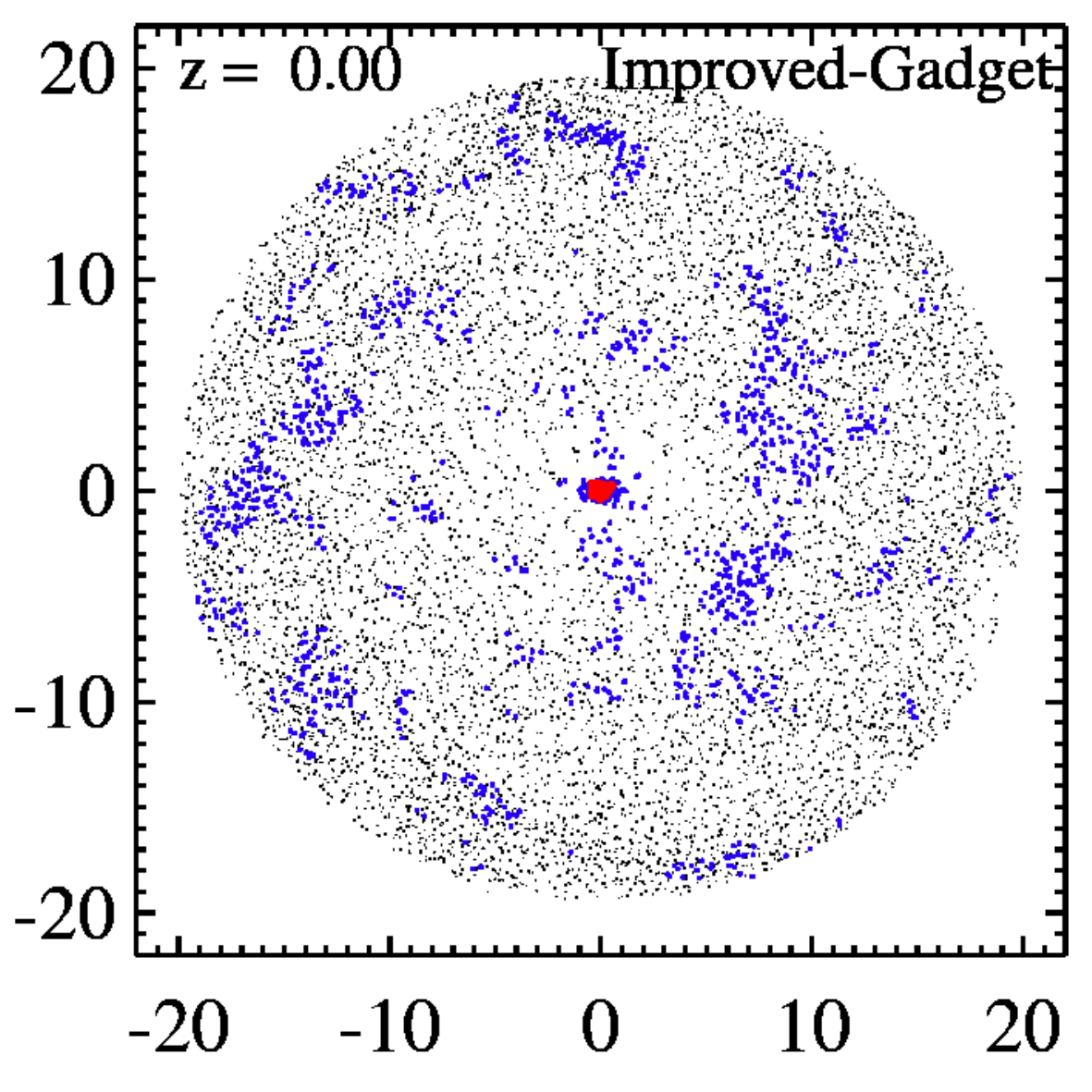}}
\resizebox{1.65in}{!}{\includegraphics{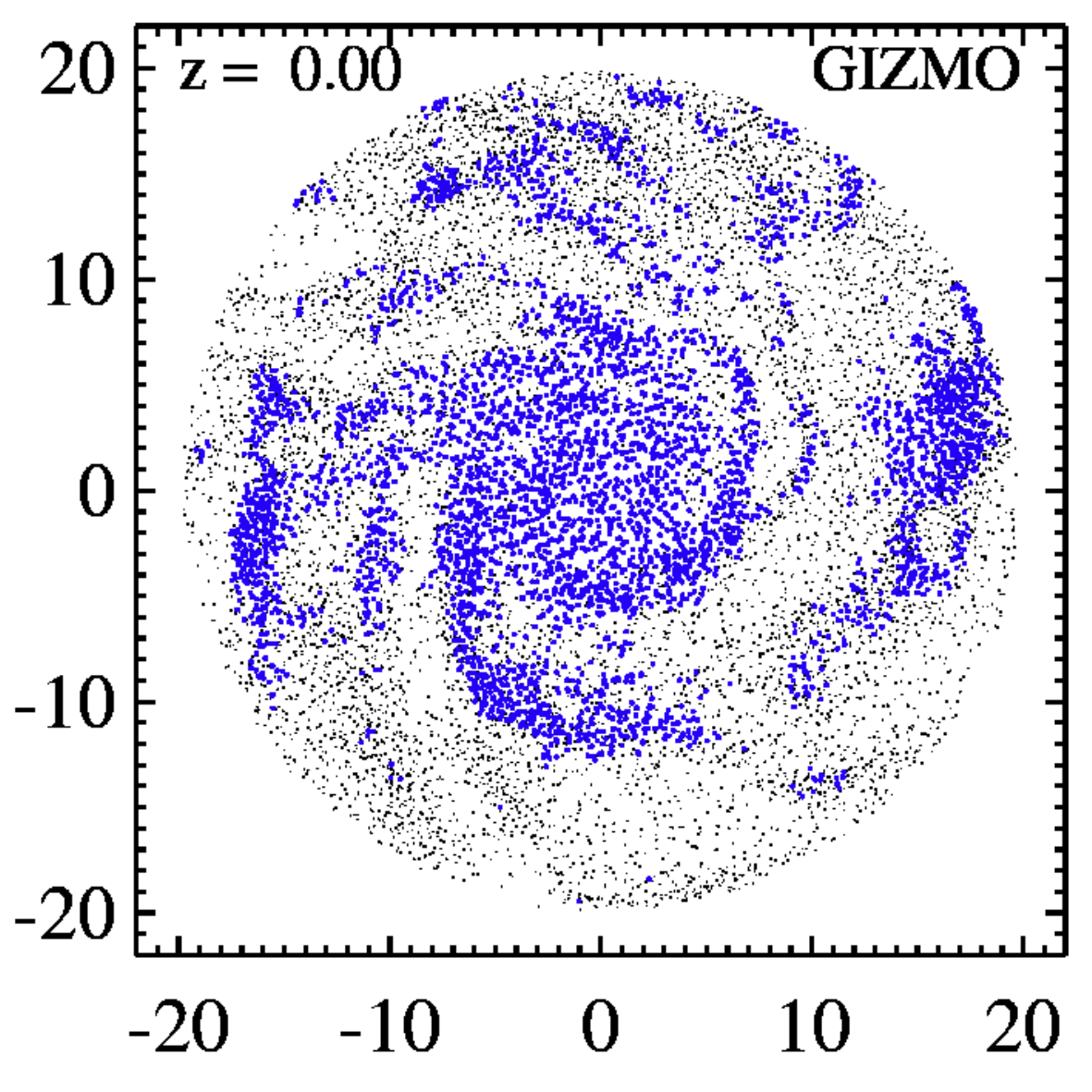}}
\end{tabular}
\caption{\label{fig:sph_gas_particle_distribution} Distribution of gas within 20 kpc at $z=0$ from both Improved-{\Gadget} and {\Gizmo} simulations, respectively.  The star-forming gas ($\rm{T < 2 \times10^5\, K}$ and $\rm{n_{H} > 0.15\, cm^{-3}}$) is shown in blue color, the very dense gas with $\rm{n_H > 10\, cm^{-3}}$ in red, and the non-star-forming gas is in black. }
\end{center}
\end{figure}

In addition, angular momentum transfer from the numerical viscosity also helps the gas lose angular momentum and move to the central region. It builds up the density and mass quickly in the large gravitational potential well. Indeed, as demonstrated in Figure~\ref{fig:sph_gas_particle_distribution}, the very dense gas clumps with $\rm{n_H > 10\, cm^{-3}}$,  {which are found to be} concentrated in the galaxy center in the Improved-Gadget simulation, are absent in the {\Gizmo} simulation. 

\subsubsection{The Star-Forming Gas}

\begin{figure}
\begin{center}
\includegraphics[scale=0.4]{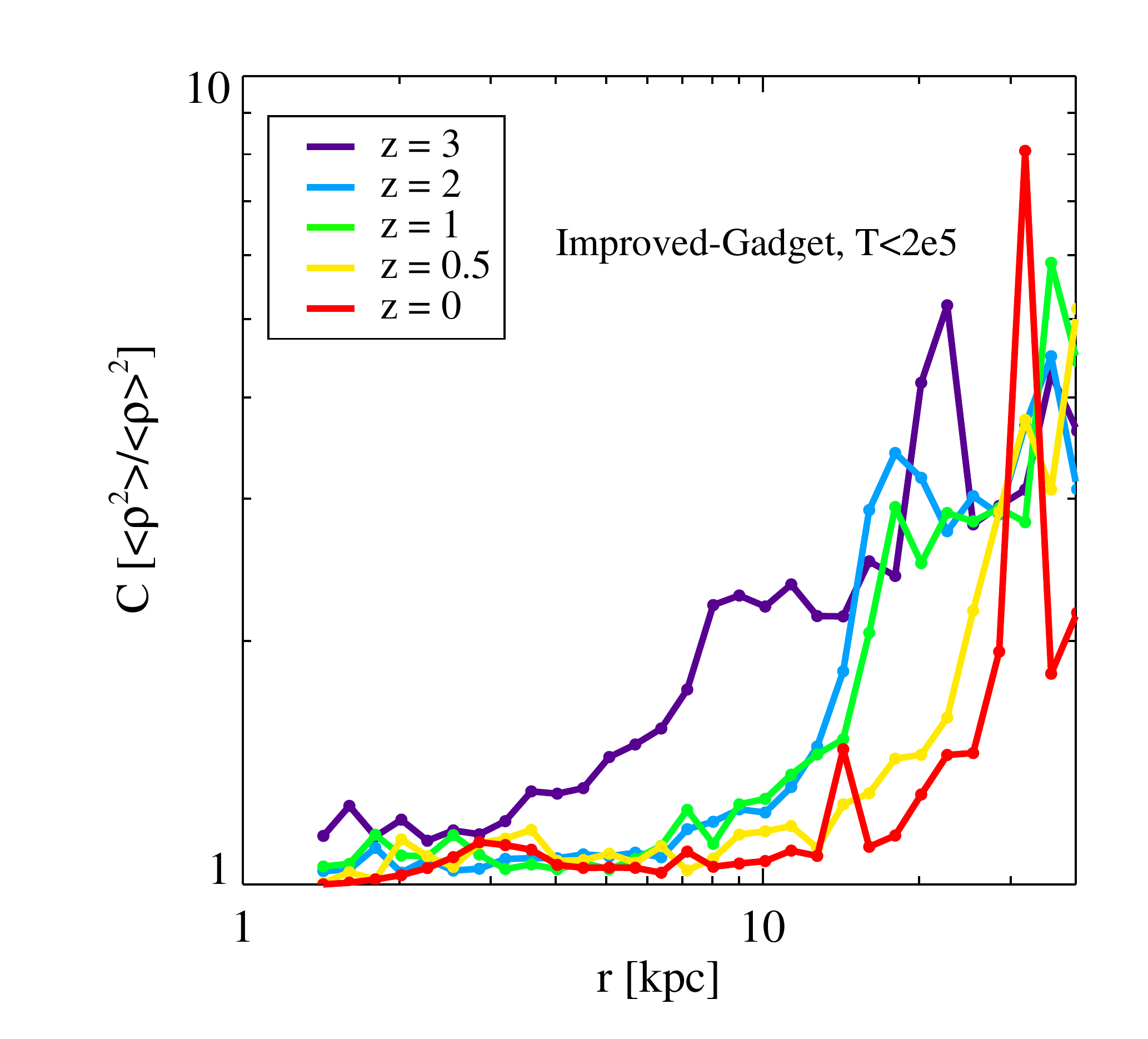}
\includegraphics[scale=0.4]{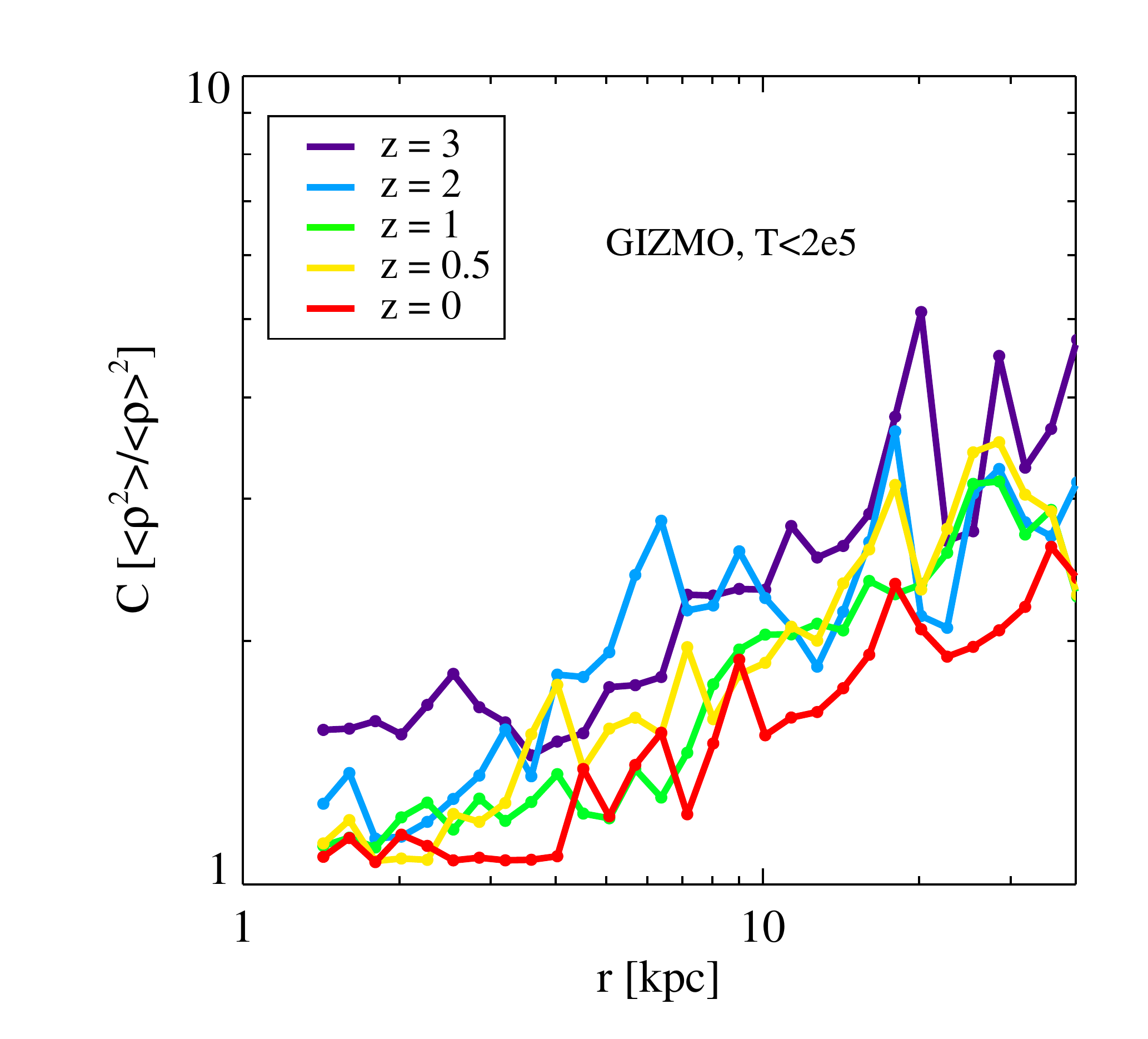}
\caption{\label{fig:ism_clump} The evolution of radial distribution of clumpiness of cold gas $C$ (with temperature $T < 2\times10^5$K) from both Improved-{\Gadget} and {\Gizmo} simulations, respectively. The different color represents different redshift. }
\end{center}
\end{figure}

To quantify gas distribution, we compute a clumping factor of gas as follows: 
\begin{equation} 
C \equiv \frac{<\rho^2>}{<\rho>^2}  = \frac{V^{-1}{\int_V\rho^2dV}}{[V^{-1}{\int_V\rho^2dV}]^2}, 
\end{equation} 
where the integral is approximated by summing up all the particles within a set of shells with volume $V$ at each galactic distance $r$. We use $m/\rho$ for the ``volume'' $dV$ associated with each particle.  This quantity measures the dispersion of gas density around the mean value.  For comparison, $C = 1$ suggests no clumping within the gas distribution while a higher value indicates a higher degree of inhomogeneity. 

Figure~\ref{fig:ism_clump} shows the radial distribution of gas clumping factor at different redshifts from both Improved-{\Gadget} and {\Gizmo} simulations. Within 10 kpc at $z = 0$, the Improved-{\Gadget} simulation shows a lower values of $C$ compared to {\Gizmo}. In addition, it shows a stronger evolution of $C$ with redshift, and there is a tendency of small scale structures being slowly smeared out in the simulation as $C$ decreased from $z = 3$ to $z = 0$. This is due to the low-order noise and artificial viscosity in SPH which damps small-scale turbulent motions. The different distribution of  $C$  in the two simulations explains the smoother distribution of the gas disk in the Improved-{\Gadget} simulation as seen in \S~{\ref{subsec:morphology}. 

\begin{figure}[h]
\begin{center}
\includegraphics[scale=0.4]{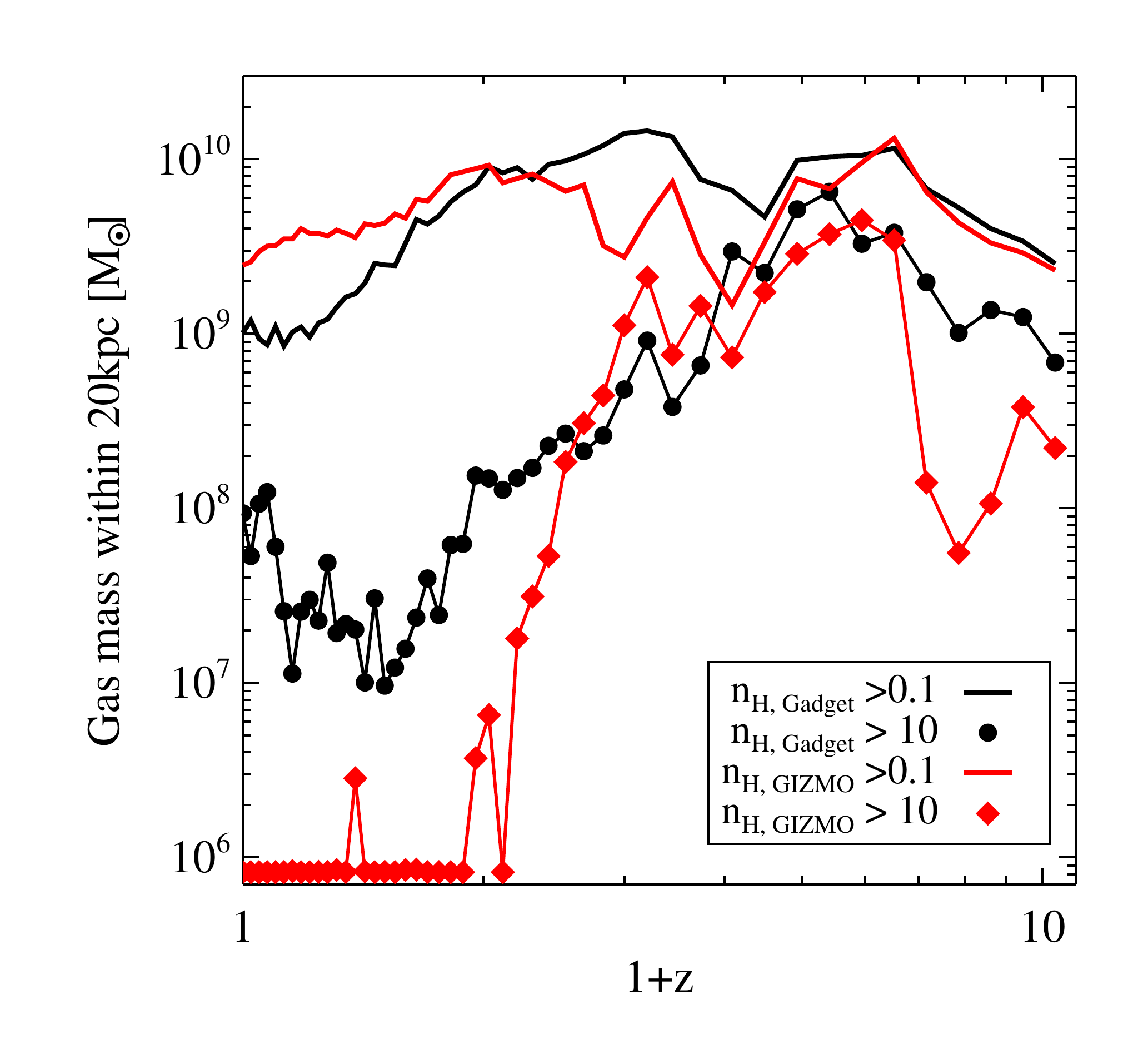}
\caption{\label{fig:cold_gas_mass} Total mass of cold gas within 20 kpc (in physical coordinates) from galaxy center as a function of redshift from both Improved-{\Gadget} and {\Gizmo} simulations, respectively. The star-forming gas ($\rm{n_{H} > 0.1 cm^{-3}}$) is represented in solid lines, while the very dense form with $\rm{n_{H} > 10 cm^{-3}}$ in lines with filled symbols.}
\end{center}
\end{figure}

\begin{figure*}
\begin{center}
\begin{tabular}{cccc}
\includegraphics[scale=0.35]{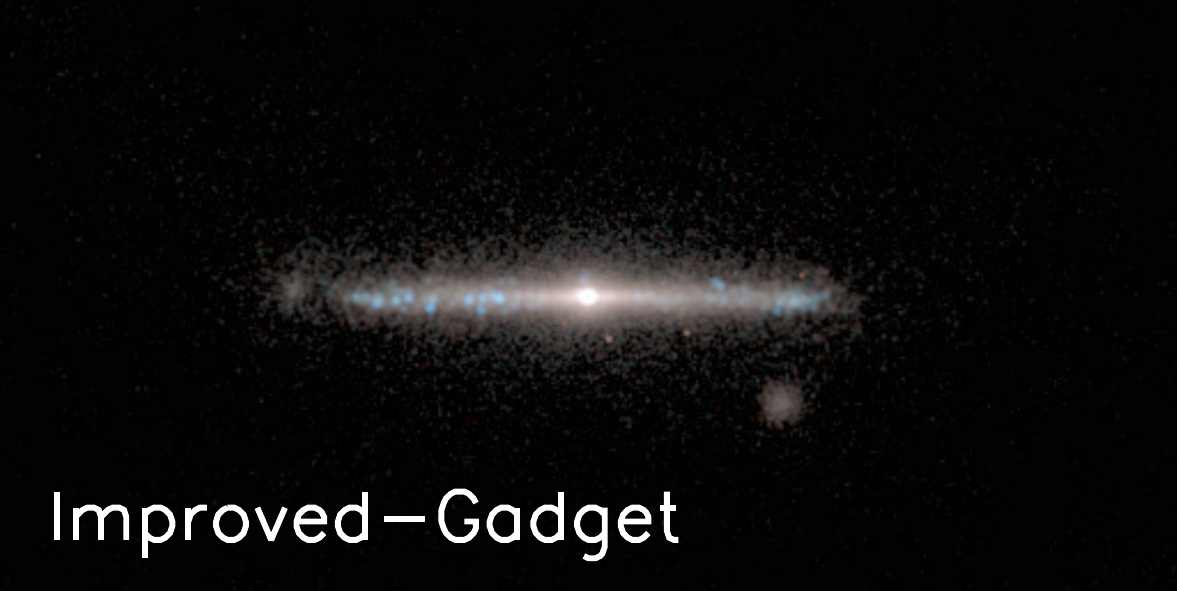}
\includegraphics[scale=0.35]{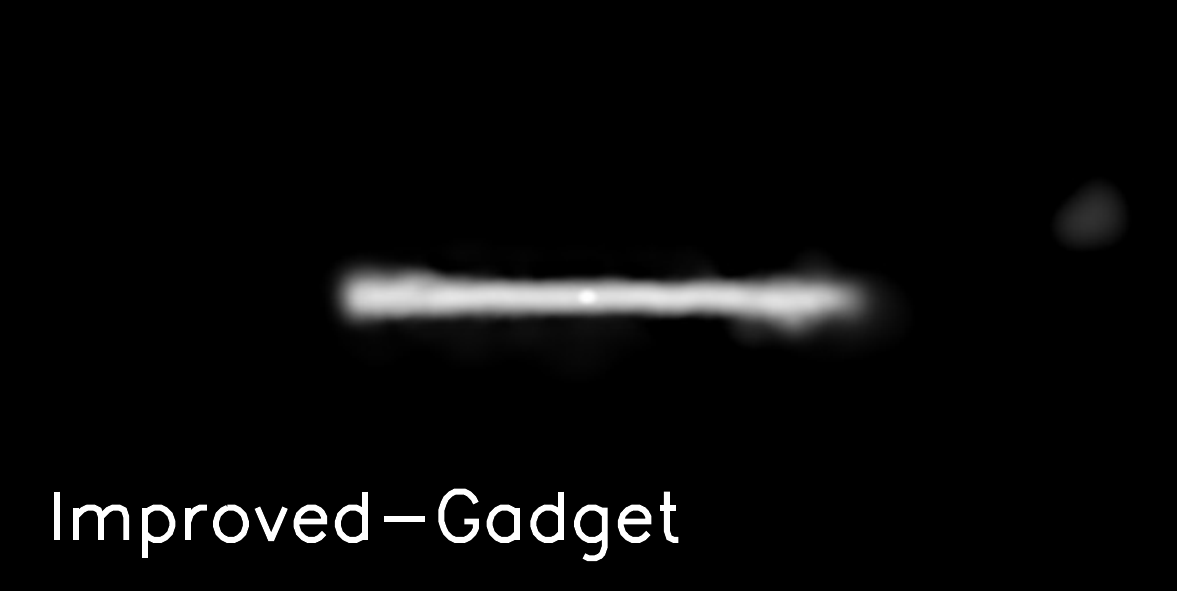}
\includegraphics[scale=0.35]{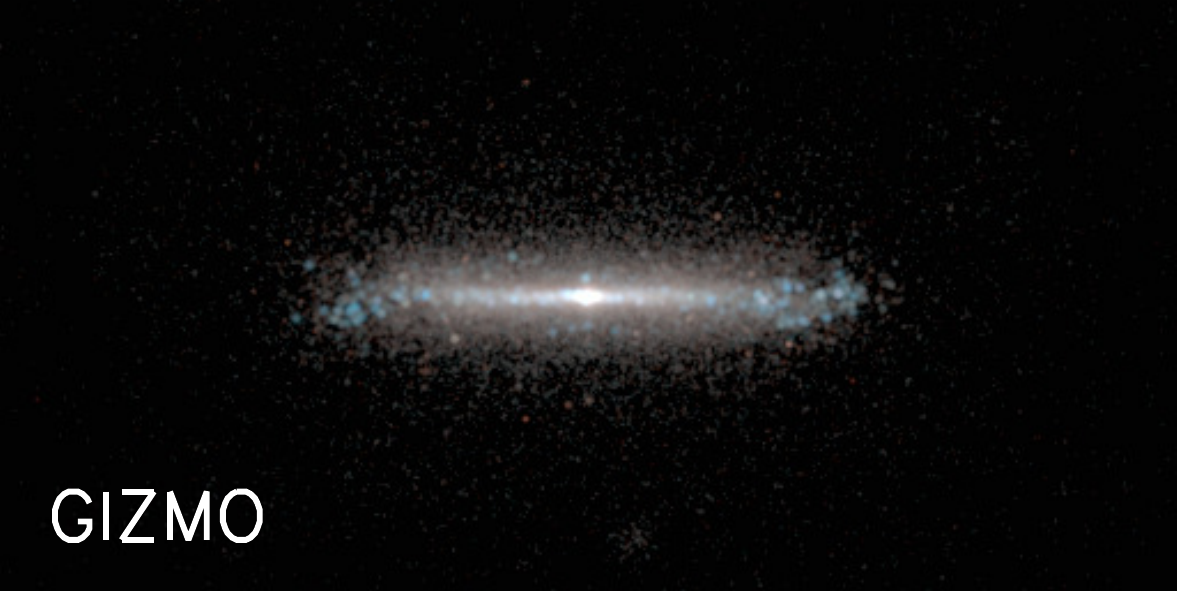}
\includegraphics[scale=0.35]{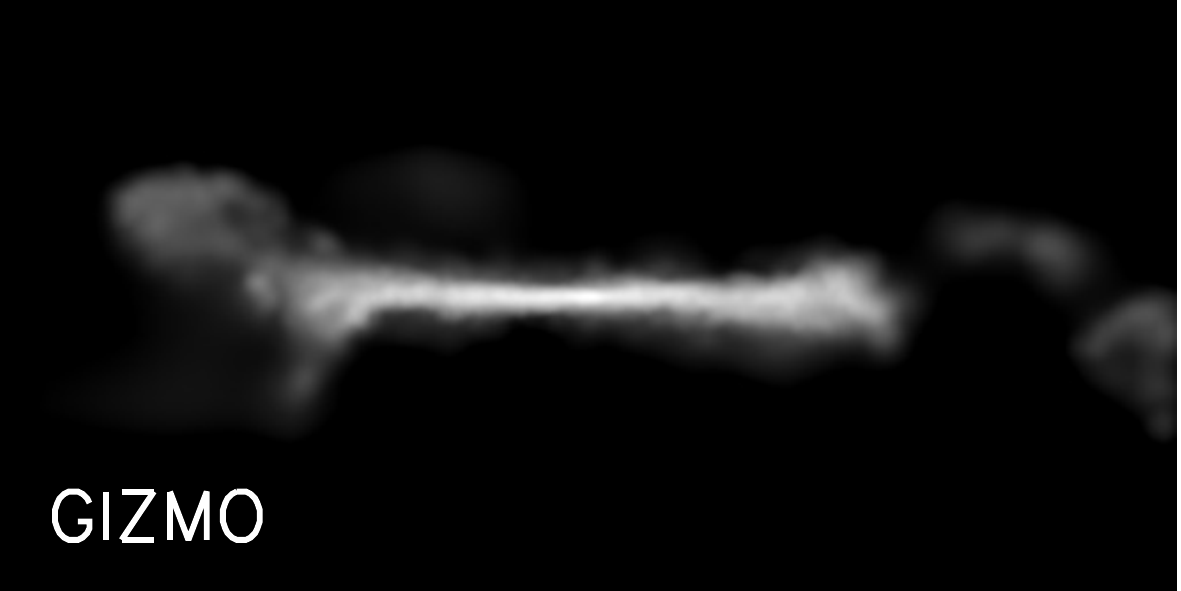}\\
\includegraphics[scale=0.35]{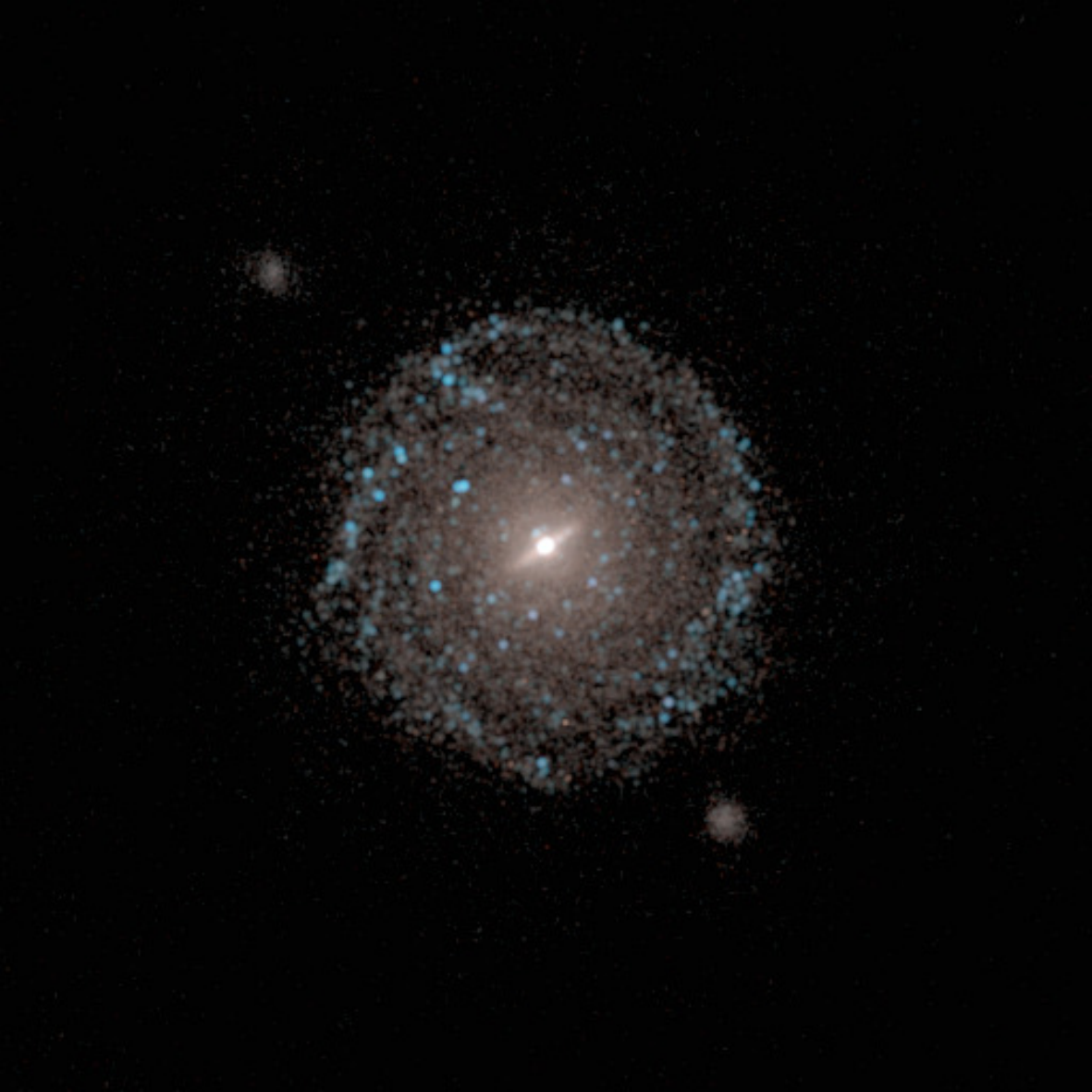}
\includegraphics[scale=0.35]{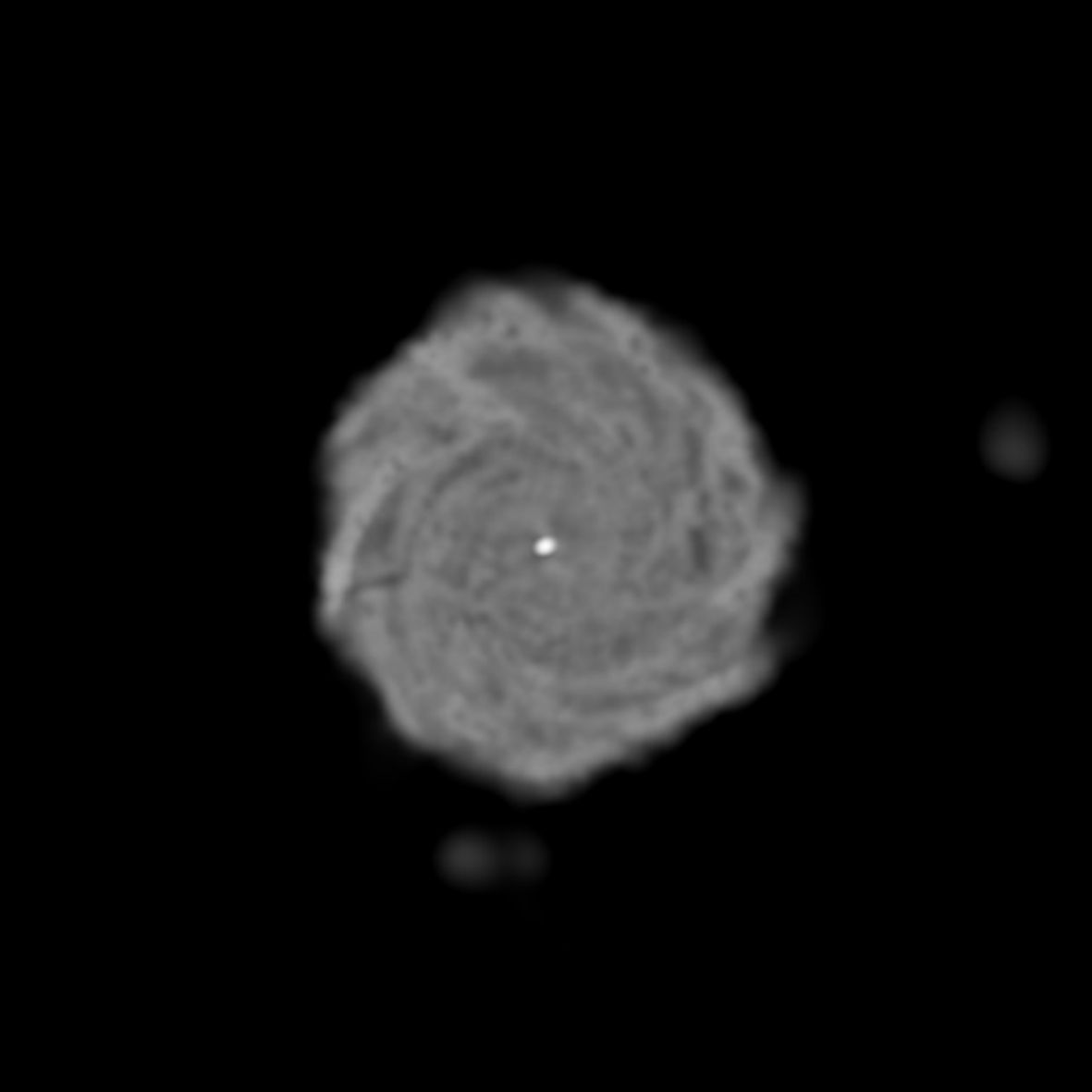}
\includegraphics[scale=0.35]{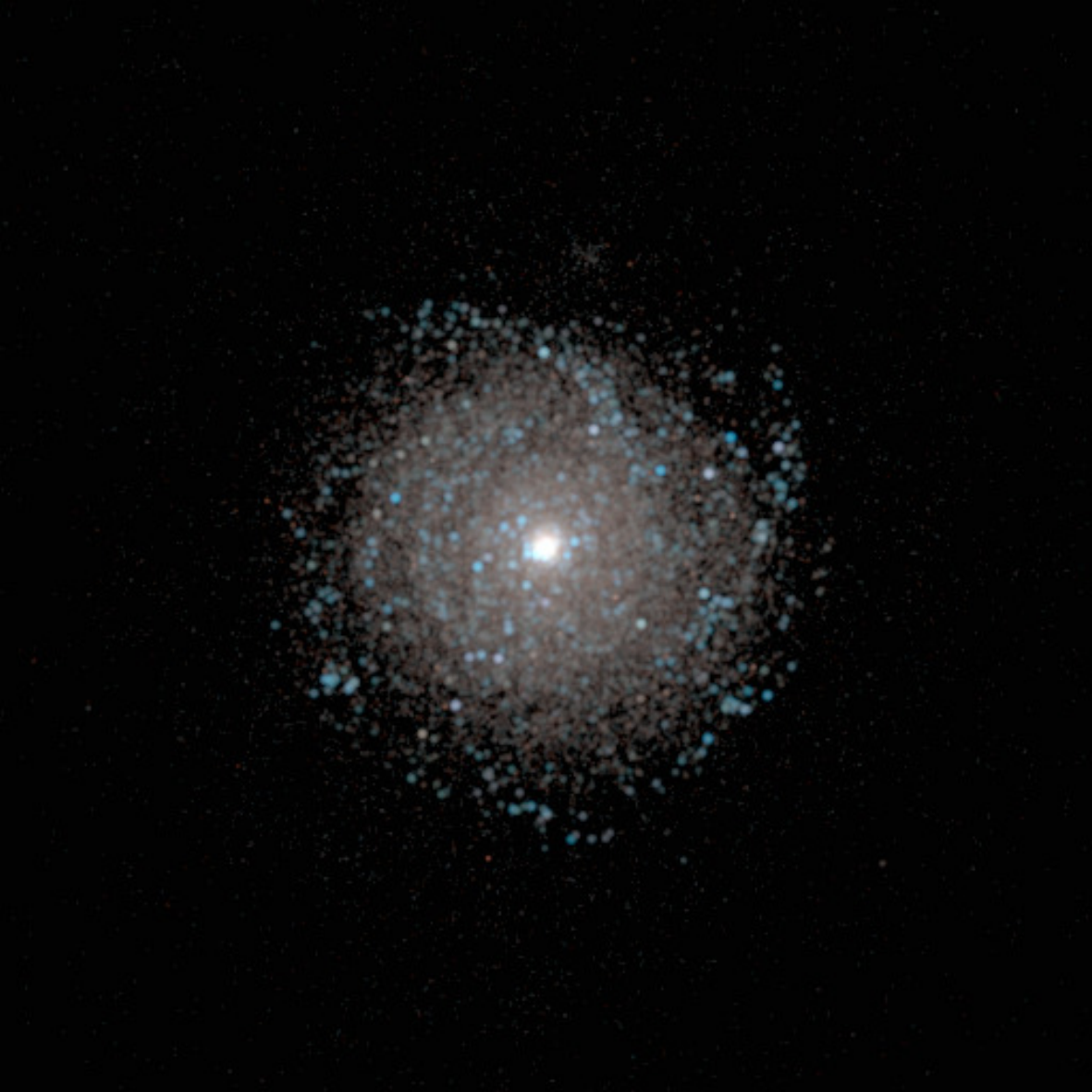}
\includegraphics[scale=0.35]{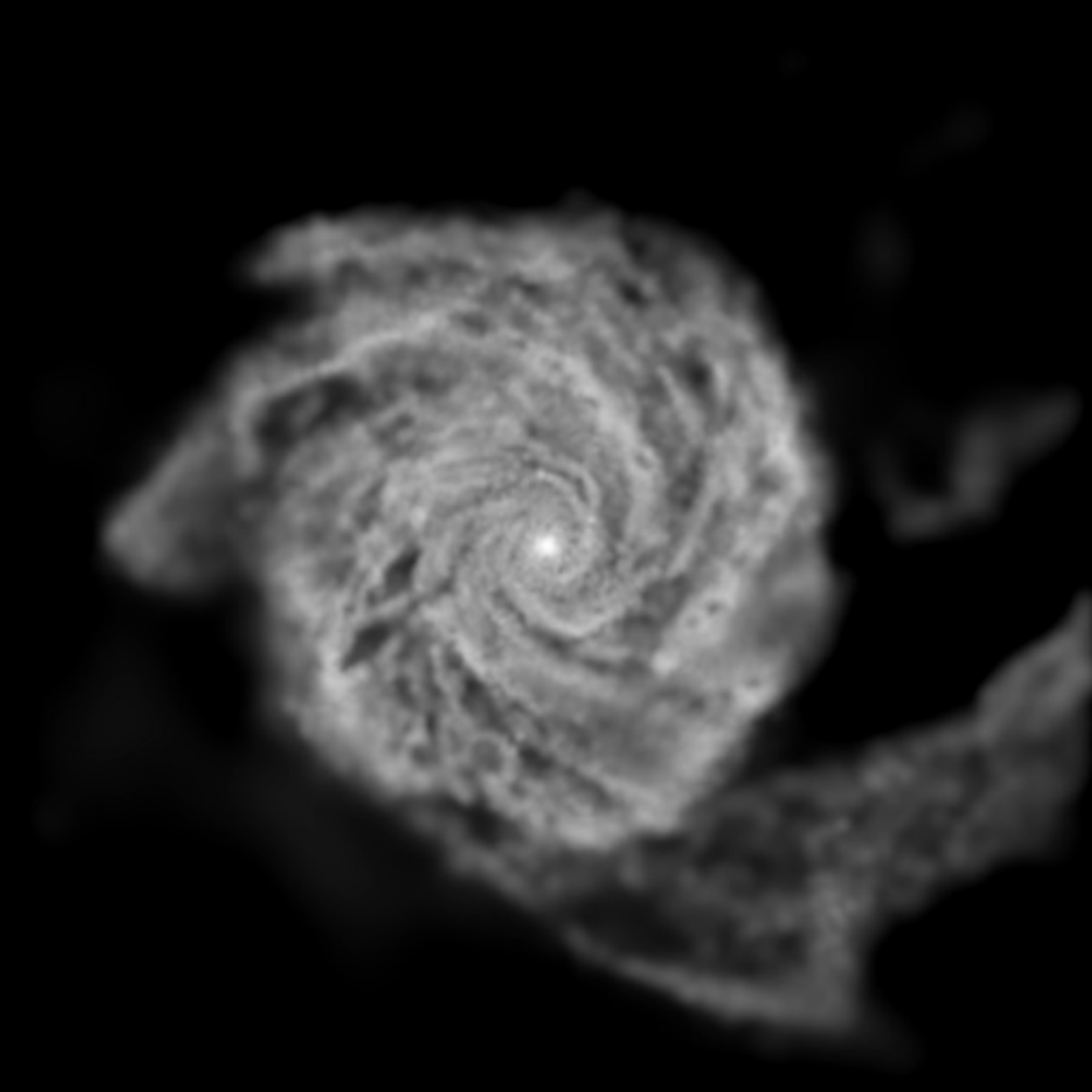}\\
\end{tabular}
\caption{\label{fig:stellar_image_edge} Edge-on and face-on views of the projected stellar and gas density maps at $z = 0$ from both Improved-{\Gadget} and {\Gizmo} simulations, respectively. The width of each panel is 100 kpc in physical coordinates. Similar to Figure~\ref{fig:star_gas_images}, the luminosity of the stars is calculated from Starburst99 based on its mass, age and metallicity, while the colors are assigned based on rest-frame $K$, $V$ and $B$-band magnitudes in $rgb$ channels, such that young stars appear in blue color, while old ones appear in red. A grey scale is used for the gas density to better visualize the contrast. The density maps are projected to rotated coordinates according to the angular momentum of stellar disk for the edge-on and face-on views.}
\end{center}
\end{figure*}

In Figure~\ref{fig:cold_gas_mass}, we compare the total mass of both star-forming gas ($ \rm{n_{H} > 0.1 cm^{-3}}$) and the highly dense gas ($ \rm{n_{H} > 10 cm^{-3}}$) within the central 20 kpc at different redshift from both Improved-{\Gadget} and {\Gizmo} simulations. Overall, the Improved-{\Gadget} produced significantly more dense gas, in particular those with $ \rm{n_{H} > 10 cm^{-3}}$, than {\Gizmo} at high redshifts. Around $z = 4$, which is the peak of star formation in {this} galaxy's history, the total mass of dense gas with $\rm{n_{H} > 10 cm^{-3}}$ is $\sim 6\times10^9\Msun$ in Improved-{\Gadget}, twice as larger than that in {\Gizmo}. Such cold, dense gas in the form of clumps resulted from inefficient mixing between the cold and hot phases of SPH, as reported in many ``cold blobs'' tests \citep{Agertz2007}.  From $z = 4$ to $z = 0$, the amount of very dense gas declined in both Improved-{\Gadget} and {\Gizmo} simulations, though more rapidly in the latter. By $z = 0$, almost all of the star-forming gas in {\Gizmo} is contributed by gas with density $\rm{n_{H} < 10 cm^{-3}}$. Our results suggest that inefficient fluid mixing in the Improved-{\Gadget} contributes to the substantially higher star formation rates at redshift $z > 1$ and the resulting higher stellar mass, while the inefficient cooling from hot halo gas explains the lower star formation rate at $z <1$. 

The concentration of dense gas in the galactic center and {an} intense starburst at high redshift produced a bar-like structure in our Improved-{\Gadget} simulation, as seen in stellar density map of Figure~\ref{fig:star_gas_images}. {A} similar structure was also reported by \cite{Okamoto2013} in {their} simulation of the same halo. However, since black holes and AGN feedback were not included in these simulations, it is unclear how they may affect the distribution of gas and stars in the galactic center.

\subsection{Properties of Galaxy Disk}
\label{subsec:disk}

\subsubsection{{Disk} Structure}

The creation of rotationally supported disks of gas and stars in galaxies has been a long-standing challenge for galaxy simulations \citep{Scannapieco2012}. With the improved numerical codes and physical models, both of our Improved-{\Gadget} and {\Gizmo} simulations produced extended gaseous and stellar disks, as shown in  Figure~\ref{fig:stellar_image_edge}. From the visual impression, the stellar and gaseous disks from Improved-{\Gadget} appear to be smaller, thinner and smoother than those from {\Gizmo}. A high concentration of {the} dense gas ($n_{\rm H} > 10$) with {a} total mass of $\sim 10^8 \Msun$ is present at the center of the Improved-{\Gadget} galaxy, as also shown in Figure~\ref{fig:sph_gas_particle_distribution}. The {\Gizmo} simulation shows a flared gas disk with inflow streams. 

\begin{figure}
\begin{center}
\includegraphics[scale=0.4]{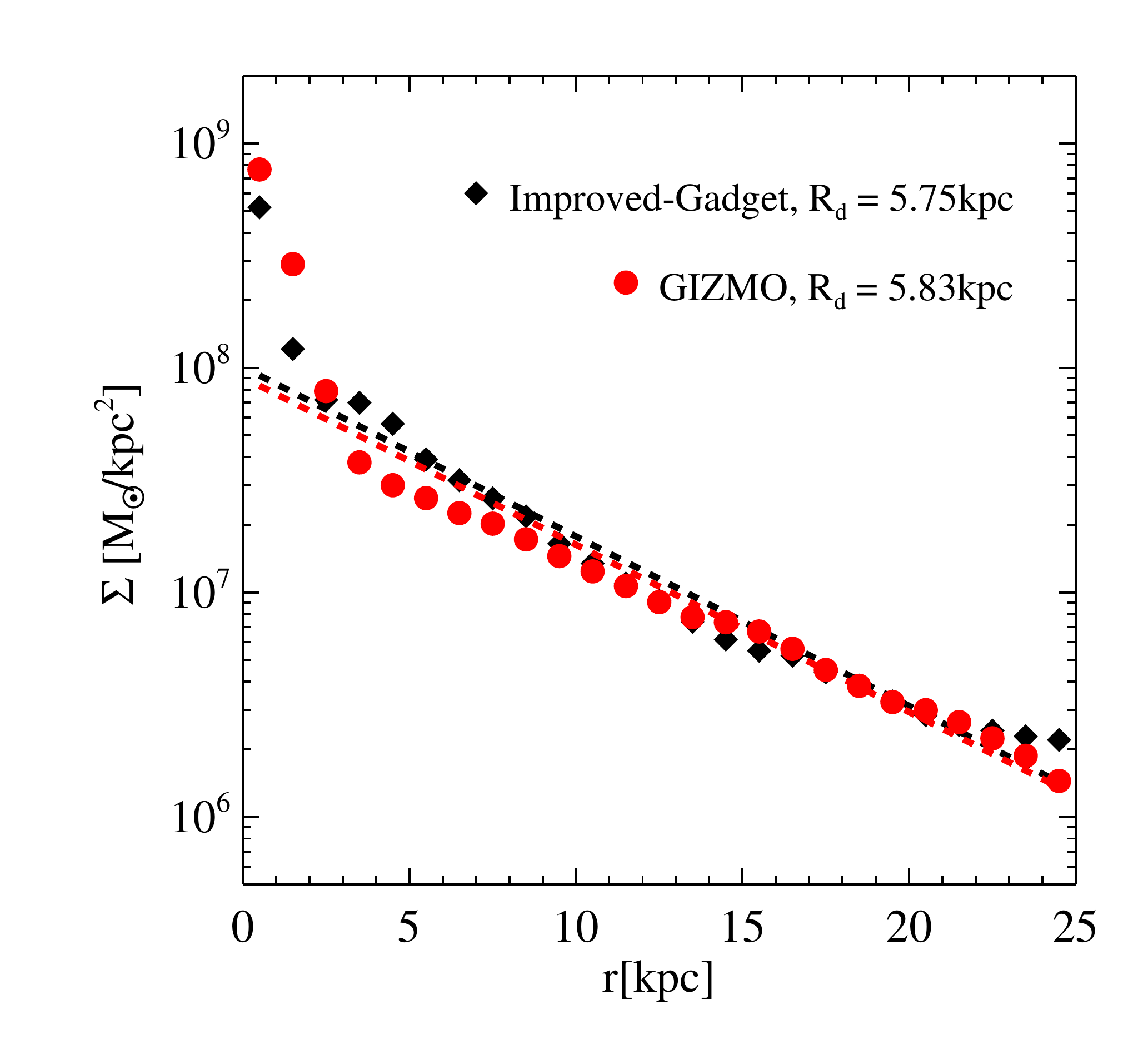}
\caption{\label{fig:surface_density}  Surface density profile of the stellar disks from both Improved-{\Gadget} and {\Gizmo} simulations, respectively. To obtain the disk scale length $R_{\rm d}$, we fit the profile with a single exponential function, $\Sigma(r) = \Sigma_0 \exp(-r/R_{\rm d})$, as represented by the dashed lines.  }
\end{center}
\end{figure}

Figure~\ref{fig:surface_density} shows the surface density profiles of the stellar disks from both Improved-{\Gadget} and {\Gizmo} simulations. Both galaxies show similar profiles. By fitting the profiles with a single exponential function, $\Sigma(r) = \Sigma_0 \exp(-r/R_{\rm d})$, we obtain the scale length of the stellar disk, $R_{\rm d} = 5.73$ kpc and $R_{\rm d} = 5.85$ kpc for the Improved-{\Gadget} and {\Gizmo} simulations, respectively.

\begin{figure}
\begin{center}
\includegraphics[scale=0.4]{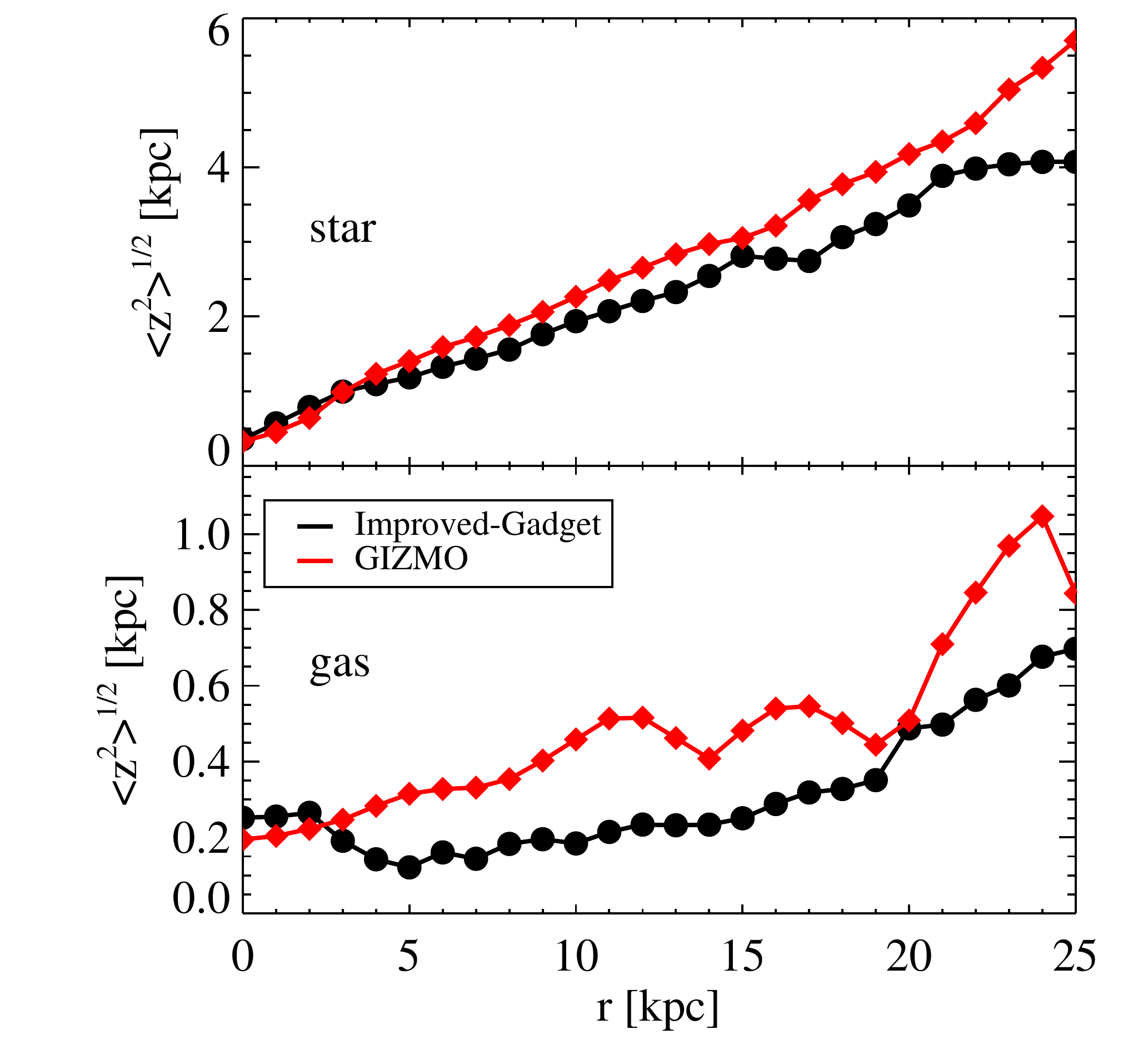}
\caption{\label{fig:vertical_height} Vertical height of the disk, $\langle z^2 \rangle^{1/2}$, as a function of radius $r$ from both Improved-{\Gadget} and {\Gizmo} simulations, respectively. The top panel shows the stellar disk while the bottom panel shows the cold gas disk. }
\end{center}
\end{figure}

The vertical structure of stellar disk is another important and numerically challenging problem in galaxy simulations, since the observed scale heights of the thin disks are of the order of 100 pc, which is significantly smaller than those from previous simulations \citep{Scannapieco2012}, and has only been achieved recently \citep[e.g.][]{Wetzel2016}. Following \cite{Marinacci2014}, we use a mass-weighted second moment of the z-coordinate as a measurement of the disk scale height, $\langle z^2 \rangle^{1/2}$. As shown in Figure~\ref{fig:vertical_height}, both stellar and gaseous disks from Improved-{\Gadget} simulation are thinner than those from {\Gizmo}. More interestingly, the gas disk from {\Gizmo} nearly doubles that from Improved-{\Gadget} in scale height, and it shows prominent ``wiggles" consistent with the spiral structures as seen in Figure~\ref{fig:stellar_image_edge}.

\subsubsection{Circular Velocity Curve}

\begin{figure}
\begin{center}
\includegraphics[scale=0.4]{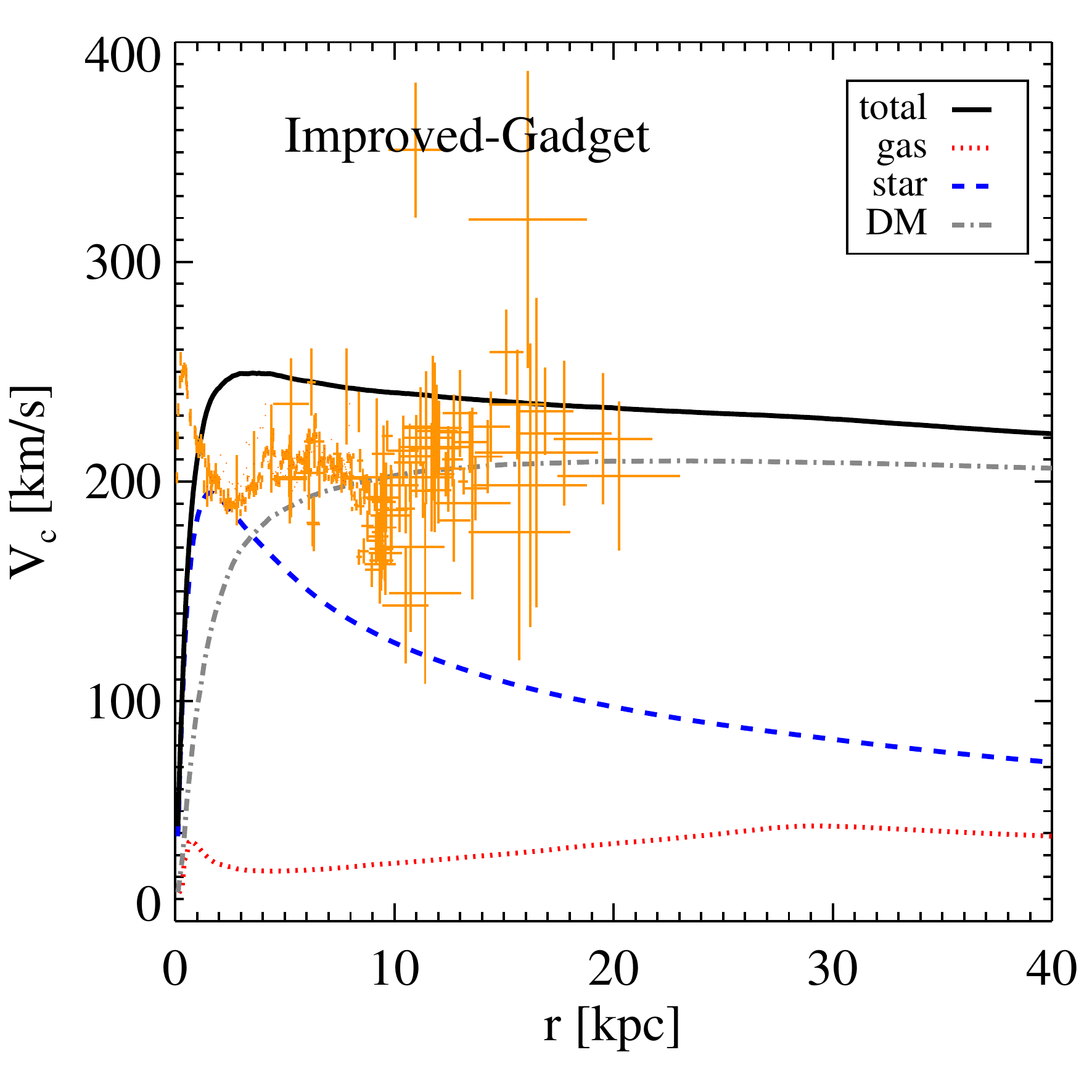}
\includegraphics[scale=0.4]{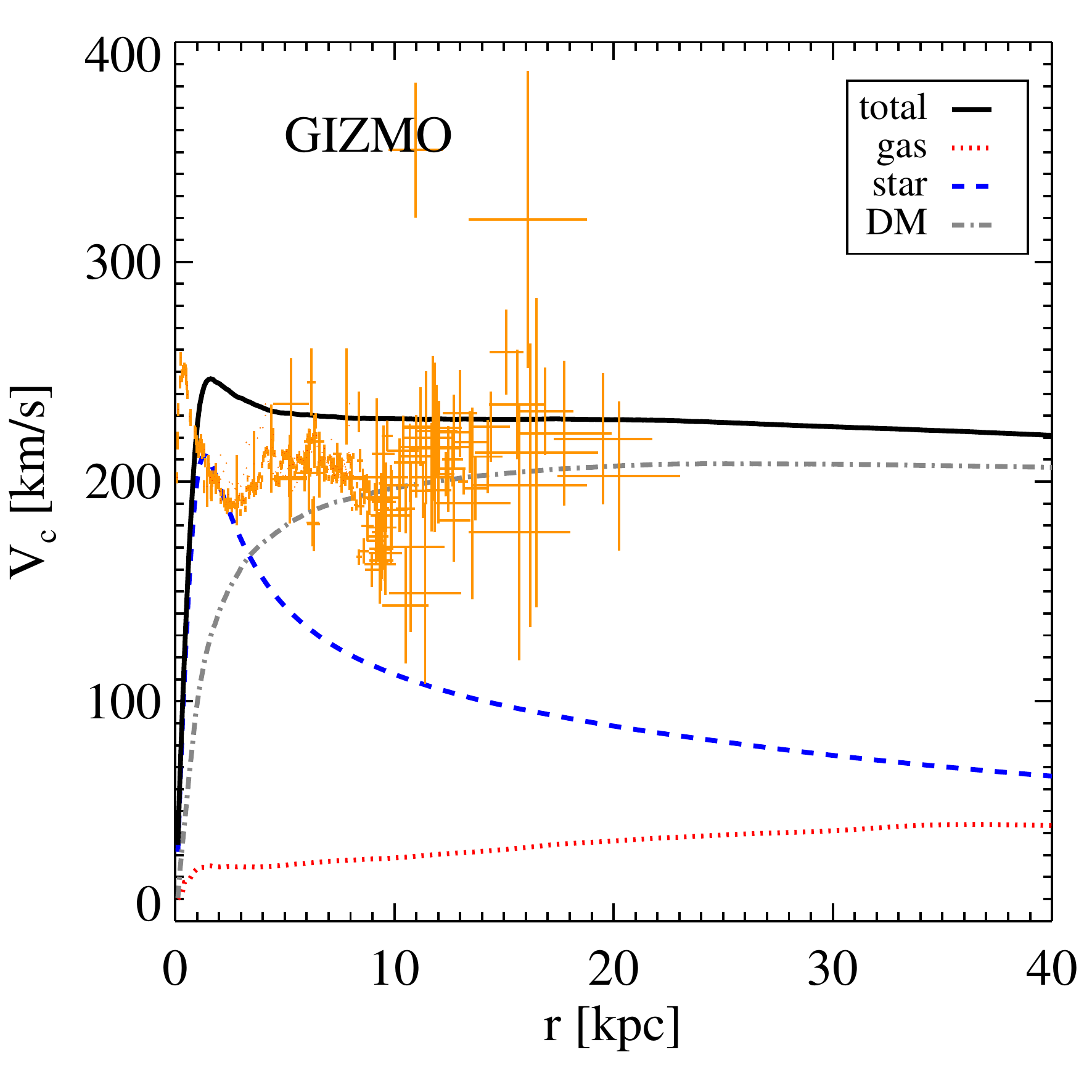}
\caption{\label{fig:rotation_curves} Circular velocity curve from the Improved-{\Gadget} (top panel) and {\Gizmo} (bottom panel), respectively. The contribution from stars, gas and dark matter component is indicated in different color as labeled in the legend. The observed data of the Milky Way from \cite{Sofue2009} are represented by orange dots with error bars. }
\end{center}
\end{figure}

{A} circular velocity curve is an important tool for the study of mass distribution in disk galaxies. Following \cite{Scannapieco2012}, we calculate the circular velocity curve as follows:
\begin{equation}
V_{c}(r) = \sqrt{GM(<r) / r}, 
\end{equation} 
where $M(<r)$ is the enclosed mass within radius $r$. 

Figure~\ref{fig:rotation_curves} shows the circular velocity curve from both Improved-{\Gadget} and {\Gizmo} simulations, respectively. Unlike most of the circular velocity curves in \cite{Scannapieco2012} which show sharp peaks around the galactic center, both of the circular velocity curves from our simulations resemble the classic ``flat" rotation curves. This is due to the feedback model which reduces the stellar mass in the central region in the early phase, as the wind mass loading depend strongly on the mass of the dark matter halo, which efficiently removes {a} large amount of gas in the early stage when {the} stellar component is relatively small. We note that our curves have higher circular velocities than the observed one of the Milky Way. This is perhaps because AGN feedback is not included in our simulations. The rotational velocity curve of the same halo from \cite{Marinacci2014}, which included AGN feedback, showed a lower maximum value $V_{c} \sim 160\, \rm{km\, s^{-1}}$, which agrees better with the observations, pointing to the importance of AGN feedback.

\subsubsection{Circularity}

\begin{figure}
\begin{center}
\includegraphics[scale=0.38]{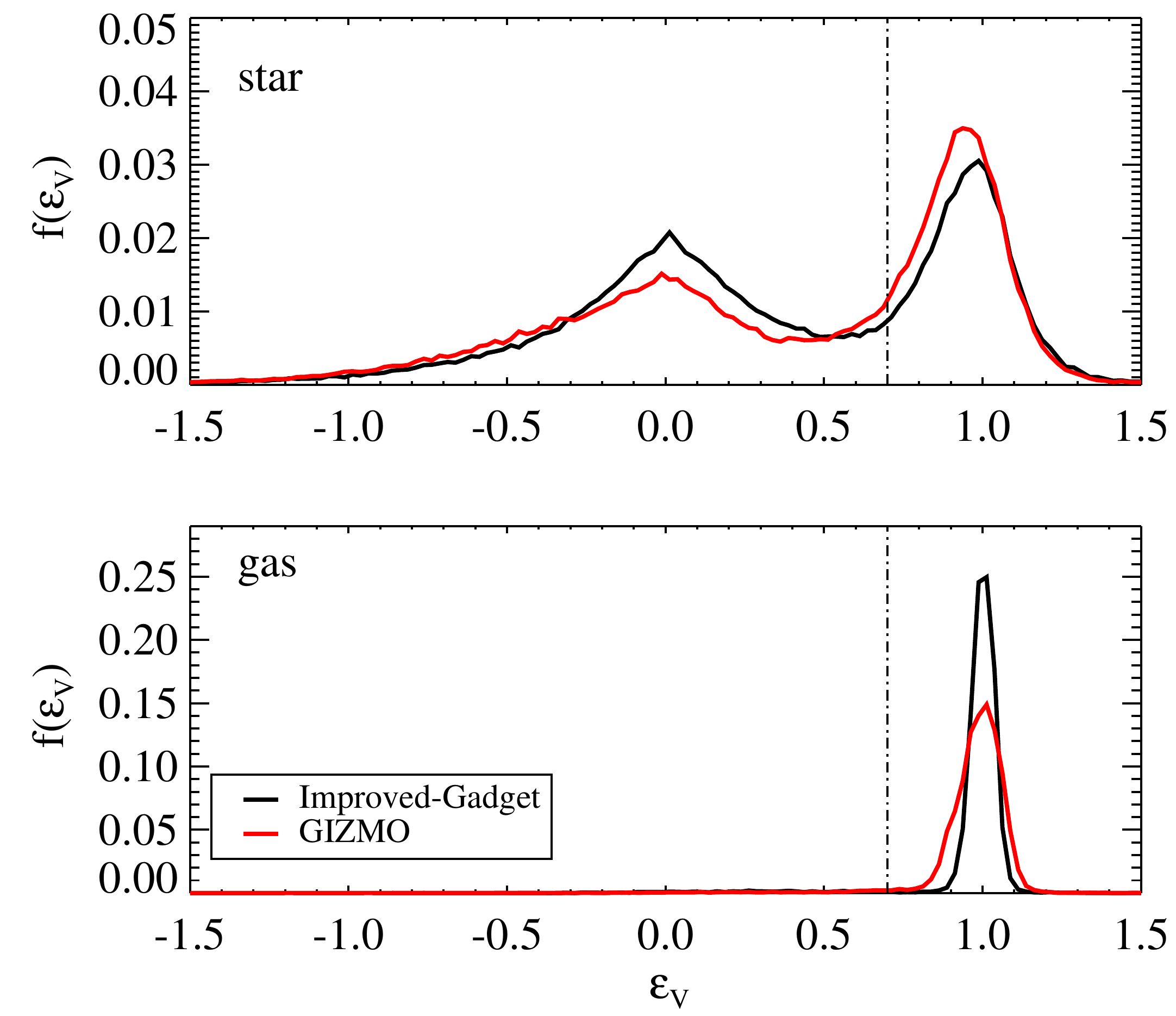}
\caption{\label{fig:circularity_distribution} The distribution of circularity $\epsilon_V=  j_{z}/(r V_{c}(r))$ for stars  (top panel) and gas (bottom panel) within 0.1$R_{200}$ at $z = 0$ from both Improved-{\Gadget} and {\Gizmo} simulations, respectively. The $z$-axis is the direction of the angular momentum for all the selected star or gas particles.  }
\end{center}
\end{figure}

The degree of rotational support can be measured by circularity $\epsilon_V$, which is defined as the ratio between the $z$ component of specific angular momentum $j_{z}(r)$ stars (or gas) and that of the circular orbit at the same radius $j_{z}(r)$, as used in \cite{Scannapieco2012}: 
\begin{equation}
\epsilon_V = j_{z}(r)/j_{c}(r) =  j_{z}/(r V_{c}(r)).
\end{equation}

A rotationally supported component has $\epsilon_V \sim 1$; if $\epsilon_V \sim 0$, then it is a non-rotating spheroidal component. 

In Figure~\ref{fig:circularity_distribution}, we show the probability distribution function $f(\epsilon_V)$ of all the stars within 25kpc ($\sim$0.1 $R_{200}$) following the definition in  \cite{Scannapieco2012}. The presence of the peak around $\epsilon_V = 1$ confirms the existence of a rotating disk for both stellar and gas components. The circularities $\epsilon_V$ of the gas in both simulations follow a much narrower distribution around 1, indicating that most of the gas is rotating in the same direction defined by the stellar disk. Moreover, the spread of the gas $\epsilon_V$ in the Improved-{\Gadget} is narrower than that in the {\Gizmo} simulation, which suggests a more coherent rotating disk. This is mainly due to suppression of turbulent motion by numerical viscosity and low-order noise in the SPH, 

We further calculated the mass fraction of both stellar and gas components in the rotational disk (defined as its circularity is above $\epsilon_V = 0.7$),  $F_{\rm star} = M(\epsilon_V > 0.7,{\rm stellar})/M_{\rm tot, stellar}$, and $F_{\rm gas}  = M(\epsilon_V > 0.7, {\rm gas})/ M_{\rm tot, gas}$. We  found  $F_{\rm star} = 0.53, 0.47$, and $F_{\rm gas}  = 0.94, 0.96 $ for Improved-{\Gadget} and {\Gizmo} simulations, respectively, as listed in Table~\ref{table:properties}. 

Compared to the results from \cite{Scannapieco2012}, the mass fraction of the disk component from our two simulations are the highest among all the SPH simulations in the Aquila Comparison Project. Our results are comparable to those of \cite{Marinacci2014}, which suggests that our disk kinematics is close to that obtained from {\Arepo}.

\begin{figure}
\begin{center}
\includegraphics[scale=0.4]{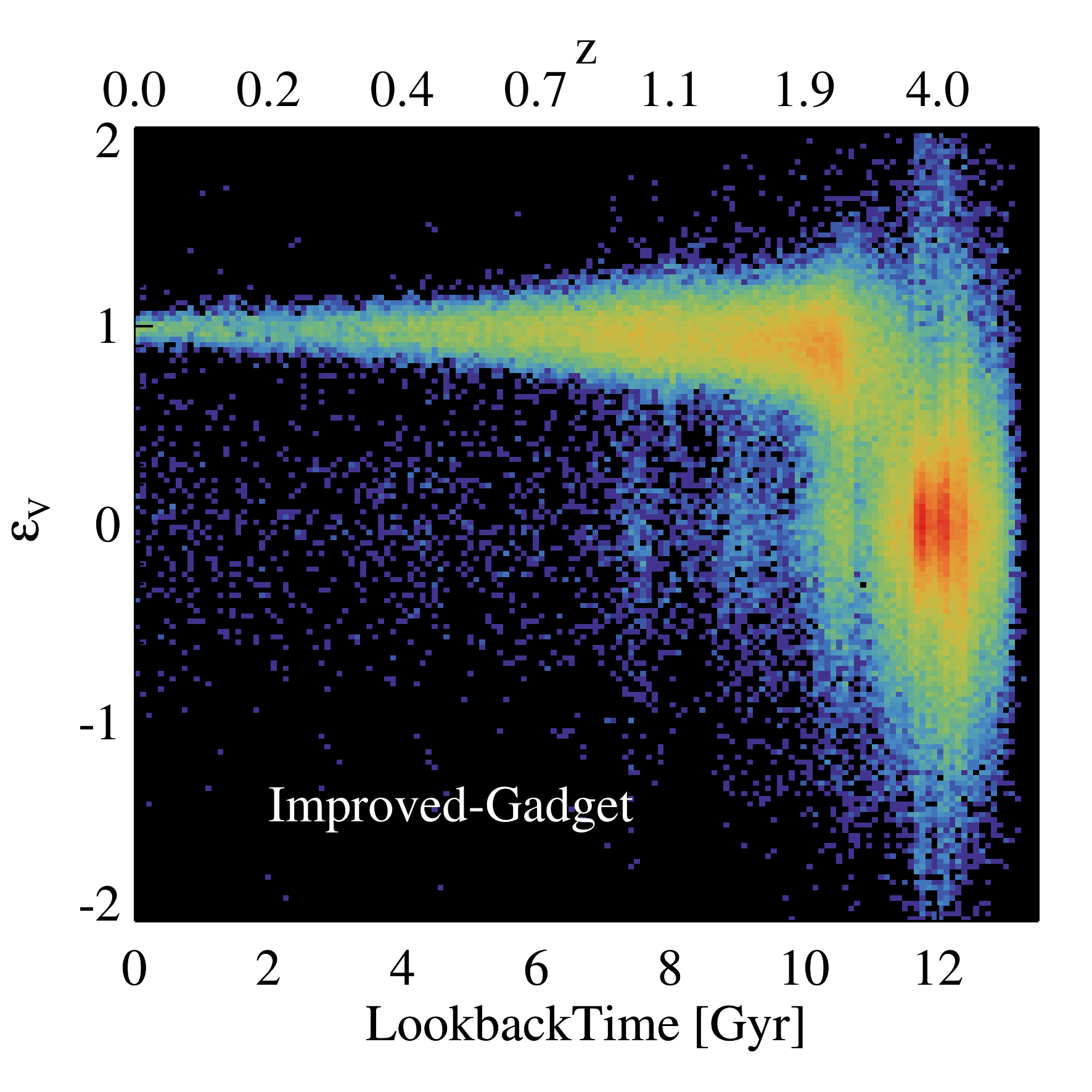}
\includegraphics[scale=0.4]{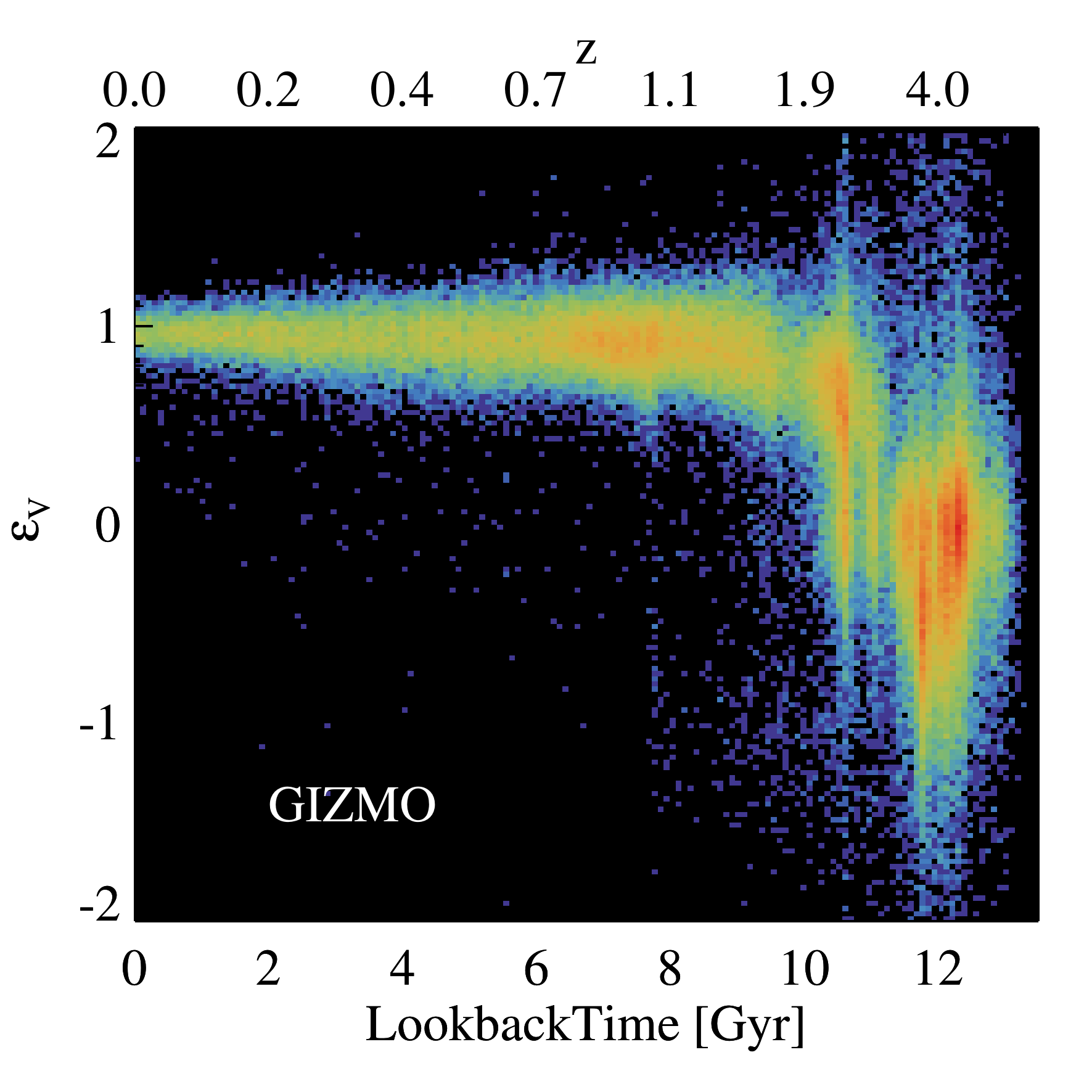}
\caption{\label{fig:circularity_age_c5} Relation between $\epsilon_V$ and age of all the stars located within 0.1 $R_{200}$ at $z = 0$ from both Improved-{\Gadget} and {\Gizmo} simulations, respectively. Rotationally supported stellar disk has the circularity distribution around  $\epsilon_V = 1$, while the spheroidal component has $\epsilon_V = 0$. }
\end{center}
\end{figure}

To investigate the buildup history of the disk, we show the relation between $\epsilon_V$ and the age of the stars in Figure ~\ref{fig:circularity_age_c5}. Interestingly, both simulations show a dramatic transition from a spheroid-dominated structure to a rotationally supported disk. The prominent disk component at $\epsilon_V = 1$ starts to form from $z\sim2$.  Before that time, the distribution of circularity of stars  lies around $\epsilon_V = 0$,  resembling a non-rotating spheroidal component. Moreover, the spheroidal component with $\epsilon_V = 0$ is mostly consist of old stellar populations.  Prominent contribution to the spheroid comes from the peaked star formation at $z = 4$, where several subunits are also visible in this plot. This is a result of the efficient gas fueling directly from the filamentary structures in Figure~\ref{fig:star_gas_images}. Our results suggest that the rotating disk started to form around $z\sim2$ and preserved until the present day while continuing to grow, in a ``inside-out" fashion, consistent with the picture from \cite{Aumer2013}.


\subsection{Comparison with Previous Studies}
\subsubsection{Gas Fraction}

\begin{figure}
\begin{center}
\includegraphics[scale=0.4]{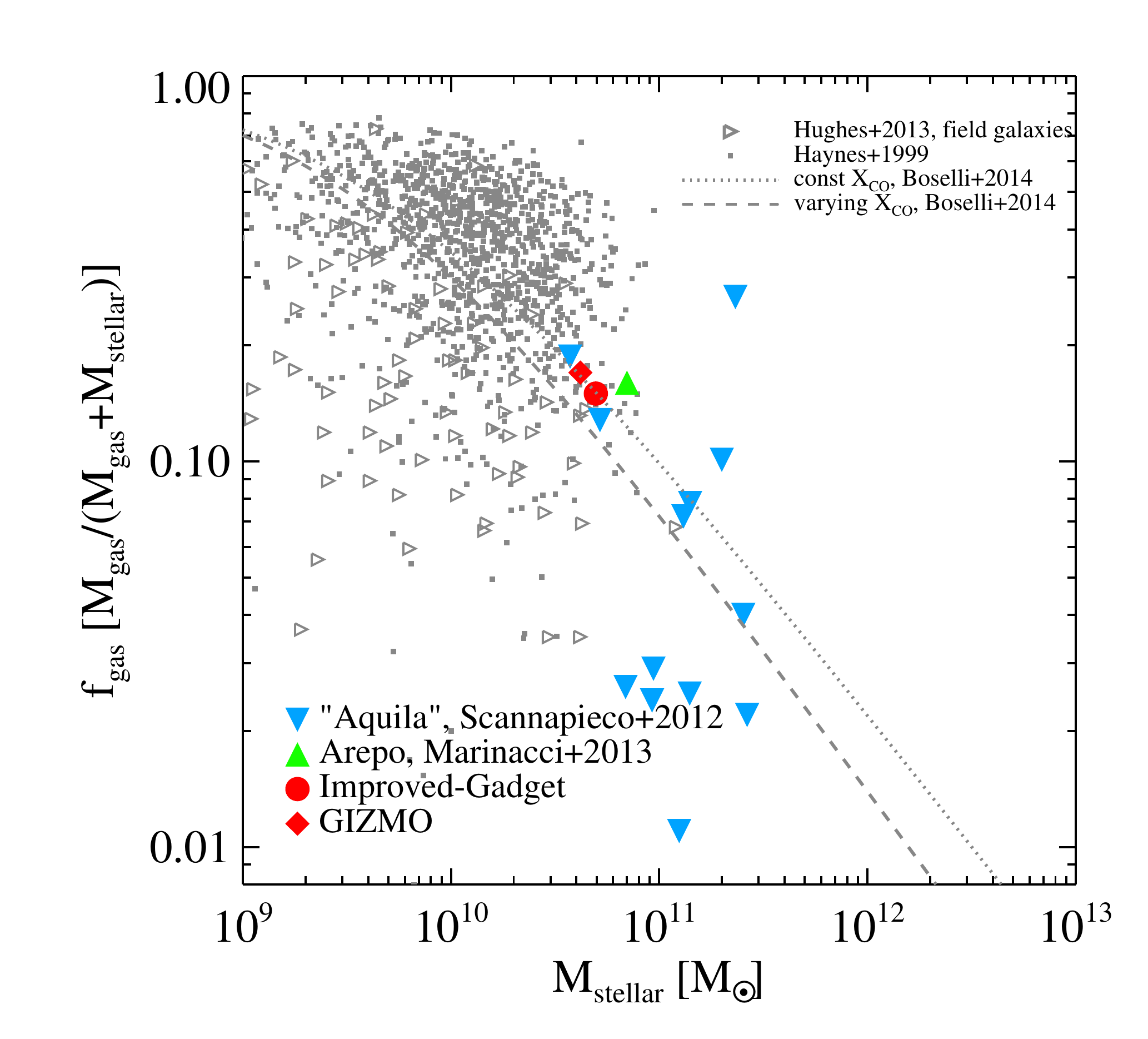}
\caption{\label{fig:fgas_mass} Gas fraction as a function of stellar mass from both Improved-{\Gadget} and {\Gizmo} simulations, respectively, in comparison with the Aquila  simulations from \cite{Scannapieco2012} (blue downward filled triangles) and the {\Arepo} simulation by \cite{Marinacci2014} (green upward filled triangle), as well as observations from \cite{Haynes1999} (gray dots), \cite{Hughes2013} (their field galaxy sample, grey open triangles),  and constrains by \cite{Boselli2014} with constant (dotted line) and varying (dashed line) $\rm{X_{CO}}$ conversion factors. The gas fraction of the simulated galaxies is computed within 0.1$\times R_{200}$ (25 kpc) from the galactic center. }
\end{center}
\end{figure}

As we show earlier, both the Improved-{\Gadget} and {\Gizmo} simulations can retain a sufficient amount of cold gas at $z = 0$ in a disk configuration to fuel star formation. To compare the gas fraction of the final galaxies with other studies, we adopt the same definition as in \cite{Marinacci2014} to compute the gas fraction as 

\begin{equation} 
f_{\rm gas} = \frac{M_{\rm gas}}{M_{\rm gas}+M_{\rm stellar}},  
\end{equation} 
where  $M_{\rm gas}$ is the total gas mass and $M_{\rm star}$ is the total stellar mass within  0.1$\times R_{200}$ (25kpc) from the galaxy center.

Figure~\ref{fig:fgas_mass} shows the gas fractions from our simulations, in comparison with previous simulations from the Aquila Comparison Project \citep{Scannapieco2012} and the {\Arepo} simulation by \cite{Marinacci2014},   as well as observations \citep{Haynes1999, Hughes2013, Boselli2014}. Both of our results are in good agreement with the observational data and the simulation by \cite{Marinacci2014}. The majority of the Aquila simulations from \cite{Scannapieco2012} have too low gas fractions, which are not consistent with observations of late-type galaxies.

\subsubsection{Galaxy Main Sequence}

\begin{figure}
\begin{center}
\includegraphics[scale=0.4]{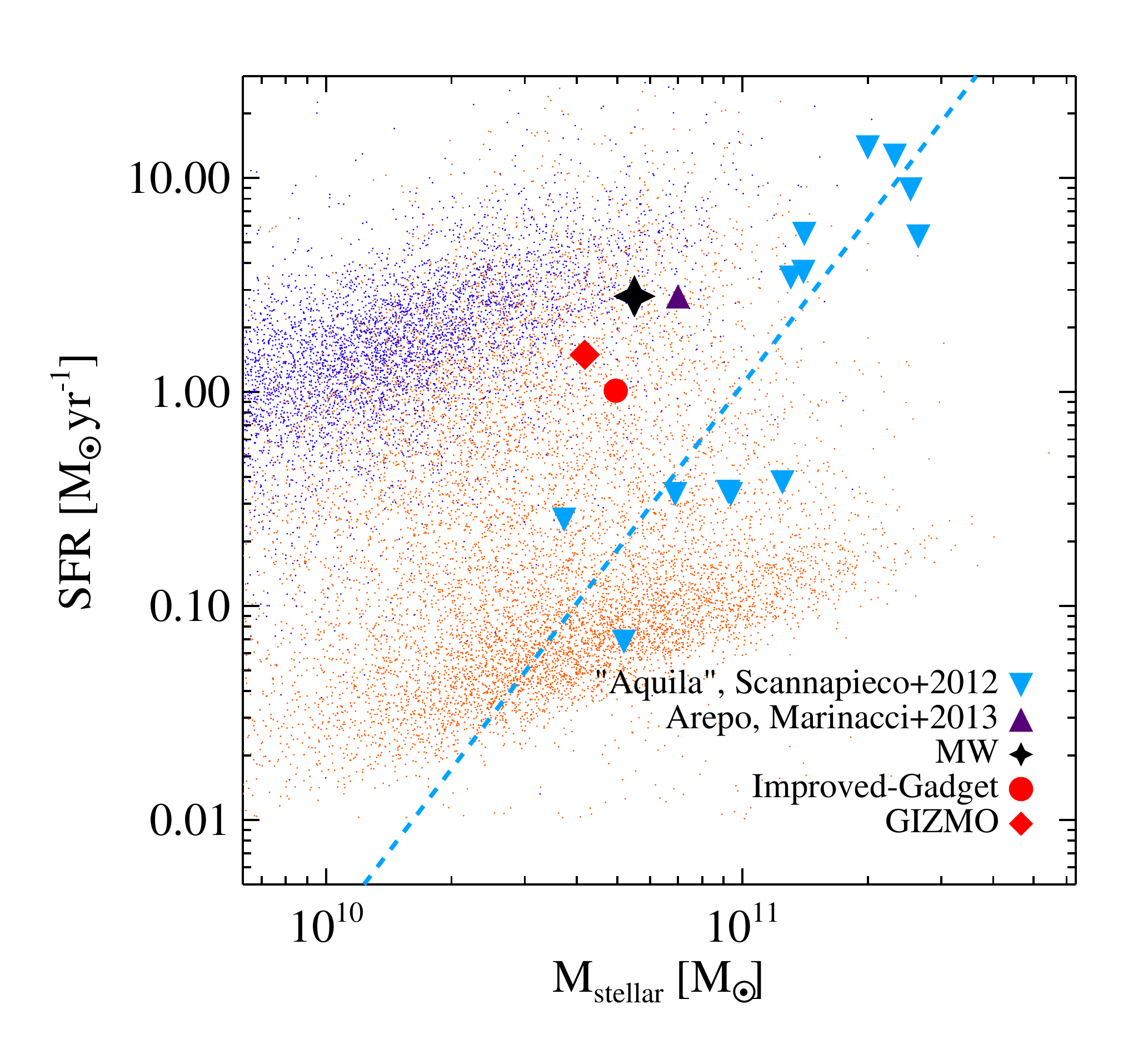}
\caption{\label{fig:star_sfr}  Star formation rate -- stellar mass relations from both Improved-{\Gadget} and {\Gizmo} simulations, respectively, in comparison with previous simulations from \cite{Scannapieco2012} (blue downward filled triangles) and \cite{Marinacci2014} (upward filled triangle), as well as the observation of nearby galaxies ($z < 0.1$) by SDSS (only  $5\%$ of the total galaxies in MPA-JHU DR 7 catalog are randomly selected and plotted here). The estimated position of the Milky Way from \cite{Oliver2010} and \cite{Leitner2011} is marked by a diamond symbol.  The blue dashed line is the fitting of data from the Aquila simulations.} 
\end{center}
\end{figure}

Over the past decade, observations of nearby galaxies by Sloan Digital Sky Survey (SDSS) have shown a remarkably tight correlation between star formation rate of a galaxy and its stellar mass, now referred to as the ``galaxy main sequence" \citep[e.g.][]{Faber2007}. The galaxies are further divided into ``blue cloud" and ``red sequence"  based on the galaxy color \citep[e.g.][]{Strateva2001, Kauffmann2003, Blanton2003}:
\begin{equation}
(g-r) = 0.59 + 0.052 \log (\rm{M_{\star}}) - 10.0, 
\end{equation}

It is believed that galaxies evolve from the ``blue cloud" to the ``red sequence", and that the MW galaxy is in the ``green valley" \citep[e.g.][]{Bell2004, Faber2007, Willett2013}.

In Figure~\ref{fig:star_sfr}, we compare our results against previous Aquila simulations in \cite{Scannapieco2012} and that of \cite{Marinacci2014}, as well as observations of a sample of the nearby SDSS galaxies ($z < 0.1$) from the MPA-JHU DR 7 catalog. As demonstrated in the figure, our simulations agree well with that of \cite{Marinacci2014} and the observation of the MW \citep[][]{Oliver2010, Leitner2011}. 

The Aquila simulations showed a different trend than the observations. Nearly half of the simulated galaxies in \cite{Scannapieco2012} are over massive with high present-day star formation rate, which {places} them on the very edge of the ``blue cloud", while  the other half have too low star formation rates which place them in the ``red sequence". Such a  discrepancy reflects the main difficulty of regulating star formation during the growth history of the galaxy. Our results suggest that the feedback processes, metal-dependent cooling, and mass return from evolved stars are critical {in reproducing} the galaxy main sequence.

\subsubsection{Tully-Fisher Relation}

\begin{figure}
\begin{center}
\includegraphics[scale=0.4]{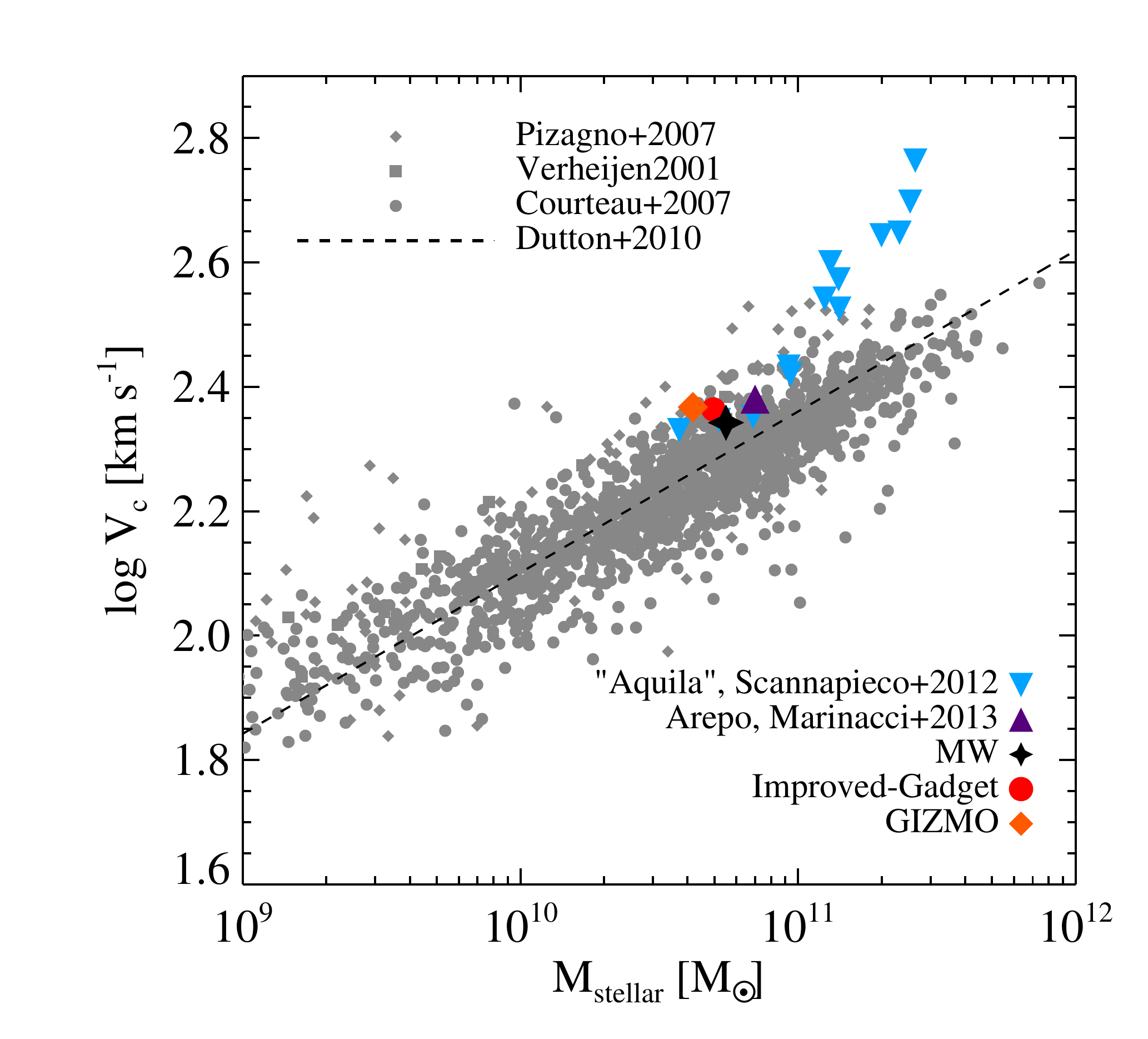}
\caption{\label{fig:vrot_mass} Stellar mass -- galaxy circular velocity relations from both Improved-{\Gadget} and {\Gizmo} simulations, respectively, in comparison with previous simulations from \cite{Scannapieco2012} (blue downward filled triangles) and \cite{Marinacci2014} (upward filled triangle), as well as the observed Tully-Fisher relation \citep{Pizagno2007, Verheijen2001, Courteau2007}. The predicted trend from \cite{Dutton2010} is shown in the dashed line. The estimated position of the Milky Way is marked by a diamond symbol. The circular velocity is measured at the stellar half-mass radius.}
\end{center}
\end{figure}

Observations of spiral galaxies show a strong correlation between the stellar mass in the disk of a galaxy and its circular velocity, known as the Tully-Fisher relation \citep{Tully1977}.  In Figure~\ref{fig:vrot_mass}, we compare the stellar mass--circular velocity relations from our simulations against previous simulations \citep{Scannapieco2012, Marinacci2014} and the observations \citep{Pizagno2007, Verheijen2001, Courteau2007}. It shows that our results are consistent with that from the {\Arepo} simulation by \cite{Marinacci2014}, and the observed Tully-Fisher relation. A majority of the Aquila simulations in \cite{Scannapieco2012} show large {deviations} from the observations due to the overly concentrated stellar disk and overly large stellar mass in those simulations. This again points to the need {for} robust numerical codes and comprehensive physical models.

\subsection{Convergence Properties}

\begin{figure}
\includegraphics[scale=0.36]{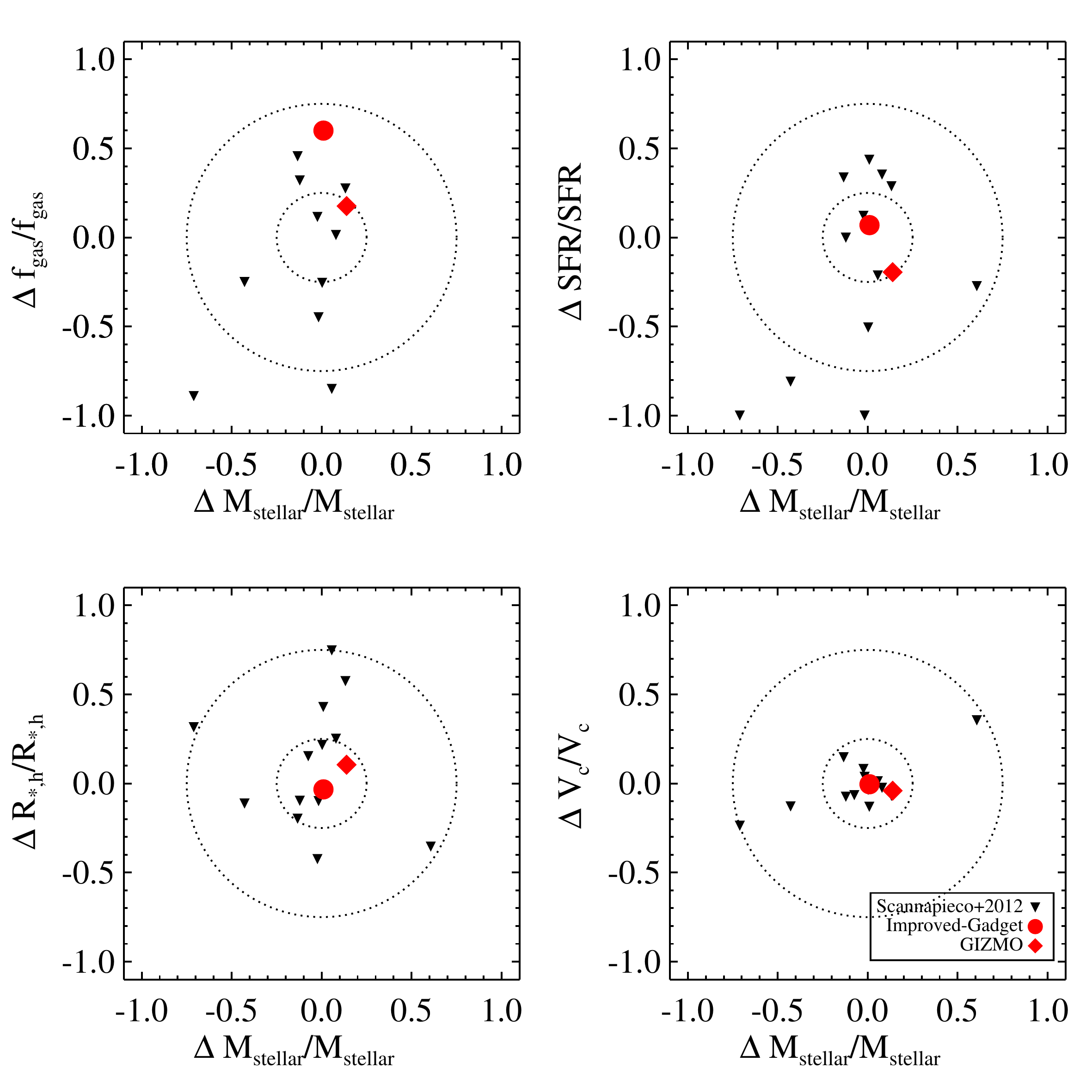}
\caption{\label{fig:resolution_aquila} Numerical convergence, represented by $\rm{\Delta Q/Q= (Q_{6} - Q_5)/Q_{5}}$, of various galaxy properties between the level-5 and level-6 resolution runs from both Improved-{\Gadget} and {\Gizmo} simulations, respectively, in comparison with previous simulations from \cite{Scannapieco2012} (black downward filled triangles). The two dotted-line circles indicate the position of $25\%$ and $75\%$ deviations, respectively (0 corresponds to full convergence). }
\end{figure}

As pointed by \cite{Zhu2015}, numerical convergence is an important and necessary criterion to access the robustness of the numerical simulations. Although some dependence on {the} resolution is inevitable, large variations across different resolutions can pose a serious challenge to the robustness of the model and even the overall correctness of the simulation. One should therefore be careful that the physical quantities such as gas density and temperature must be ``well resolved" by simulations. More importantly, when resolution changes, input parameters which control the behavior such as gas cooling, star formation and feedback should stay the {\it same}. If this condition is not met, it is impossible to unravel the entanglement between physical processes and numerical artifacts. From a mathematical perspective, galaxy formation and evolution essentially {become} an ill-posed problem if the final galaxy properties are extremely sensitive to resolution or other numerical parameters in any model.

To investigate the convergence of our simulations, we follow \cite{Scannapieco2012} and quantify the difference between level 5 and level 6 for a given quantity $Q$ as 
\begin{equation}
\Delta Q/Q = (Q_{6} - Q_5)/Q_{5}.
\end{equation}

In Figure~\ref{fig:resolution_aquila}, we compare the convergence properties of our simulations against those of the Aquila simulations from \cite{Scannapieco2012}. Overall, the {\Gizmo} simulation demonstrates better numerical convergence than the Improved-{\Gadget} in our study. It is encouraging seeing that a number o properties such as circular velocities and stellar mass show good convergence against numerical resolution across the different simulations. However, {a} significant deviation is present in other properties such as gas fraction, star formation rate and galaxy size, where $\rm{\Delta Q/Q}$ is larger than $25\%$ in many of the simulations. In particular, the gas fraction suffers from the worst convergence in most of the simulations including our Improved-{\Gadget} run.

The poor convergence in SPH simulations is due to the zeroth order error in both of the density estimate and in the equation of motion. In \cite{Zhu2015}, we showed that an analytical second-order convergence rate of SPH can be achieved if we increase the number of neighbors systematically for higher resolutions. If a $fixed$ number of neighbors is used for different resolutions, the fluctuation in the density estimate and volume estimate in SPH will effectively slow the convergence of SPH making it behave just as a Monte Carlo method. Varying the number of neighbors is necessary from a {purely} mathematical point of view. Moreover, we showed that formal numerical convergence is possible in SPH only if the total number of particles and the number of neighbor particles are infinitely large and the smoothing length is infinitely small, which is impossible to achieve in realistic simulations.

Figure~\ref{fig:resolution_aquila} demonstrates the challenge SPH methods face in numerical convergence in galaxy formation simulations. We suggest that {\Gizmo} is a promising alternative given its better convergence properties and lower computational cost.

\begin{table}[htdp]
\caption{Properties of simulated galaxies at $z = 0$}
\begin{center}
\begin{tabular}{ccc}
\hline
 \,\,   &  Improved-{\Gadget} & {\sc Gizmo} \\
\hline
$M_{200} [10^{10} \rm{M_{\odot}}]$ & 154.55 & 159.26 \\
\hline
$R_{200}$ [kpc] & 231.88 & 233.76 \\
\hline
$M_{\rm stellar} [10^{10} \rm{M_{\odot}}]$ & 4.96 & 4.18 \\
\hline
SFR $[\rm{M_{\odot}}/yr]$ & 1.01 & 1.49 \\
\hline
$f_{\rm gas}$ & 0.15 & 0.17 \\
\hline
$V_{\rm max} [{\rm km/s}]$ & 248.47 & 247.43 \\
\hline
$R_{\rm d}$ [kpc] & 5.75 & 5.83 \\
\hline
$F^{\rm star}(\epsilon_V>0.7)$ & 0.53 & 0.47 \\
\hline
$F^{\rm gas}(\epsilon_V>0.7)$ & 0.94 & 0.96 \\
\hline
\end{tabular}
\end{center}
\label{table:properties}
{Galaxy properties include: Virial mass $M_{200}$, Virial radius $R_{200}$,  total stellar mass, present-day star formation rate, gas fraction, 
maximum circular velocity, stellar disk scale length, stellar mass fraction in the disk $F_{\rm star} = M(\epsilon_V > 0.7,{\rm stellar})/M_{\rm tot, stellar} $ and gas mass fraction in the disk $F_{\rm gas}  = M(\epsilon_V > 0.7, {\rm gas})/ M_{\rm tot, gas}$.}
\end{table}

The present-day galaxy properties from both Improved-{\Gadget} and {\Gizmo} simulations are summarized in Table~\ref{table:properties}. These include the mass, size, star formation rate, gas fraction, maximum circular velocity, stellar disk scale length, and mass fractions of stars and gas in the disk. These properties are discussed in the previous sections.

\section{Discussion}
\label{sec:discussion}

\subsection{Metal-dependent Gas Cooling}

{Since metals are simply advected with particles in both Improved-{\Gadget} and {\Gizmo}, we have the same ``enrichment sampling problem'' raised by \cite{Wiersma2009b} when distributing metals from stars to gas particles. The use of a smoothed metallicity, in principle, helps alleviate this problem. Nevertheless, our choice of Wendland function for Improved-{\Gadget} may introduce additional sources of uncertainty when comparing the performances of the two codes. To assess the impact on the metal-dependent cooling induced by two different smoothing kernels, we carry out two additional MW galaxy simulations with a fixed metallicity, $0.1Z_{\sun}$, similar to the average final metallicity from the above simulations, for gas cooling rate throughout the simulation. }

\begin{figure}
\begin{center}
\includegraphics[scale=0.4]{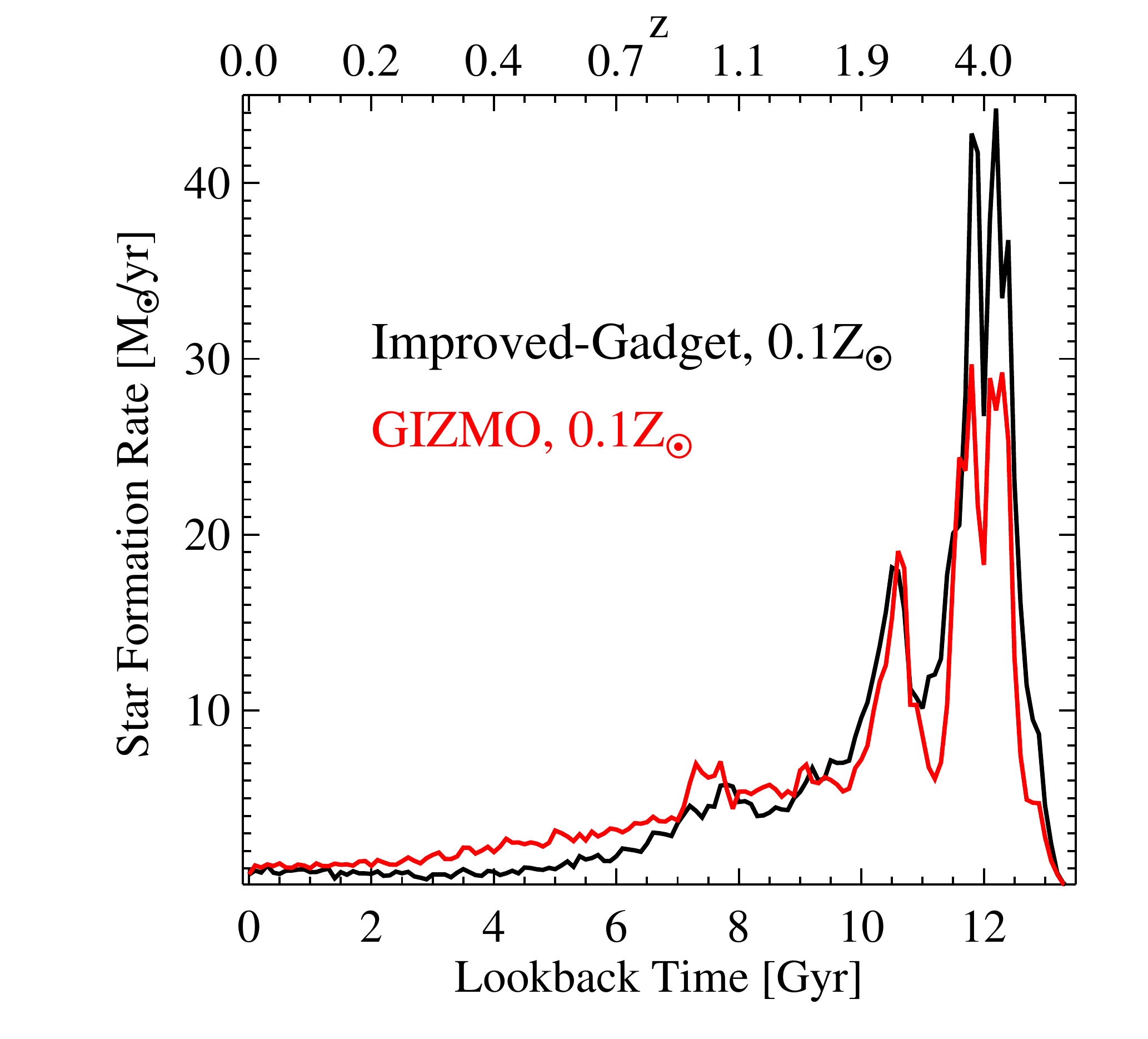}
\caption{\label{fig:sfr_cons_z} {Same as Figure~\ref{fig:sfr}, but for Improved-{\Gadget} (black line) and {\Gizmo} (red line) simulations with a constant metallicity $0.1 Z_{\sun}$.}}
\end{center}
\end{figure}

\begin{figure}
\begin{center}
\includegraphics[scale=0.4]{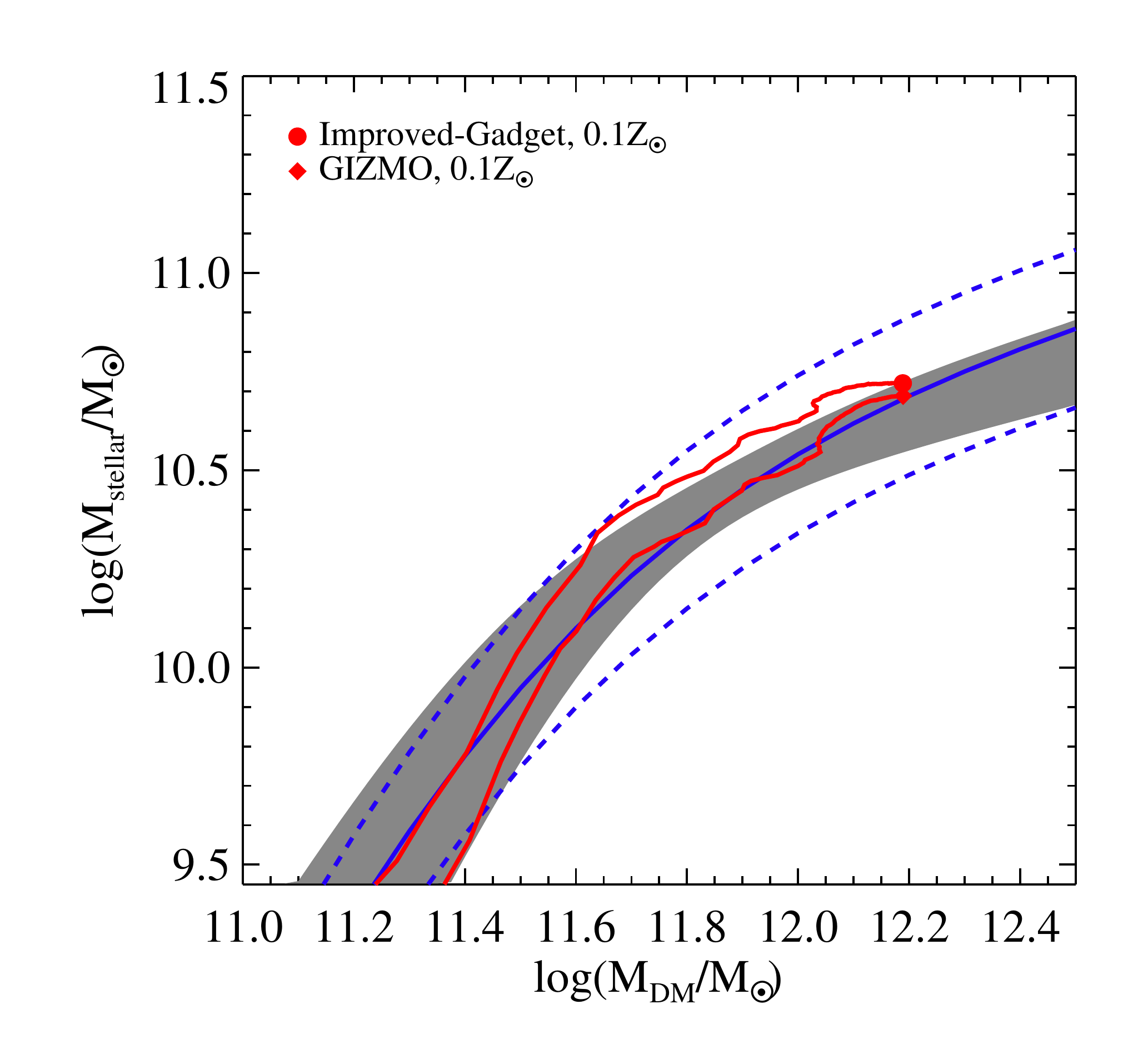}
\caption{\label{fig:halo_star_cons_z} {Same as Figure~\ref{fig:halo_star}, but for stellar -- dark matter mass relations with a constant metallicity $0.1 Z_{\sun}$.}}
\end{center}
\end{figure}

\begin{figure}
\begin{center}
\begin{tabular}{cc}
\includegraphics[scale=0.35]{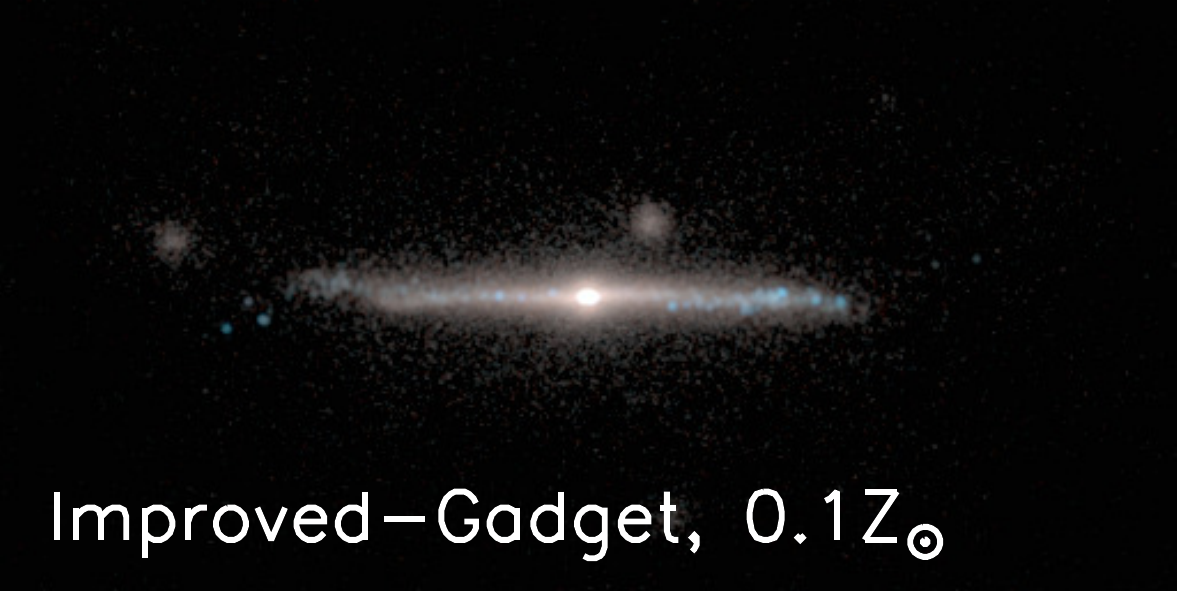}
\includegraphics[scale=0.35]{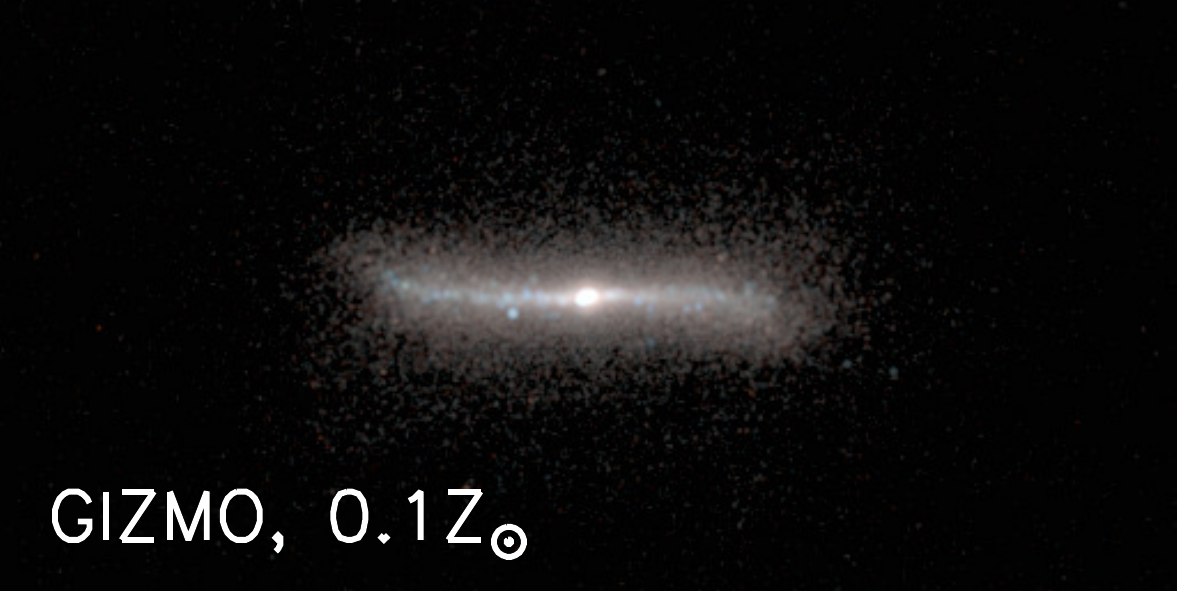}\\
\includegraphics[scale=0.35]{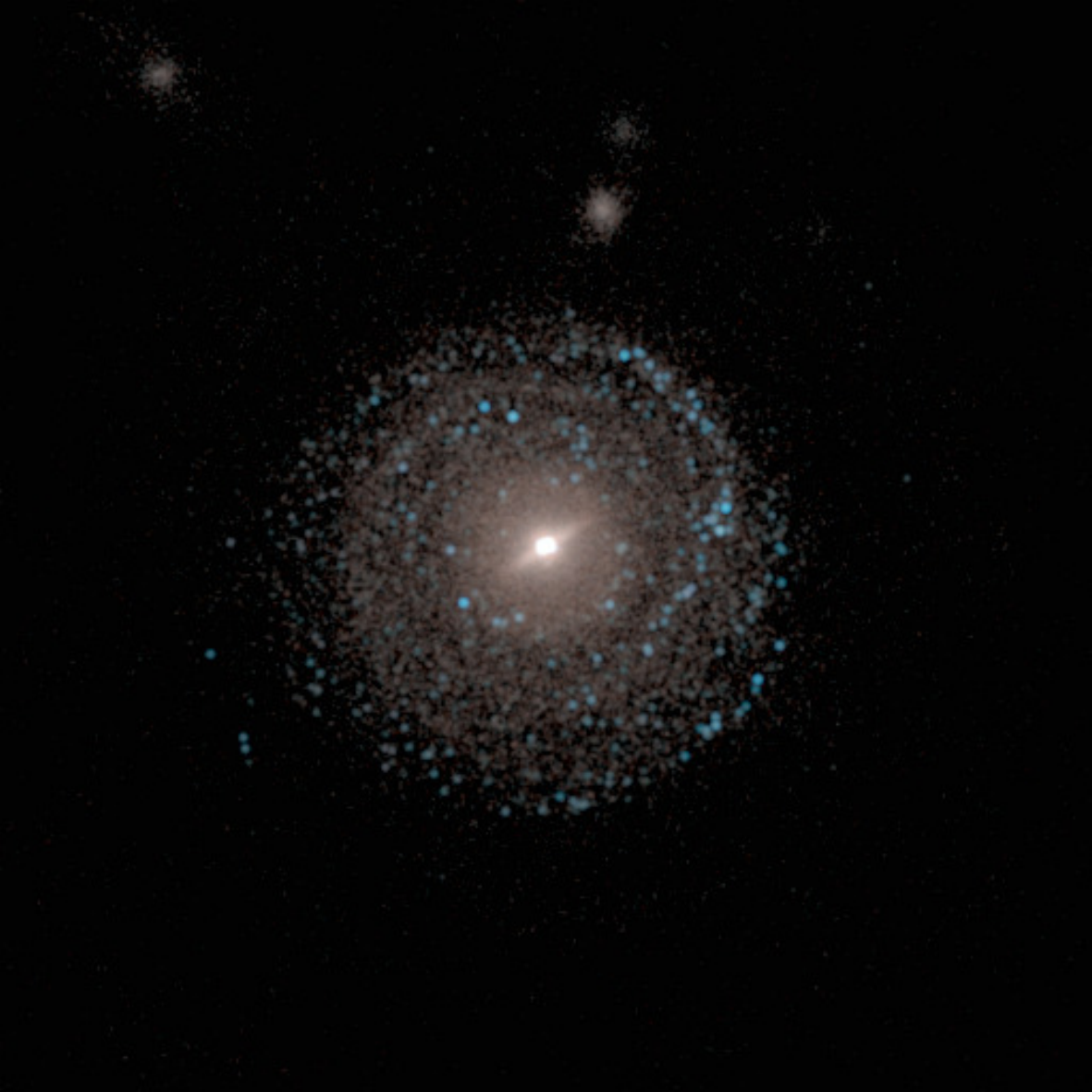}
\includegraphics[scale=0.35]{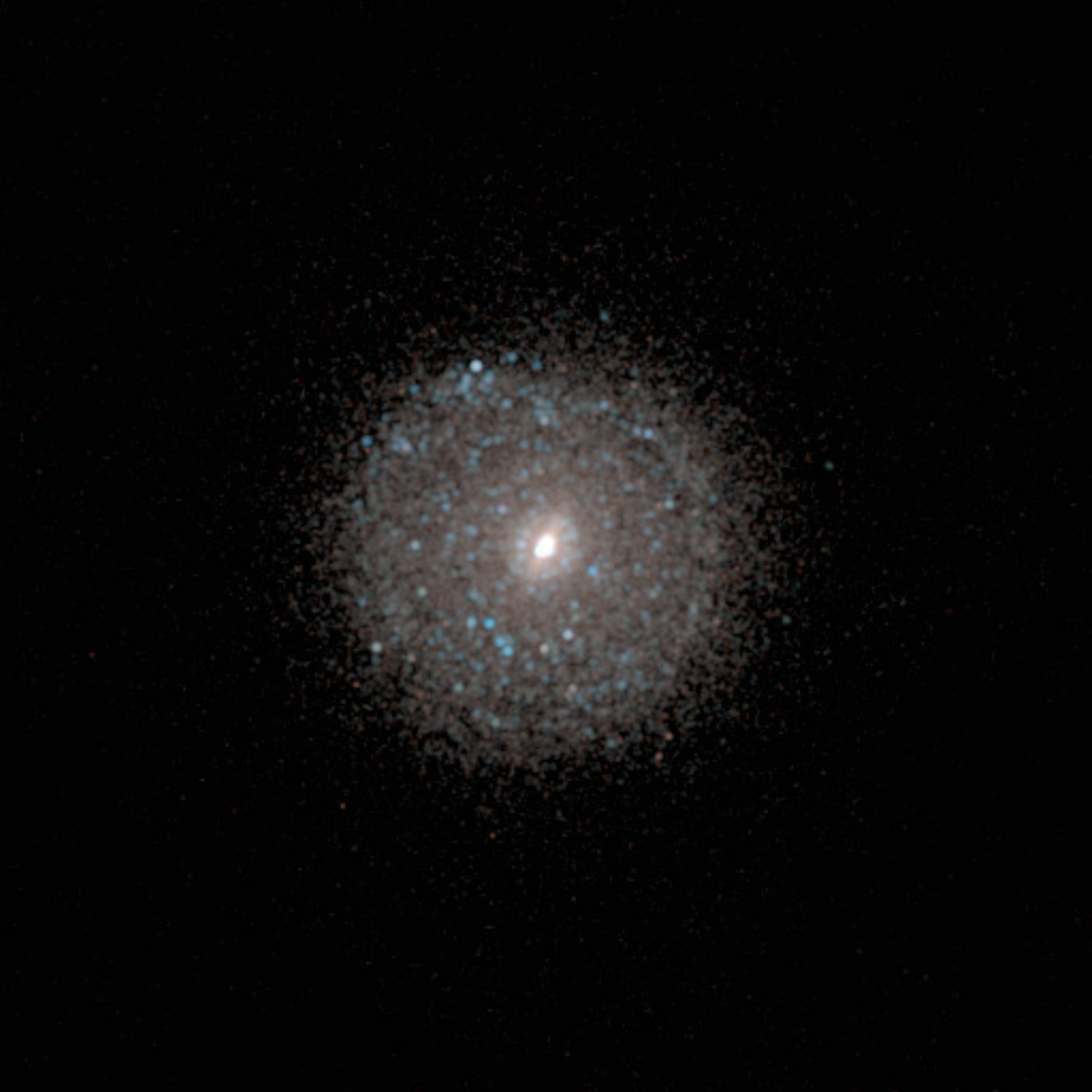}\\
\end{tabular}
\caption{\label{fig:stellar_image_edge_cons_z} {Same as Figure~\ref{fig:stellar_image_edge}, edge-on and face-on views of the projected stellar maps at $z = 0$ for Improved-{\Gadget} and {\Gizmo} simulations using a constant $0.1Z_{\sun}$ when calculating gas cooling rate.}}
\end{center}
\end{figure}

{The resulting star formation histories from the test simulations are shown in Figure~\ref{fig:sfr_cons_z}. Compared to Figure~\ref{fig:sfr}, the overall star formation histories are similar but with some notable differences between the original and the new test runs: star formation rate with the fixed metallicity is higher than that with realistic metal enrichment at high redshifts $z > 1$ in both Improved-{\Gadget} and {\Gizmo} simulations, but is slightly lower at low redshifts $z < 1$.  The evolutionary tracks of stellar and total masses from the two tests, as shown in Figure~\ref{fig:halo_star_cons_z}, closely resemble those in Figure~\ref{fig:halo_star}.}

{The stellar disks from the constant-metallicity test simulations are shown in Figure~\ref{fig:stellar_image_edge_cons_z}. Compared to Figure~\ref{fig:stellar_image_edge}, the disks are similar to those from the original simulations with more realistic metal enrichment, which suggests that the galaxy disk is a robust feature of both simulations.}

{These results suggest that despite some differences between the simulations using different smoothing kernels,  the remarkable differences in galaxy properties between Improved-{\Gadget} and {\Gizmo} simulations described in \S~\ref{sec:results} are mainly due to the different hydrodynamics solvers in these two codes.}

\subsection{SPH vs Alternatives}

SPH has been a popular method in numerical simulations to investigate galaxy formation and evolution over the years. This is largely due to its automatic adaptive nature, its simple formulation, and the availability of a number of well-documented SPH codes. And thanks to vigorous efforts in improving this method in recent years, the difference between SPH and non-SPH codes has become smaller. For example, the SPH Eagle Simulation \citep{Schaye2015} and the moving-mesh Illustris Simulation \citep{Vogelsberger2014b} show a broad agreement over a wide range of galaxy properties on large scales such as the cosmic star formation history and galaxy mass functions. 

However, remarkable differences remain in specific test problems and detailed galaxy simulations using different numerical methods, as reported by previous works \citep{Sijacki2012, Keres2012, Vogelsberger2012, Torrey2012, Nelson2013, Hayward2014, Zhu2015}, and as we found in this study. For example, the main difference between our Improved-{\Gadget} and the sophisticated ``Anarchy SPH" used in the Eagle Simulation \citep{Schaye2015} is that the latter employs a Wenland $C^2$ function with a number of neighbors $\rm{N_{nb}  = 58}$, while we use a bias-corrected Wendland $C^4$ function \citep{Dehnen2012} with $\rm{N_{nb}  = 200}$. In our test problems, we found using a low number of neighbors can cause significantly more noises along contact discontinuities. Moreover, using the same initial conditions and the same physical models, we found that {\Gizmo} achieved better convergence properties than the Improved-{\Gadget} and other codes in the previous comparison projects, and it produced a MW-size disk galaxy in good  agreement with recent simulations using {\Arepo}. 

Furthermore, the higher computational cost and slower convergence rate of SPH become a barrier for high-resolution simulations. In the detailed usage of CPU timing, a substantial amount is spent on finding SPH neighbors in the density calculation stage. In theory, one would expect the complexity of neighbors finding scales as $\rm{N_{gb}\ln(N_{gb})}$. In zoom-in simulations in this work, hydro forces typically {take} up $40\%$ of the CPU time. Thus, one would expect that {\Gizmo} runs more efficiently since it does not need a large $\rm{N_{gb}}$ as in the Improved-{\Gadget} code. Indeed, {\Gizmo} simulation finished faster than the improved-{\Gadget} code at both C-5 and C-6 levels. 

Therefore, another advantage to use the new method in {\Gizmo} is that one can achieve same or better results with \textit{less} computational effort. This is due to the faster convergence in {\Gizmo} method. On the other hand with the same problems, SPH code shows {a} worse result for higher resolution if fixed number of neighbors is used. A theoretical argument would require such neighbors scales as the resolution goes higher, as demonstrated in \cite{Zhu2015}. We expect that SPH code will not stay as competitive as other non-SPH alternatives in the coming  high-resolution cosmological simulations targeting parsec-scales to resolve the interstellar medium.

\subsection{Numerical Methods vs Physical Models}

The accuracy of numerical methods, compared to the uncertainties in the physical processes of galaxy formation and evolution, may be a less dominant factor. Although SPH inevitably shows a Monte Carlo behavior, final galaxy properties seems to be less sensitive to the fluctuations in the density and velocity field. This is due to the fact that the fluctuations in the density and velocity field in SPH simulations are on the order of a percent level, while the magnitude of other sources of errors, such as gas cooling rate and supernova feedback, are much larger \citep{Zhu2015}. For example, \cite{Schaller2015} compared the results obtained with ``Anarchy SPH'' and the original {\Gadget} formulation. They conclude that the properties of the most galaxies such as stellar masses and galaxy size are not significantly affected by the choice of SPH solver. Moreover, the fact that our simulations produce similar extended gaseous and stellar disks as those by \cite{Marinacci2014} using similar physical processes suggests that the physical models do not depend sensitively on the hydrodynamics solver. {Recently, similar numerical issues related to hydro solvers have being investigated by a code comparison project on galaxy cluster scales (nIFTy galaxy cluster simulations, \cite{Sembolini2016, Elahi2016, Cui2016}). It remains a priority to identify the sources of discrepancies evidently in the simulations with different numerical methods.}

However, due to the limited scope of this study, it is unclear what the important processes are in the formation of disks in galaxies. A number of models have been proposed, such as stellar feedbacks \citep[e.g.][]{Scannapieco2012, Hopkins2013c, Hopkins2013b, Agertz2013} and the galactic winds \citep{Okamoto2010, Brook2011, Puchwein2013, Ubler2014}. We will explore effects of these different processes on the growth of disk component in a companion comparison project focusing on the physical models. 

\section{Conclusions}
\label{sec:summary}

We have conducted a code comparison project focusing on two particle-based hydrodynamic solvers for galaxy simulations: an  SPH code Improved-{\Gadget} and an MFM code  {\Gizmo}.  We have implemented the same physical models, which include physics of both dark matter and baryons,  star formation,  ``energy-driven'' outflows, {metal-dependent} cooling, stellar evolution and metal enrichment from stellar evolution, into both Improved-{\Gadget} and {\Gizmo}, and used them to perform a set of cosmological hydrodynamic simulations of the formation and evolution of a Milky Way-size galaxy. We found that both numerical schemes produce a broadly similar assembly history of the modeled galaxy, in which a rotationally supported gas disk started to form at redshift $z \sim 2$, and it evolved to become an extended disk of gas and stars at the present day. Both codes are able to produce galaxies with realistic stellar mass, gas fraction and galaxy size. However, there are significant differences in a number of galaxy properties between the two simulations, as summarized below:

\begin{itemize}

\item Gas properties:  On the one hand, the Improved-{\Gadget} produced substantially more cold, dense gas with density $\rm{n_{H} > 10\, cm^{-3}}$ than {\sc Gizmo}, by up to two orders of magnitude in mass, due to less efficient mixing between the cold and hot phases. On the other hand, the temperature of the hot-phase gas at low redshift in the Improved-{\Gadget} simulation is slightly higher, due to less efficient cooling from hot gas. However, despite the existence of dense clumps, the overall distribution of cold gas in the Improved-{\Gadget} simulation is smoother due to the damping of turbulent motion by large numerical viscosity in the SPH algorithm.

\item  Star formation history: The availability of more cold, dense gas in the Improved-{\Gadget} simulation fueled stronger star formation by nearly a factor of 2 up to redshifts $z \sim 1$, while the hotter gas at low redshift produced a lower star formation rate. As a result, Improved-{\Gadget} produced a higher stellar mass by $20\%$ and a lower gas fraction by $10\%$ at $z = 0$ compared to {\Gizmo}.

\item Disk properties: Both simulations produce a stellar disk with comparable radial scale length and radial density profile. However, both gaseous and stellar disks produced by Improved-{\Gadget} are thinner and smoother, lacking the prominent spiral structures as seen in the {\sc Gizmo} simulation. Moreover,  due to low-order noise and suppression of turbulence by numerical viscosity, star-forming gas in Improved-{\Gadget} is more aligned towards the direction of disk rotation, leading to a more coherent rotating disk. The dense gas clumps in Improved-{\Gadget} simulation easily lose angular momentum and settle into galactic center, forming bar-like structure. 

\item Convergence properties: {\Gizmo} showed a faster convergence rate and outperformed the Improved-{\Gadget} in resolving fine gas structures at the same computational cost, even though the latter used more than 3 times larger number of neighbors. 

\end{itemize}

Based on these findings, we conclude that {\Gizmo} provides a promising alternative approach to model galaxy formation in a cosmological context. A major weak point of SPH algorithm is its short-scale noise and the required high CPU expenses to fix the classic SPH problems. Although the drastic difference among different hydrodynamics solvers seen in the original Aquila comparison project is much reduced in this study, a number of characteristic SPH behaviors still persist to some observable level. Such numerical artifact could convolve with galaxy formation model and artificially give a stronger starburst in the early stage while a lower star formation rate during quiescent disk growth epoch for late-type $L^{*}$ galaxies. Moreover, it remains a significant challenge to disentangle the coupled effects between physical processes and numerical methods in galaxy simulations. In a following paper, we will focus on the comparison of different physical models in order to determine the crucial mechanisms that shape the formation of a disk galaxy like the Milky Way using the tools developed in this study.

 \section{Acknowledgement}
We thank Lars Hernquist for valuable discussions that inspired this project. We are greatly indebted to Phil Hopkins for providing us a private version of {\Gizmo}, and to Volker Springel for his {\Gadget} code and his feedback on the original draft. We thank Mark Vogelsberger and Federico Marinacci for their patience in explaining the feedback model in detail, Paul Torrey,  Andreas Bauer, Jim Stone, Steinn Sigurdsson, Yohan Dubroi and Evghenii Gaburov for their useful discussions on the various aspects related with numerical methods, and the anonymous referee for constructive comments that have helped improve the manuscript. YL acknowledges support from NSF grants AST-0965694, AST-1009867 and AST-1412719. We acknowledge the Institute For CyberScience at The Pennsylvania State University for providing computational resources and services that have contributed to the research results reported in this paper. The Institute for Gravitation and the Cosmos is supported by the Eberly College of Science and the Office of the Senior Vice President for Research at The Pennsylvania State University.


\begin{thebibliography}{}
\expandafter\ifx\csname natexlab\endcsname\relax\def\natexlab#1{#1}\fi

\bibitem[{{Agertz} {et~al.}(2013){Agertz}, {Kravtsov}, {Leitner}, \&
  {Gnedin}}]{Agertz2013}
{Agertz}, O., {Kravtsov}, A.~V., {Leitner}, S.~N., \& {Gnedin}, N.~Y. 2013,
  \apj, 770, 25

\bibitem[{{Agertz} {et~al.}(2011){Agertz}, {Teyssier}, \& {Moore}}]{Agertz2011}
{Agertz}, O., {Teyssier}, R., \& {Moore}, B. 2011, \mnras, 410, 1391

\bibitem[{{Agertz} {et~al.}(2007){Agertz}, {Moore}, {Stadel}, {Potter},
  {Miniati}, {Read}, {Mayer}, {Gawryszczak}, {Kravtsov}, {Nordlund}, {Pearce},
  {Quilis}, {Rudd}, {Springel}, {Stone}, {Tasker}, {Teyssier}, {Wadsley}, \&
  {Walder}}]{Agertz2007}
{Agertz}, O., {Moore}, B., {Stadel}, J., {et~al.} 2007, \mnras, 380, 963

\bibitem[{{Aumer} {et~al.}(2013){Aumer}, {White}, {Naab}, \&
  {Scannapieco}}]{Aumer2013}
{Aumer}, M., {White}, S.~D.~M., {Naab}, T., \& {Scannapieco}, C. 2013, \mnras,
  434, 3142

\bibitem[{{Bauer} \& {Springel}(2012)}]{Bauer2012}
{Bauer}, A., \& {Springel}, V. 2012, \mnras, 423, 2558

\bibitem[{{Bell} {et~al.}(2004){Bell}, {Wolf}, {Meisenheimer}, {Rix}, {Borch},
  {Dye}, {Kleinheinrich}, {Wisotzki}, \& {McIntosh}}]{Bell2004}
{Bell}, E.~F., {Wolf}, C., {Meisenheimer}, K., {et~al.} 2004, \apj, 608, 752

\bibitem[{{Blanton} {et~al.}(2003){Blanton}, {Hogg}, {Bahcall}, {Baldry},
  {Brinkmann}, {Csabai}, {Eisenstein}, {Fukugita}, {Gunn}, {Ivezi{\'c}},
  {Lamb}, {Lupton}, {Loveday}, {Munn}, {Nichol}, {Okamura}, {Schlegel},
  {Shimasaku}, {Strauss}, {Vogeley}, \& {Weinberg}}]{Blanton2003}
{Blanton}, M.~R., {Hogg}, D.~W., {Bahcall}, N.~A., {et~al.} 2003, \apj, 594,
  186

\bibitem[{{Boselli} {et~al.}(2014){Boselli}, {Cortese}, {Boquien}, {Boissier},
  {Catinella}, {Lagos}, \& {Saintonge}}]{Boselli2014}
{Boselli}, A., {Cortese}, L., {Boquien}, M., {et~al.} 2014, \aap, 564, A66

\bibitem[{{Brook} {et~al.}(2011){Brook}, {Governato}, {Ro{\v s}kar}, {Stinson},
  {Brooks}, {Wadsley}, {Quinn}, {Gibson}, {Snaith}, {Pilkington}, {House}, \&
  {Pontzen}}]{Brook2011}
{Brook}, C.~B., {Governato}, F., {Ro{\v s}kar}, R., {et~al.} 2011, \mnras, 415,
  1051

\bibitem[{{Courteau} {et~al.}(2007){Courteau}, {Dutton}, {van den Bosch},
  {MacArthur}, {Dekel}, {McIntosh}, \& {Dale}}]{Courteau2007}
{Courteau}, S., {Dutton}, A.~A., {van den Bosch}, F.~C., {et~al.} 2007, \apj,
  671, 203

\bibitem[{{Cui} {et~al.}(2016){Cui}, {Power}, {Knebe}, {Kay}, {Sembolini},
  {Elahi}, {Yepes}, {Pearce}, {Cunnama}, {Beck}, {Dalla Vecchia}, {Dav{\'e}},
  {February}, {Huang}, {Hobbs}, {Katz}, {McCarthy}, {Murante}, {Perret},
  {Puchwein}, {Read}, {Saro}, {Teyssier}, \& {Thacker}}]{Cui2016}
{Cui}, W., {Power}, C., {Knebe}, A., {et~al.} 2016, \mnras, 458, 4052

\bibitem[{{Cullen} \& {Dehnen}(2010)}]{Cullen2010}
{Cullen}, L., \& {Dehnen}, W. 2010, \mnras, 408, 669

\bibitem[{{Dehnen} \& {Aly}(2012)}]{Dehnen2012}
{Dehnen}, W., \& {Aly}, H. 2012, \mnras, 425, 1068

\bibitem[{{Di Matteo} {et~al.}(2005){Di Matteo}, {Springel}, \&
  {Hernquist}}]{DiMatteo2005}
{Di Matteo}, T., {Springel}, V., \& {Hernquist}, L. 2005, \nat, 433, 604

\bibitem[{{Diemand} {et~al.}(2008){Diemand}, {Kuhlen}, {Madau}, {Zemp},
  {Moore}, {Potter}, \& {Stadel}}]{Diemand2008}
{Diemand}, J., {Kuhlen}, M., {Madau}, P., {et~al.} 2008, \nat, 454, 735

\bibitem[{{Durier} \& {Dalla Vecchia}(2012)}]{Durier2012}
{Durier}, F., \& {Dalla Vecchia}, C. 2012, \mnras, 419, 465

\bibitem[{{Dutton} {et~al.}(2010){Dutton}, {Conroy}, {van den Bosch}, {Prada},
  \& {More}}]{Dutton2010}
{Dutton}, A.~A., {Conroy}, C., {van den Bosch}, F.~C., {Prada}, F., \& {More},
  S. 2010, \mnras, 407, 2

\bibitem[{{Elahi} {et~al.}(2016){Elahi}, {Knebe}, {Pearce}, {Power}, {Yepes},
  {Cui}, {Cunnama}, {Kay}, {Sembolini}, {Beck}, {Dav{\'e}}, {February},
  {Huang}, {Katz}, {McCarthy}, {Murante}, {Perret}, {Puchwein}, {Saro}, \&
  {Teyssier}}]{Elahi2016}
{Elahi}, P.~J., {Knebe}, A., {Pearce}, F.~R., {et~al.} 2016, \mnras, 458, 1096

\bibitem[{{Faber} {et~al.}(2007){Faber}, {Willmer}, {Wolf}, {Koo}, {Weiner},
  {Newman}, {Im}, {Coil}, {Conroy}, {Cooper}, {Davis}, {Finkbeiner}, {Gerke},
  {Gebhardt}, {Groth}, {Guhathakurta}, {Harker}, {Kaiser}, {Kassin},
  {Kleinheinrich}, {Konidaris}, {Kron}, {Lin}, {Luppino}, {Madgwick},
  {Meisenheimer}, {Noeske}, {Phillips}, {Sarajedini}, {Schiavon}, {Simard},
  {Szalay}, {Vogt}, \& {Yan}}]{Faber2007}
{Faber}, S.~M., {Willmer}, C.~N.~A., {Wolf}, C., {et~al.} 2007, \apj, 665, 265

\bibitem[{{Furlong} {et~al.}(2015){Furlong}, {Bower}, {Theuns}, {Schaye},
  {Crain}, {Schaller}, {Dalla Vecchia}, {Frenk}, {McCarthy}, {Helly},
  {Jenkins}, \& {Rosas-Guevara}}]{Furlong2015}
{Furlong}, M., {Bower}, R.~G., {Theuns}, T., {et~al.} 2015, \mnras, 450, 4486

\bibitem[{{Gaburov} \& {Nitadori}(2011)}]{Gaburov2010}
{Gaburov}, E., \& {Nitadori}, K. 2011, \mnras, 414, 129

\bibitem[{{Genel} {et~al.}(2014){Genel}, {Vogelsberger}, {Springel}, {Sijacki},
  {Nelson}, {Snyder}, {Rodriguez-Gomez}, {Torrey}, \& {Hernquist}}]{Genel2014}
{Genel}, S., {Vogelsberger}, M., {Springel}, V., {et~al.} 2014, \mnras, 445,
  175

\bibitem[{{Guedes} {et~al.}(2011){Guedes}, {Callegari}, {Madau}, \&
  {Mayer}}]{Guedes2011}
{Guedes}, J., {Callegari}, S., {Madau}, P., \& {Mayer}, L. 2011, \apj, 742, 76

\bibitem[{{Guo} {et~al.}(2010){Guo}, {White}, {Li}, \&
  {Boylan-Kolchin}}]{Guo2010}
{Guo}, Q., {White}, S., {Li}, C., \& {Boylan-Kolchin}, M. 2010, \mnras, 404,
  1111

\bibitem[{{Haynes} {et~al.}(1999){Haynes}, {Giovanelli}, {Chamaraux}, {da
  Costa}, {Freudling}, {Salzer}, \& {Wegner}}]{Haynes1999}
{Haynes}, M.~P., {Giovanelli}, R., {Chamaraux}, P., {et~al.} 1999, \aj, 117,
  2039

\bibitem[{{Hayward} {et~al.}(2014){Hayward}, {Torrey}, {Springel}, {Hernquist},
  \& {Vogelsberger}}]{Hayward2014}
{Hayward}, C.~C., {Torrey}, P., {Springel}, V., {Hernquist}, L., \&
  {Vogelsberger}, M. 2014, \mnras, 442, 1992

\bibitem[{{Hopkins}(2013)}]{Hopkins2013}
{Hopkins}, P.~F. 2013, \mnras, 428, 2840

\bibitem[{{Hopkins}(2015)}]{Hopkins2015}
---. 2015, \mnras, 450, 53

\bibitem[{{Hopkins} {et~al.}(2013){Hopkins}, {Cox}, {Hernquist}, {Narayanan},
  {Hayward}, \& {Murray}}]{Hopkins2013c}
{Hopkins}, P.~F., {Cox}, T.~J., {Hernquist}, L., {et~al.} 2013, \mnras, 430,
  1901

\bibitem[{{Hopkins} {et~al.}(2014){Hopkins}, {Kere{\v s}}, {O{\~n}orbe},
  {Faucher-Gigu{\`e}re}, {Quataert}, {Murray}, \& {Bullock}}]{Hopkins2013b}
{Hopkins}, P.~F., {Kere{\v s}}, D., {O{\~n}orbe}, J., {et~al.} 2014, \mnras,
  445, 581

\bibitem[{{Hu} {et~al.}(2014){Hu}, {Naab}, {Walch}, {Moster}, \&
  {Oser}}]{Hu2014}
{Hu}, C.-Y., {Naab}, T., {Walch}, S., {Moster}, B.~P., \& {Oser}, L. 2014,
  \mnras, 443, 1173

\bibitem[{{Hughes} {et~al.}(2013){Hughes}, {Cortese}, {Boselli}, {Gavazzi}, \&
  {Davies}}]{Hughes2013}
{Hughes}, T.~M., {Cortese}, L., {Boselli}, A., {Gavazzi}, G., \& {Davies},
  J.~I. 2013, \aap, 550, A115

\bibitem[{{Iwamoto} {et~al.}(1999){Iwamoto}, {Brachwitz}, {Nomoto},
  {Kishimoto}, {Umeda}, {Hix}, \& {Thielemann}}]{Iwamoto1999}
{Iwamoto}, K., {Brachwitz}, F., {Nomoto}, K., {et~al.} 1999, \apjs, 125, 439

\bibitem[{{Kauffmann} {et~al.}(2003){Kauffmann}, {Heckman}, {White}, {Charlot},
  {Tremonti}, {Peng}, {Seibert}, {Brinkmann}, {Nichol}, {SubbaRao}, \&
  {York}}]{Kauffmann2003}
{Kauffmann}, G., {Heckman}, T.~M., {White}, S.~D.~M., {et~al.} 2003, \mnras,
  341, 54

\bibitem[{{Kere{\v s}} {et~al.}(2012){Kere{\v s}}, {Vogelsberger}, {Sijacki},
  {Springel}, \& {Hernquist}}]{Keres2012}
{Kere{\v s}}, D., {Vogelsberger}, M., {Sijacki}, D., {Springel}, V., \&
  {Hernquist}, L. 2012, \mnras, 425, 2027

\bibitem[{{Kim} {et~al.}(2014){Kim}, {Abel}, {Agertz}, {Bryan}, {Ceverino},
  {Christensen}, {Conroy}, {Dekel}, {Gnedin}, {Goldbaum}, {Guedes}, {Hahn},
  {Hobbs}, {Hopkins}, {Hummels}, {Iannuzzi}, {Keres}, {Klypin}, {Kravtsov},
  {Krumholz}, {Kuhlen}, {Leitner}, {Madau}, {Mayer}, {Moody}, {Nagamine},
  {Norman}, {Onorbe}, {O'Shea}, {Pillepich}, {Primack}, {Quinn}, {Read},
  {Robertson}, {Rocha}, {Rudd}, {Shen}, {Smith}, {Szalay}, {Teyssier},
  {Thompson}, {Todoroki}, {Turk}, {Wadsley}, {Wise}, {Zolotov}, \& {AGORA
  Collaboration29}}]{Kim2014}
{Kim}, J.-h., {Abel}, T., {Agertz}, O., {et~al.} 2014, \apjs, 210, 14

\bibitem[{{Kuhlen} {et~al.}(2012){Kuhlen}, {Vogelsberger}, \&
  {Angulo}}]{Kuhlen2012}
{Kuhlen}, M., {Vogelsberger}, M., \& {Angulo}, R. 2012, Physics of the Dark
  Universe, 1, 50

\bibitem[{Lanson \& Vila(2008)}]{Lanson2008}
Lanson, N., \& Vila, J.-P. 2008, SIAM Journal on Numerical Analysis, 46, pp.
  1912

\bibitem[{{Leitherer} {et~al.}(1999){Leitherer}, {Schaerer}, {Goldader},
  {Delgado}, {Robert}, {Kune}, {de Mello}, {Devost}, \&
  {Heckman}}]{Leitherer1999}
{Leitherer}, C., {Schaerer}, D., {Goldader}, J.~D., {et~al.} 1999, \apjs, 123,
  3

\bibitem[{{Leitner} \& {Kravtsov}(2011)}]{Leitner2011}
{Leitner}, S.~N., \& {Kravtsov}, A.~V. 2011, \apj, 734, 48

\bibitem[{{Marinacci} {et~al.}(2014){Marinacci}, {Pakmor}, \&
  {Springel}}]{Marinacci2014}
{Marinacci}, F., {Pakmor}, R., \& {Springel}, V. 2014, \mnras, 437, 1750

\bibitem[{{Marri} \& {White}(2003)}]{Marri2003}
{Marri}, S., \& {White}, S.~D.~M. 2003, \mnras, 345, 561

\bibitem[{{Mocz} {et~al.}(2015){Mocz}, {Vogelsberger}, {Pakmor}, {Genel},
  {Springel}, \& {Hernquist}}]{Mocz2015}
{Mocz}, P., {Vogelsberger}, M., {Pakmor}, R., {et~al.} 2015, \mnras, 452, 3853

\bibitem[{{Moster} {et~al.}(2013){Moster}, {Naab}, \& {White}}]{Moster2013}
{Moster}, B.~P., {Naab}, T., \& {White}, S.~D.~M. 2013, \mnras, 428, 3121

\bibitem[{{Nelson} {et~al.}(2013){Nelson}, {Vogelsberger}, {Genel}, {Sijacki},
  {Kere{\v s}}, {Springel}, \& {Hernquist}}]{Nelson2013}
{Nelson}, D., {Vogelsberger}, M., {Genel}, S., {et~al.} 2013, \mnras, 429, 3353

\bibitem[{{Okamoto}(2013)}]{Okamoto2013}
{Okamoto}, T. 2013, \mnras, 428, 718

\bibitem[{{Okamoto} {et~al.}(2010){Okamoto}, {Frenk}, {Jenkins}, \&
  {Theuns}}]{Okamoto2010}
{Okamoto}, T., {Frenk}, C.~S., {Jenkins}, A., \& {Theuns}, T. 2010, \mnras,
  406, 208

\bibitem[{{Okamoto} {et~al.}(2003){Okamoto}, {Jenkins}, {Eke}, {Quilis}, \&
  {Frenk}}]{Okamoto2003}
{Okamoto}, T., {Jenkins}, A., {Eke}, V.~R., {Quilis}, V., \& {Frenk}, C.~S.
  2003, \mnras, 345, 429

\bibitem[{{Oliver} {et~al.}(2010){Oliver}, {Frost}, {Farrah},
  {Gonzalez-Solares}, {Shupe}, {Henriques}, {Roseboom}, {Alfonso-Luis},
  {Babbedge}, {Frayer}, {Lencz}, {Lonsdale}, {Masci}, {Padgett}, {Polletta},
  {Rowan-Robinson}, {Siana}, {Smith}, {Surace}, \& {Vaccari}}]{Oliver2010}
{Oliver}, S., {Frost}, M., {Farrah}, D., {et~al.} 2010, \mnras, 405, 2279

\bibitem[{{Owen} {et~al.}(1998){Owen}, {Villumsen}, {Shapiro}, \&
  {Martel}}]{Owen1998}
{Owen}, J.~M., {Villumsen}, J.~V., {Shapiro}, P.~R., \& {Martel}, H. 1998,
  \apjs, 116, 155

\bibitem[{{Pakmor} {et~al.}(2015){Pakmor}, {Springel}, {Bauer}, {Mocz},
  {Munoz}, {Ohlmann}, {Schaal}, \& {Zhu}}]{Pakmor2015}
{Pakmor}, R., {Springel}, V., {Bauer}, A., {et~al.} 2015, ArXiv e-prints,
  arXiv:1503.00562

\bibitem[{{Pizagno} {et~al.}(2007){Pizagno}, {Prada}, {Weinberg}, {Rix},
  {Pogge}, {Grebel}, {Harbeck}, {Blanton}, {Brinkmann}, \&
  {Gunn}}]{Pizagno2007}
{Pizagno}, J., {Prada}, F., {Weinberg}, D.~H., {et~al.} 2007, \aj, 134, 945

\bibitem[{{Price}(2008)}]{Price2008}
{Price}, D.~J. 2008, Journal of Computational Physics, 227, 10040

\bibitem[{{Price}(2012{\natexlab{a}})}]{Price2012turbulence}
---. 2012{\natexlab{a}}, \mnras, 420, L33

\bibitem[{{Price}(2012{\natexlab{b}})}]{Price2012review}
---. 2012{\natexlab{b}}, Journal of Computational Physics, 231, 759

\bibitem[{{Puchwein} \& {Springel}(2013)}]{Puchwein2013}
{Puchwein}, E., \& {Springel}, V. 2013, \mnras, 428, 2966

\bibitem[{{Rasio}(2000)}]{Rasio2000}
{Rasio}, F.~A. 2000, Progress of Theoretical Physics Supplement, 138, 609

\bibitem[{{Read} {et~al.}(2010){Read}, {Hayfield}, \& {Agertz}}]{Read2010}
{Read}, J.~I., {Hayfield}, T., \& {Agertz}, O. 2010, \mnras, 405, 1513

\bibitem[{{Ritchie} \& {Thomas}(2001)}]{Ritchie2001}
{Ritchie}, B.~W., \& {Thomas}, P.~A. 2001, \mnras, 323, 743

\bibitem[{{Saitoh} \& {Makino}(2009)}]{Saitoh2009}
{Saitoh}, T.~R., \& {Makino}, J. 2009, \apjl, 697, L99

\bibitem[{{Saitoh} \& {Makino}(2013)}]{Saitoh2013}
---. 2013, \apj, 768, 44

\bibitem[{{Sawala} {et~al.}(2012){Sawala}, {Scannapieco}, \&
  {White}}]{Sawala2012}
{Sawala}, T., {Scannapieco}, C., \& {White}, S. 2012, \mnras, 420, 1714

\bibitem[{{Scannapieco} {et~al.}(2006){Scannapieco}, {Tissera}, {White}, \&
  {Springel}}]{Scannapieco2006}
{Scannapieco}, C., {Tissera}, P.~B., {White}, S.~D.~M., \& {Springel}, V. 2006,
  \mnras, 371, 1125

\bibitem[{{Scannapieco} {et~al.}(2012){Scannapieco}, {Wadepuhl}, {Parry},
  {Navarro}, {Jenkins}, {Springel}, {Teyssier}, {Carlson}, {Couchman}, {Crain},
  {Dalla Vecchia}, {Frenk}, {Kobayashi}, {Monaco}, {Murante}, {Okamoto},
  {Quinn}, {Schaye}, {Stinson}, {Theuns}, {Wadsley}, {White}, \&
  {Woods}}]{Scannapieco2012}
{Scannapieco}, C., {Wadepuhl}, M., {Parry}, O.~H., {et~al.} 2012, \mnras, 423,
  1726

\bibitem[{{Schaller} {et~al.}(2015){Schaller}, {Dalla Vecchia}, {Schaye},
  {Bower}, {Theuns}, {Crain}, {Furlong}, \& {McCarthy}}]{Schaller2015}
{Schaller}, M., {Dalla Vecchia}, C., {Schaye}, J., {et~al.} 2015, \mnras, 454,
  2277

\bibitem[{{Schawinski} {et~al.}(2007){Schawinski}, {Thomas}, {Sarzi},
  {Maraston}, {Kaviraj}, {Joo}, {Yi}, \& {Silk}}]{Schawinski2007}
{Schawinski}, K., {Thomas}, D., {Sarzi}, M., {et~al.} 2007, \mnras, 382, 1415

\bibitem[{{Schaye} {et~al.}(2015){Schaye}, {Crain}, {Bower}, {Furlong},
  {Schaller}, {Theuns}, {Dalla Vecchia}, {Frenk}, {McCarthy}, {Helly},
  {Jenkins}, {Rosas-Guevara}, {White}, {Baes}, {Booth}, {Camps}, {Navarro},
  {Qu}, {Rahmati}, {Sawala}, {Thomas}, \& {Trayford}}]{Schaye2015}
{Schaye}, J., {Crain}, R.~A., {Bower}, R.~G., {et~al.} 2015, \mnras, 446, 521

\bibitem[{{Sembolini} {et~al.}(2016){Sembolini}, {Yepes}, {Pearce}, {Knebe},
  {Kay}, {Power}, {Cui}, {Beck}, {Borgani}, {Dalla Vecchia}, {Dav{\'e}},
  {Elahi}, {February}, {Huang}, {Hobbs}, {Katz}, {Lau}, {McCarthy}, {Murante},
  {Nagai}, {Nelson}, {Newton}, {Perret}, {Puchwein}, {Read}, {Saro}, {Schaye},
  {Teyssier}, \& {Thacker}}]{Sembolini2016}
{Sembolini}, F., {Yepes}, G., {Pearce}, F.~R., {et~al.} 2016, \mnras, 457, 4063

\bibitem[{{Shen} {et~al.}(2010){Shen}, {Wadsley}, \& {Stinson}}]{Shen2010}
{Shen}, S., {Wadsley}, J., \& {Stinson}, G. 2010, \mnras, 407, 1581

\bibitem[{{Sijacki} {et~al.}(2012){Sijacki}, {Vogelsberger}, {Kere{\v s}},
  {Springel}, \& {Hernquist}}]{Sijacki2012}
{Sijacki}, D., {Vogelsberger}, M., {Kere{\v s}}, D., {Springel}, V., \&
  {Hernquist}, L. 2012, \mnras, 424, 2999

\bibitem[{{Silk} \& {Rees}(1998)}]{Silk1998}
{Silk}, J., \& {Rees}, M.~J. 1998, \aap, 331, L1

\bibitem[{{Sofue} {et~al.}(2009){Sofue}, {Honma}, \& {Omodaka}}]{Sofue2009}
{Sofue}, Y., {Honma}, M., \& {Omodaka}, T. 2009, \pasj, 61, 227

\bibitem[{{Sparre} {et~al.}(2015){Sparre}, {Hayward}, {Springel},
  {Vogelsberger}, {Genel}, {Torrey}, {Nelson}, {Sijacki}, \&
  {Hernquist}}]{Sparre2015}
{Sparre}, M., {Hayward}, C.~C., {Springel}, V., {et~al.} 2015, \mnras, 447,
  3548

\bibitem[{{Springel}(2005)}]{Gadget2}
{Springel}, V. 2005, \mnras, 364, 1105

\bibitem[{{Springel}(2010{\natexlab{a}})}]{Springel2010arepo}
---. 2010{\natexlab{a}}, \mnras, 401, 791

\bibitem[{{Springel}(2010{\natexlab{b}})}]{Springel2010review}
---. 2010{\natexlab{b}}, \araa, 48, 391

\bibitem[{{Springel}(2012)}]{Springel2012}
---. 2012, Astronomische Nachrichten, 333, 515

\bibitem[{{Springel} {et~al.}(2005){Springel}, {Di Matteo}, \&
  {Hernquist}}]{Springel2005}
{Springel}, V., {Di Matteo}, T., \& {Hernquist}, L. 2005, \mnras, 361, 776

\bibitem[{{Springel} \& {Hernquist}(2003)}]{Springel2003}
{Springel}, V., \& {Hernquist}, L. 2003, \mnras, 339, 289

\bibitem[{{Springel} {et~al.}(2001){Springel}, {Yoshida}, \& {White}}]{Gadget1}
{Springel}, V., {Yoshida}, N., \& {White}, S.~D.~M. 2001, \na, 6, 79

\bibitem[{{Springel} {et~al.}(2008){Springel}, {Wang}, {Vogelsberger},
  {Ludlow}, {Jenkins}, {Helmi}, {Navarro}, {Frenk}, \& {White}}]{Springel2008}
{Springel}, V., {Wang}, J., {Vogelsberger}, M., {et~al.} 2008, \mnras, 391,
  1685

\bibitem[{{Stadel} {et~al.}(2009){Stadel}, {Potter}, {Moore}, {Diemand},
  {Madau}, {Zemp}, {Kuhlen}, \& {Quilis}}]{Stadel2009}
{Stadel}, J., {Potter}, D., {Moore}, B., {et~al.} 2009, \mnras, 398, L21

\bibitem[{{Stinson} {et~al.}(2013){Stinson}, {Brook}, {Macci{\`o}}, {Wadsley},
  {Quinn}, \& {Couchman}}]{Stinson2013a}
{Stinson}, G.~S., {Brook}, C., {Macci{\`o}}, A.~V., {et~al.} 2013, \mnras, 428,
  129

\bibitem[{{Strateva} {et~al.}(2001){Strateva}, {Ivezi{\'c}}, {Knapp},
  {Narayanan}, {Strauss}, {Gunn}, {Lupton}, {Schlegel}, {Bahcall}, {Brinkmann},
  {Brunner}, {Budav{\'a}ri}, {Csabai}, {Castander}, {Doi}, {Fukugita}, {Gy{\H
  o}ry}, {Hamabe}, {Hennessy}, {Ichikawa}, {Kunszt}, {Lamb}, {McKay},
  {Okamura}, {Racusin}, {Sekiguchi}, {Schneider}, {Shimasaku}, \&
  {York}}]{Strateva2001}
{Strateva}, I., {Ivezi{\'c}}, {\v Z}., {Knapp}, G.~R., {et~al.} 2001, \aj, 122,
  1861

\bibitem[{{Torrey} {et~al.}(2012){Torrey}, {Vogelsberger}, {Sijacki},
  {Springel}, \& {Hernquist}}]{Torrey2012}
{Torrey}, P., {Vogelsberger}, M., {Sijacki}, D., {Springel}, V., \&
  {Hernquist}, L. 2012, \mnras, 427, 2224

\bibitem[{{Trayford} {et~al.}(2015){Trayford}, {Theuns}, {Bower}, {Schaye},
  {Furlong}, {Schaller}, {Frenk}, {Crain}, {Dalla Vecchia}, \&
  {McCarthy}}]{Trayford2015}
{Trayford}, J.~W., {Theuns}, T., {Bower}, R.~G., {et~al.} 2015, \mnras, 452,
  2879

\bibitem[{{Tully} \& {Fisher}(1977)}]{Tully1977}
{Tully}, R.~B., \& {Fisher}, J.~R. 1977, \aap, 54, 661

\bibitem[{{{\"U}bler} {et~al.}(2014){{\"U}bler}, {Naab}, {Oser}, {Aumer},
  {Sales}, \& {White}}]{Ubler2014}
{{\"U}bler}, H., {Naab}, T., {Oser}, L., {et~al.} 2014, \mnras, 443, 2092

\bibitem[{{Verheijen}(2001)}]{Verheijen2001}
{Verheijen}, M.~A.~W. 2001, \apj, 563, 694

\bibitem[{{Vogelsberger} {et~al.}(2013){Vogelsberger}, {Genel}, {Sijacki},
  {Torrey}, {Springel}, \& {Hernquist}}]{Vogelsberger2013}
{Vogelsberger}, M., {Genel}, S., {Sijacki}, D., {et~al.} 2013, \mnras, 436,
  3031

\bibitem[{{Vogelsberger} {et~al.}(2012){Vogelsberger}, {Sijacki}, {Kere{\v s}},
  {Springel}, \& {Hernquist}}]{Vogelsberger2012}
{Vogelsberger}, M., {Sijacki}, D., {Kere{\v s}}, D., {Springel}, V., \&
  {Hernquist}, L. 2012, \mnras, 425, 3024

\bibitem[{{Vogelsberger} {et~al.}(2014){Vogelsberger}, {Genel}, {Springel},
  {Torrey}, {Sijacki}, {Xu}, {Snyder}, {Nelson}, \&
  {Hernquist}}]{Vogelsberger2014b}
{Vogelsberger}, M., {Genel}, S., {Springel}, V., {et~al.} 2014, ArXiv e-prints,
  arXiv:1405.2921

\bibitem[{{Wadepuhl} \& {Springel}(2011)}]{Wadepuhl2011}
{Wadepuhl}, M., \& {Springel}, V. 2011, \mnras, 410, 1975

\bibitem[{{Wadsley} {et~al.}(2008){Wadsley}, {Veeravalli}, \&
  {Couchman}}]{Wadsley2008}
{Wadsley}, J.~W., {Veeravalli}, G., \& {Couchman}, H.~M.~P. 2008, \mnras, 387,
  427

\bibitem[{{Wetzel} {et~al.}(2016){Wetzel}, {Hopkins}, {Kim}, {Faucher-Giguere},
  {Keres}, \& {Quataert}}]{Wetzel2016}
{Wetzel}, A.~R., {Hopkins}, P.~F., {Kim}, J.-h., {et~al.} 2016, ArXiv e-prints,
  arXiv:1602.05957

\bibitem[{{Wiersma} {et~al.}(2009{\natexlab{a}}){Wiersma}, {Schaye}, \&
  {Smith}}]{Wiersma2009}
{Wiersma}, R.~P.~C., {Schaye}, J., \& {Smith}, B.~D. 2009{\natexlab{a}},
  \mnras, 393, 99

\bibitem[{{Wiersma} {et~al.}(2009{\natexlab{b}}){Wiersma}, {Schaye}, {Theuns},
  {Dalla Vecchia}, \& {Tornatore}}]{Wiersma2009b}
{Wiersma}, R.~P.~C., {Schaye}, J., {Theuns}, T., {Dalla Vecchia}, C., \&
  {Tornatore}, L. 2009{\natexlab{b}}, \mnras, 399, 574

\bibitem[{{Willett} {et~al.}(2013){Willett}, {Lintott}, {Bamford}, {Masters},
  {Simmons}, {Casteels}, {Edmondson}, {Fortson}, {Kaviraj}, {Keel}, {Melvin},
  {Nichol}, {Raddick}, {Schawinski}, {Simpson}, {Skibba}, {Smith}, \&
  {Thomas}}]{Willett2013}
{Willett}, K.~W., {Lintott}, C.~J., {Bamford}, S.~P., {et~al.} 2013, \mnras,
  435, 2835

\bibitem[{{Zhu} {et~al.}(2015){Zhu}, {Hernquist}, \& {Li}}]{Zhu2015}
{Zhu}, Q., {Hernquist}, L., \& {Li}, Y. 2015, \apj, 800, 6

\end{thebibliography}

\end{document}